\documentclass[10pt]{article}
\usepackage{amsmath,amsfonts,amssymb,amsthm,bbm}
\usepackage{pdfsync}
\usepackage{graphicx}
\usepackage[latin1]{inputenc}
\usepackage{hyperref}
\usepackage[all]{xy}
\usepackage{stmaryrd}

\newcommand{\bit}{\begin{itemize}}
\newcommand{\eit}{\end{itemize}}
\newcommand{\bd}{\begin{description}}
\newcommand{\ed}{\end{description}}

\newcommand{\bc}{\begin{center}}
\newcommand{\ec}{\end{center}}
\newcommand{\Ref}[1]{(\ref{#1})}

\newcommand{\C}{{\mathbb C}}
\newcommand{\N}{{\mathbb N}}
\newcommand{\R}{{\mathbb R}}
\newcommand{\Z}{{\mathbb Z}}

\newcommand{\SU}{\mathrm{SU}}

\newcommand{\SL}{\mathrm{SL}}

\newcommand{\be}{\begin{equation}}
\newcommand{\ee}{\end{equation}}
\newcommand{\bea}{\begin{eqnarray}}
\newcommand{\eea}{\end{eqnarray}}
\newcommand{\bs}{\begin{subequations}}
\newcommand{\es}{\end{subequations}}
\newcommand{\nn}{\nonumber}

\newcommand{\f}{\frac}
\newcommand{\tl}{\tilde}

\newcommand{\ra}{\rangle}

\newcommand{\bra}[1]{\langle {#1}|}
\newcommand{\ket}[1]{|{#1}\rangle}

\renewcommand{\a}{\alpha} \renewcommand{\b}{\beta} \newcommand{\g}{\gamma}
\renewcommand{\d}{\delta}  \newcommand{\eps}{\epsilon}  \newcommand{\z}{\zeta}
       
\let\m=\mu    \renewcommand{\r}{\rho} \newcommand{\s}{\sigma}       \let\om=\omega
\let\G=\Gamma \let\D=\Delta

\newcommand{\Wthree}[6]{\left(\begin{array}{ccc} #1 & #2 & #3 \\ #4 & #5 & #6 \end{array}\right)}
\newcommand{\Wfour}[9]{\left(\begin{array}{cccc} #1 & #2 & #3 & #4 \\ #5 & #6 & #7 & #8 \end{array}\right)^{(#9)}}

\begin{document}

\title{\bf Boosting Wigner's $nj$-symbols}

\author{\Large{Simone Speziale}
\smallskip \\ \small{Aix Marseille Univ., Univ. de Toulon, CNRS, CPT, UMR 7332, 13288 Marseille, France}
}
\date{v1: September 7, 2016. v2: December 8, 2016}

\maketitle

\begin{abstract}
\noindent 
We study the $\SL(2,\C)$ Clebsch-Gordan coefficients appearing in the lorentzian EPRL spin foam amplitudes for loop quantum gravity. We show how the amplitudes decompose into SU(2) $nj-$symbols at the vertices and integrals over boosts at the edges. The integrals define edge amplitudes that can be evaluated analytically using and adapting results in the literature, leading to a pure state sum model formulation. This procedure introduces virtual representations which, in a manner reminiscent to virtual momenta in Feynman amplitudes, are off-shell of the simplicity constraints present in the theory, but with the integrands that peak at the on-shell values.
We point out some properties of the edge amplitudes which are helpful for numerical and analytical evaluations of spin foam amplitudes, and suggest among other things a simpler model useful for calculations of certain lowest order amplitudes.
As an application, we estimate the large spin scaling behaviour of the simpler model, on a closed foam with all 4-valent edges and Euler characteristic $\chi$, to be $N^{\chi - 5E+V/2}$. The paper contains a review and an extension of results on $\SL(2,\C)$ Clebsch-Gordan coefficients among unitary representations of the principal series that can be useful beyond their application to quantum gravity considered here.

\end{abstract}

\vspace{.3cm}

{\footnotesize{\emph{Key words:} spin foam models, loop quantum gravity, Clebsch-Gordan coefficients, SL(2,C) unitary irreps}}

\vspace{.5cm}

\begin{flushright}
\emph{Dedicated to Carlo Rovelli for his 60th birthday}
\end{flushright}

\tableofcontents

\section{Introduction}

A compelling approach to the dynamics of loop quantum gravity is the spin foam formalism (for reviews, see \cite{PerezLR,RovelliVidotto}); It defines transition amplitudes for the spin network states of the canonical theory in the form of a sum over histories of quantum geometries, providing a regularised version of the quantum gravity path integral. The state of the art is the model proposed by Engle, Pereira, Rovelli and Livine (EPRL) \cite{EPRL} (see also \cite{EPR,LS,LS2,FK}), notably for its semiclassical properties \cite{BarrettEPRAsymp,BarrettLorAsymp} and the fact that it provides transition amplitudes for all possible spin networks \cite{KKL,CarloGenSF}.
Spin foam models in general have the mathematical structure of a gauge theory on an arbitrary lattice, which represents a discretisation of spacetime.\footnote{The continuum limit has then to be reached either via refining (e.g. \cite{RovelliMagic}), which may require renormalisation, see e.g. \cite{Dittrich:2014mxa,Dittrich:2014ala,BahrSteinhaus15} for work in this direction, or via a resummation over the 2-complexes, defined for instance using group field theory/random tensor models, see e.g. \cite{Bonzom:2012hw,Benedetti:2011nn,Carrozza:2013wda,Geloun:2016qyb}.} The key object is however not the plaquette, but the vertex amplitude. This is defined by integrals of tensor products of irreducible representations over the gauge group. For Euclidean signature, the relevant group is SU(2), and these integrals can be evaluated in terms of its well-known Clebsch-Gordan coefficients. For Lorentzian signature, the group is $\SL(2,\C)$, and the expression of the amplitudes in terms of $\SL(2,\C)$ Clebsch-Gordan coefficients has not appeared yet, perhaps due to the fact that the latter are less known. 
Filling this gap is the main goal of this paper. In doing so we review and extend existing results on $\SL(2,\C)$ Clebsch-Gordan coefficients, and provide many analytical and numerical insights on the amplitudes, which we hope will be of use for the notoriously difficult problem of explicitly computing physical processes such as \cite{IoCarloTunnelling16}.

The 
$\SL(2,\C)$ Clebsch-Gordan coefficients we are interested in are those among the unitary, infinite-dimensional irreducible representations (irreps) of the principal series. These have been studied at length in the mathematical literature, starting from the seminal work of Naimark \cite{Naimark}, and their formal properties are summarised in the reference monograph by Ruhl \cite{Ruhl}. Explicit values, symmetry properties and recurrence relations have been studied by a number of authors, e.g. \cite{Rashid70I,Rashid70II,Kerimov,Kerimov75} which are the ones most relevant to our work. The literature is rather uniform insofar as the norm of the coefficients is concerned, less so for the phase. Clarifying the phase differences and amending at places existing results was part of the work for the present paper. In particular, we show following \cite{Kerimov,Kerimov75} how to fix phase conventions such that all Clebsch-Gordan's are real, and thus also all invariant tensors, which is practical for applications and numerical studies (These conventions are, unfortunately, not those of \cite{Ruhl}, which lead to complex Clebsch-Gordan's, and differ also from those of \cite{Rashid70I,Rashid70II}, which lead to Clebsch-Gordan's and invariant tensors which are either real or purely imaginary).

The crucial property of the  Clebsch-Gordan's for $\SL(2,\C)$ is that 
they can be decomposed in terms of those for SU(2), times coefficients defined by integrations along a single boost direction. The same property applies to tensor product and graph invariants, and shows up in the EPRL model: it is possible to 
 factorise its quantum amplitude in terms of SU(2) $nj$-symbols at the vertices, `boosted' by edges amplitudes carrying the non-compact integrations over the rapidities $r$'s. 
The qualitative structure of this factorisation is probably already familiar to experts in spin foams (it follows directly using Cartan's decomposition of $\SL(2,\C)$), however going into the details here allows us to define a number of relevant objects and to highlight a few important properties that are new. In particular, the factorisation suggests the definition of a simplified model, in which only the intertwiner labels and not the spins are boosted. 
The simplified model is significantly faster to evaluate numerically, and turns out to provide a useful first order approximation for certain large spin regimes.

Next, we show how the boost integrals defining the edge amplitudes can be performed exactly in terms of finite sums of Gamma functions, using and amending a key result by Kerimov and Verdier \cite{Kerimov}. Together with the factorisation, this leads to a pure state sum formulation of the model, in which all the group integrals have been completely eliminated. When intertwiners are present, the application of the finite sums formula requires a recoupling scheme and introduces a new feature in the EPRL model, namely virtual Lorentz irreps off-shell of the simplicity constraints. Remarkably, the amplitudes are strongly peaked at values of the virtual irreps satisfying the constraints. 

To complete the analysis, we study numerically the edge amplitudes in the special case of 3-valent (no intertwiners) and 4-valent (single intertwiner) cases, to identify the scaling behaviours. Our investigations show a few characteristic features. First, the amplitudes are generically peaked on the minimal spin configurations used in the simplified model, thus supporting its relevance. 
The peakedness is power-law in the shift away from the minimal configuration, and its details depend on the actual irrep considered. 
We also investigate the large spin scaling. For the minimal configurations, we find (oscillating) power laws with a universal scaling $N^{-3/2}$ (except for degenerate cases in the 3-valent amplitude) and peakedness on diagonal intertwiner labels, confirming the results of \cite{Puchta:2013lza} based on a saddle point analysis. For non minimal configurations, we find faster power laws or exponential decays, as well as peakedness on non-equal intertwiners.
On general grounds, the various peakedness properties tend to be sharper for smaller values of the Immirzi parameter $\gamma$.

As a simple application of our results, in the final Section \ref{SecSimplified} we use the factorisation and the scaling properties to estimate the scaling behaviour of the simplified model in two examples, obtaining $N^{\chi - 5/2E+V/2}$ for a foam with only 3-valent edges, and $N^{\chi - 5E+V/2}$ for a foam with all edges 4-valent.

Although mainly motivated by their applications to quantum gravity, our review and extension of results on $\SL(2,\C)$ Clebsch-Gordan coefficients are general and can be of interest beyond the applications considered here. In particular, Sections 3,4 and 5 and the Appendices can be read without any reference to spin foam models.

\section{EPRL spin foam amplitudes and their factorisation}\label{SecEPRL}

We assume that the reader is familiar with the EPRL model, and refer to the original literature \cite{EPRL,LS,LS2,FK} and existing reviews (e.g.  \cite{PerezLR,RovelliVidotto}) for motivations, details and its relation to loop quantum gravity. The only technical aspect that we need to recall here is the interplay between SU(2) and $\SL(2,\C)$, which plays a major role. While the local semi-simple gauge group of general relativity is the non-compact Lorentz group, the use of real Ashtekar-Barbero introduces an auxiliary group SU(2), embedded non-trivially in $\SL(2,\C)$. The embedding is determined by the simplicity constraints relating general relativity to the topological BF theory, and depends on the Immirzi parameter $\g$. 
The Lorentzian EPRL model provides an implementation of this set-up at the non-perturbative quantum level: 
The partition function of the theory is defined via a summation over SU(2) spins only, as typical in spin foam models, 
but the SU(2) labels are non-trivially embedded in the unitary irreducible representations of the Lorentz group.

\subsection{Definition of the model}

The partition function of the EPRL model on a closed 2-complex $\cal C$ is a state sum over $\SU(2)$ spins $j_f$ and intertwiners $i_e$, associated respectively with faces $f$ and edges $e$ of the 2-complex,
\be\label{Z}
Z^{EPRL}_{\cal C} = \sum_{\{j_f,i_e\}} \prod_f d_{j_f} \prod_e d_{i_e} \prod_v A_v(j_f,i_e).
\ee
Here $d_{j}:= 2j+1$, the face weights are chosen requiring the convolution property of the path integral at fixed boundary graph, and the edge weights can be reabsorbed in the vertex amplitude, so that the expression in which they do not appear is also found in the literature.
In the following our main object of interest will be the vertex amplitude $A_v$, 
constructed from unitary irreducible representations of the principal series. 
These are labelled by a pair 
$(\r \in \R,k\in \N/2)$, 
with canonical basis chosen to diagonalise the operators $L^2$ and $L_z$ of the rotation subgroup $SU(2)$ of $\SL(2,\C)$ \cite{Naimark,Ruhl}. The basis vectors thus read $\ket{\r,k;j,m}$, and the group elements $h$ are represented by the infinite dimensional unitary matrices
\be\label{Dh}
D^{(\r,k)}_{jmln}(h), \qquad (j,l)\geq |k|, \quad -j\leq m \leq j, \quad -l\leq n \leq l, \qquad h\in\SL(2,\C).
\ee
Notice that because of the symmetries of these matrices, see Appendix \ref{AppBoost}, attention can often be restricted to positive labels. 
Among these irreps, the EPRL model selects a special class satisfying the following conditions,
\be\label{simple}
\r = \g k, \qquad k=j.
\ee
This crucial restriction on the labels, denoted $Y$-map in \cite{EPRL}, is what implements the primary simplicity constraints linking general relativity to topological BF theory. Without these constraints, the vertex amplitude of the spin foam model would correspond to $\SL(2,\C)$ BF theory. 

A brief comment on the constraints, referring the interest reader to the cited literature for details on the origin and geometrical meaning of the constraints. The first restriction is Lorentz invariant; the second one is not, as it relates the Lorentz-invariant irrep label $k$ to the matrix element label $j$ invariant only under the canonical SU(2) subgroup. 
This fixed `time gauge' formalism is convenient to define the quantum theory, 
and the amplitudes have been shown to be perfectly gauge-invariant \cite{IoCarloCov,IoTwistorNet}. 
The restricted matrix elements satisfying \Ref{simple} are the discrete equivalent of the non-trivial embedding of the Ashtekar-Barbero SU(2) connection in $\SL(2,\C)$ variables, as mentioned above. See e.g. \cite{IoMiklos} for a recent detailed discussion of this point. 

Then, to write $A_v$ in a compact form for a $N$-valent vertex, it is customary to consider the boundary graph to the vertex, and denote by $a,b=1,\ldots, N$ its nodes, dual to the edges at the vertex, and by $ab$ its links, dual to the faces; so that spins are associated with links $ab$ and the 2-complex orientation can be used to distinguish the source magnetic numbers $m_{ab}$ and the target ones $m_{ba}$. The vertex amplitude in the magnetic index basis is\footnote{Another common representation uses the holonomy basis, see \cite{RovelliVidotto}.}
\be\label{Av1}
A_v[j_{ab}, m_{ab},m_{ba}] := 
\int \prod_{a=1}^{N-1} dh_a \prod_{(ab)} D^{(\g j_f,j_f)}_{j_f m_{ab} j_f m_{ba}}(h^{-1}_a h_b),
\ee
with the integrations being over the group manifold $\SL(2,\C)$, and the simplicity constraints \Ref{simple} are imposed on each face of the 2-complex.
The key property of the model is that for a 4-simplex vertex graph, a saddle point approximation of the above integrals at large spins $j_f$ gives exponentials of the Regge action \cite{BarrettEPRAsymp,BarrettLorAsymp}, thus supporting the geometric interpretation of the model as a sampling of path integral of quantum geometries.
 The quantum numbers being summed over describe fuzzy polyhedra, each dual to a half-edge and forming the boundary of a flat polytope dual to the vertex; $h_a^{-1}h_b$ carries the extrinsic curvature of this boundary, and the $Y$ map imposes that the polyhedron shared by two adjacent polytopes lives in the same space-like hyperplane.
 Thanks to this property, the intrinsic curvature of the discrete spacetime can be described via a deficit angle \`a la Regge.
The geometric interpretation has been further investigated in a number of papers (e.g. \cite{Bonzom:2009wm,BaratinOriti,EteraHoloEucl,HanZhangLor,HellmannFlatness,Thiemann:2013lka,BahrOperatorSF,Engle:2015zqa,Wieland:2014nka}), and it is in our biased opinion much clarified using twistors and twisted geometries \cite{twigeo,IoWolfgang,IoMiklos}.

The original model was defined for the special case of a 2-complex dual to a triangulation of spacetime with flat 4-simplices \cite{EPRL}, and later generalised to an arbitrary 2-complex \cite{CarloGenSF,KKL,BahrOperatorSF}.\footnote{And also to constantly curved 4-simplices via the use of quantum groups, e.g. \cite{Haggard:2015yda}.} Care is needed however in considering general 2-complexes: first of all, arbitrary 2-complexes can correspond to very singular manifolds, whose relevance for quantum gravity is debatable; Secondly, not all graphs are integrable, because of the unboundness of the group integrals, unlike in Euclidean models. For the amplitude to be well-defined, one has to eliminate a redundant $\SL(2,\C)$ integration per vertex, which is the reason why in the above formula we only have $N-1$ integrals; And even so, some graphs lead to divergent amplitudes and have thus to be excluded in the definition of the generalised model. A sufficient condition for integrability, as argued in \cite{Baez:2001fh} and proved in \cite{Kaminski:2010qb}, is for the graph to be 3-link-connected, meaning that any partition of the nodes of the graph can not be disjointed by cutting only two links.\footnote{The integrability condition is the same for both the EPRL and its predecessor the Barrett-Crane model \cite{BarrettCraneLor}. The two differ in fact only in the restriction of the labels, which does not affect the leading divergent behaviour in the integrals.}

In dealing with the model, it is practical to use a graphical notation, where each representation matrix is represented by an oriented line, with rows associated to the end point and columns associated to the starting point.\footnote{This is the convention associated with action from the right, and it is the one used in the twistorial papers. The opposite convention, rows on source and columns on target can also be commonly found and corresponds to the action from the left.} Each edge of the foam is split into a multi-strand line according to its valency (i.e. the number of faces sharing that edge), and a vertex is identified by the intersection of the edges; a squared box indicates an integration over $\SL(2,\C)$, and a blue thin and filled box a $Y$ map. To give an explicit example, for a vertex $\s$ dual to a 4-simplex, we have (neglecting the orientation of the links not to clutter the picture)
\be
A_\s[j_{ab}, m_{ab}] =
 \parbox[2cm]{1.8cm}{\begin{picture}(0,55) (0,0) \includegraphics[width=2cm]{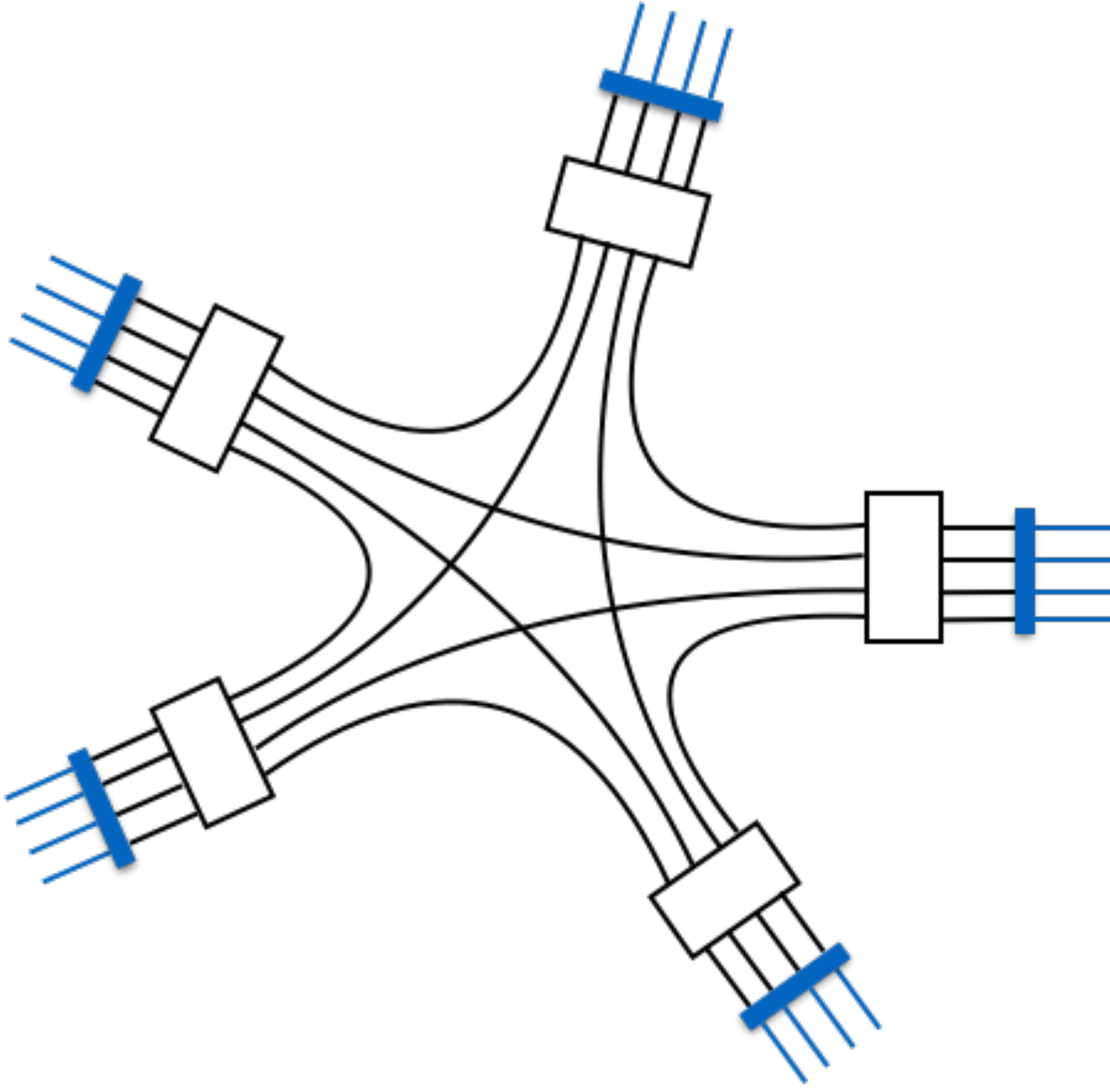} \end{picture}} \hspace{.3cm}.
\ee
The outgoing blue lines represent SU(2) magnetic indices $m_i$, which are glued to the next vertex amplitude and summed over. The integration over $\SL(2,\C)$ induces SU(2)-gauge-invariance of the boundary, therefore the magnetic indices can be contracted for free with Wigner's generalised $3jm$-symbols, giving the vertex amplitude in the form 
\be\label{As}
A_\s(j_f,i_e) =  \parbox[2cm]{1.8cm}{\begin{picture}(0,55) (0,0) \includegraphics[width=2cm]{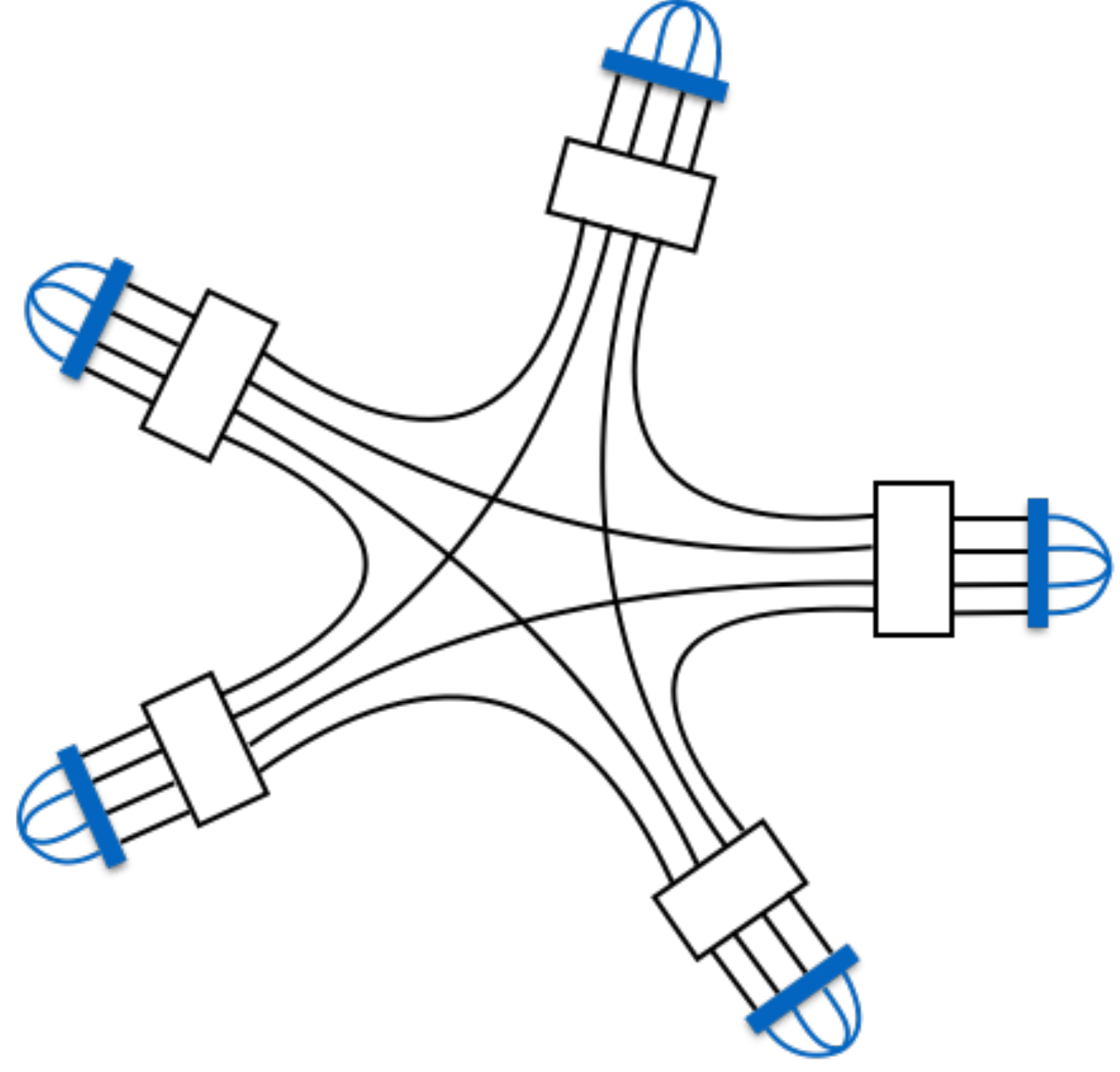} \end{picture}} \hspace{.3cm}.
\ee
In this way, the summations over magnetic indices at each edge can be replaced by summations over intertwiners, once we have chosen a basis for each recoupling, and one obtains the state sum formula \Ref{Z}.
When boundaries are present, there is a single integration at the boundary edge, and spins and intertwiners of faces and edges at the boundary are not summed over, but become the spins and intertwiners of links and nodes of the boundary EPRL projected spin networks \cite{DupuisLivine,IoCarloCov}.

Before concluding this brief review, notice that the $Y$-map is present only on one side of each group element in \Ref{Av1}, the one reaching out to the next vertex:
the two group elements joining at the vertex are instead multiplied together without the $Y$ map, a property that should be clear in the graphical notation. Accordingly, the infinite-dimensional matrix product $h^{-1}_a h_b$ contains an infinite summation over SU(2) spins (which are just like magnetic numbers from the perspective of the infinite-dimensional $\SL(2,\C)$ irrep), 
\be\label{Dsplit}
D^{(\g j,j)}_{j m j m'}(h^{-1}_a h_b) = 
\sum_{l\geq j} \sum_{n=-l}^l D^{(\g j,j)}_{j m l n}(h^{-1}_a) D^{(\g j,j)}_{l n j m'}( h_b).
\ee
If we make these summations explicit at each vertex, the partition function will have as many new spins $l_{f_v}$ per face as its valence.\footnote{These implicit extra variables are often neglected in Euclidean models, but they are present from the start in the general definition of spin foam models based on wedges \cite{Reisenberger:1997sk}.}

\subsection{Factorisation in SU(2) vertex amplitudes}

As a first step to deal with the $\SL(2,\C)$ integrals, we exploit the Cartan decomposition of $h$ as
\be\label{g_par}
h = u e^{\f r2\s_3} v^{-1}, 
\ee
where $u$ and $v$ are arbitrary rotations, and $r\in[0,\infty)$ is the rapidity parameter of a boost along the $z$ axis. The parametrization has a U(1) gauge of common rotations of $u$ and $v$ along the $z$ axis, but this redundancy is harmless since the orbits are compact, and can be easily taken into account normalising the Haar measure as in \cite{Ruhl},
\be\label{dh}
dh = d\mu(r) \, du \, dv, \qquad  d\mu(r)=\f1{4\pi}\sinh^2r \, dr,
\ee
where $du$ and $dv$ are Haar measures for SU(2).
In an arbitrary unitary irrep of the principal series, \Ref{g_par} reads
\be\label{gparj}
D^{(\r,k)}_{jmln}(h) = \sum_p D^{(j)}_{mp}(u) d^{(\r,k)}_{jlp}(r) D^{(l)}_{pn}(v^{-1}),
\ee
where $D^{(j)}$ are Wigner's matrices for SU(2) and the boost matrix elements $d^{(\r,k)}$ are given as sums of hypergeometric functions in Appendix \ref{AppBoost}.
Using this decomposition it is easy to see that the partition function \Ref{Z} factorises into SU(2) $nj$-symbols at the vertices, times integrals over the boost parameter associated with the edges.

Consider an edge in the bulk of the spin foam, and the two $\SL(2,\C)$ integrations along it. For concreteness, let us fix
a four-stranded edge with all strands oriented left to right, part of a simplicial spin foam as in \Ref{As}.\footnote{If a  strand has a different orientation the formulae below are modified as in the usual graphical calculus for SU(2), inverting the relevant group element and intertwiner.} 
Denote as in \Ref{Dsplit} by $\{ j,m \}$ the labels in the middle of the edge, $\{l,n\}$ and $\{l',n'\}$ those respectively on the side of the source and target vertices of the edge. The integration on the first half-edge gives
\begin{align}\label{B4calc}
\int dh \ \bigotimes_{i=1}^4 \, D^{(\g j_i,j_i)}_{j_im_il_in_i}(h) & = 
\int du \, dv\, d\m(r)\ \bigotimes_{i=1}^4 \, \sum_{p_i} D^{(j_i)}_{m_ip_i}(u) d^{(\g j_i,j_i)}_{j_il_ip_i}(r) D^{(l_i)}_{p_in_i}(v^{-1})
\\ \nn & 
= \sum_{i,k} d_{i} d_{k} \left(\begin{array}{c} j_i \\ m_i \end{array}\right)^{(i)} \left(\begin{array}{c} l_i \\ n_i \end{array}\right)^{(k)}
B^{\g}_4(j_i,l_i;i,k),
\end{align}
where we defined the following half-edge weight, or \emph{dipole graph} amplitude,
\begin{align}\label{B4}
B^\g_4(j_i,l_i;i, k) &:= \int d\m(r) \ \sum_{p_i}
 \left(\begin{array}{c} j_i \\ p_i \end{array}\right)^{(i)} \left(\begin{array}{c} l_i \\ p_i \end{array}\right)^{(k)}
\bigotimes_{i=1}^4 \ d^{(\g j_i,j_i)}_{j_il_ip_i}(r),
\end{align}
and we used twice the SU(2) result
\be\label{W4}
\int du \ \bigotimes_{i=1}^4 \,D^{(j_i)}_{m_ip_i}(u)
= \sum_{i} d_{i} \left(\begin{array}{c} j_i \\ m_i \end{array}\right)^{(i)} \left(\begin{array}{c} j_i \\ p_i \end{array}\right)^{(i)}, 
\ee
in terms of Wigner's $4jm$ symbols, see Appendix \ref{AppSU2}. The dipole graph amplitude describes two of these SU(2) symbols averaged over all possible $z$-boosts relating them, and it will be a central object of interest of the paper.

\begin{figure}
\be\nn
 \parbox[2cm]{6.2cm}{\begin{picture}(0,40) (0,0) \includegraphics[width=6cm]{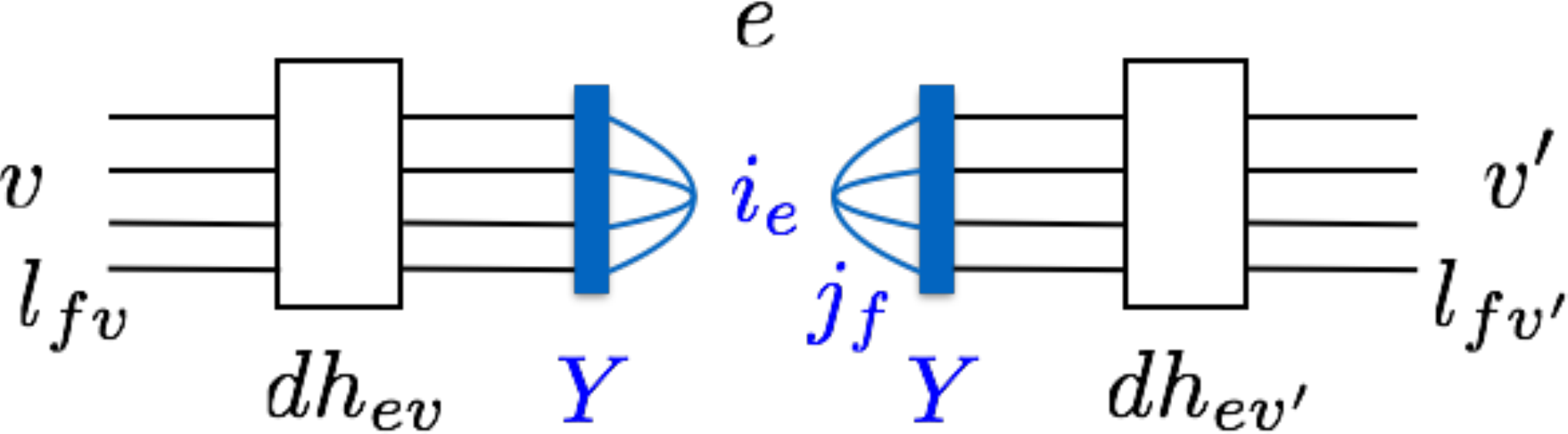} \end{picture}} 
 = \ \parbox[2cm]{7.5cm}{\begin{picture}(0,40) (0,0) \includegraphics[width=8.5cm]{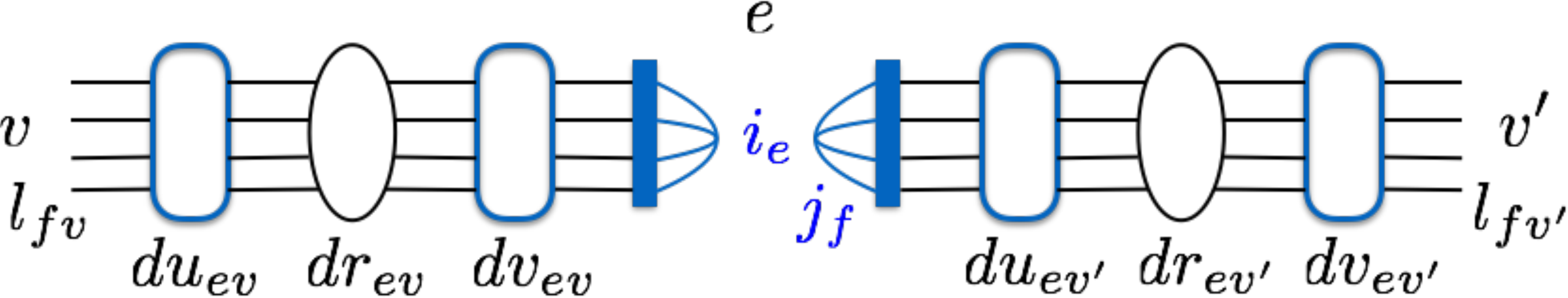} \end{picture}} 
\ee
\be\nn
= \parbox[2cm]{7.5cm}{\begin{picture}(0,40) (0,0) \includegraphics[width=8.5cm]{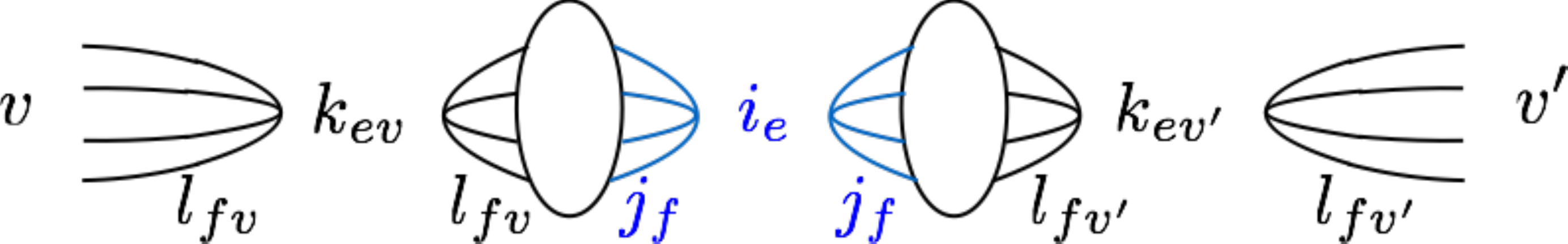} \end{picture}} 
\ee
\caption{\label{Fig_edge} {\small{\emph{The steps leading to \Ref{A4}, in graphical notation. Each edge contains two group integrations, represented by boxes, joined by a $Y$ map, represented by the blue bar. Using \Ref{dh}, we split each $\SL(2,\C)$ integration in three: first, an SU(2) integral on the vertex side, whose evaluation leads to an SU(2) $nj$-symbol labelled by the spins $l$; second, an integral over the boost parameter $r$, bridging between the $l$ and $j$ vectors; third, another SU(2) integral absorbed by the SU(2) intertwiner gluing the two half-edges. Over the whole edge, we have two integrals over the boost parameter on either side of the edge intertwiner, $i_e$, connected via summations over new SU(2) intertwiners, say $k_{ev}$ and $k_{ev'}$, to the SU(2) vertex amplitudes labelled with spins $l_{fv}$. The graphical notation motivates the name dipole amplitude for $B^\g_4$.
In the last picture we have kept just the blue colour in the dipole diagrams to mark the $Y$ map still present on one side of the $dr$ integrals. Finally, attaching the edge to its source and target vertices gives a graphical representation like in Fig.\ref{Figfactorfoam}. }}} }
\end{figure}

\begin{figure}[ht]
\begin{center} \includegraphics[width=16cm]{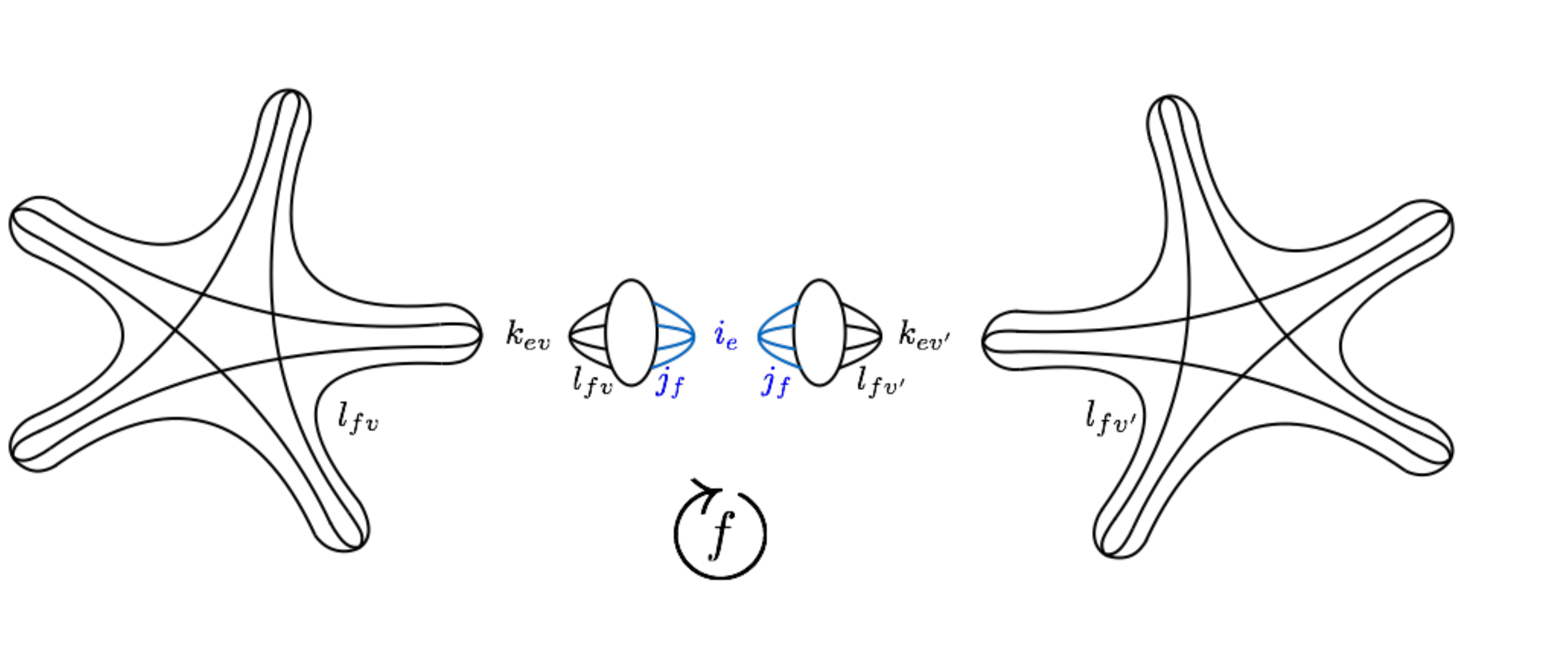} \end{center}
\caption{\label{Figfactorfoam} {\small{\emph{The EPRL spin foam after the factorisation, assuming a 4-valent edge connecting two 5-valent vertices. The original spins and intertwiners, in blue, are carried only by the $\g$-dependent edge amplitudes, and the vertex amplitudes are pure SU(2) $nj$-symbols like in SU(2) BF theory. Summations over the spin and intertwiner numbers are implicitly understood, as well as $d_j$ weight factors.}}}}
\end{figure}

Next, gluing this expression to the integral on the second half of the same edge integral, and using the orthogonality of the $4jm$ symbols on the $m_i$ indices (see \Ref{CG4ortho}), we obtain
\begin{align}
& \int d\tl h \ \bigotimes_{i=1}^4 \, D^{(\g j_i,j_i)}_{l'_in'_ij_im_i}(\tl h) \int dh \ \bigotimes_{i=1}^4 \, D^{(\g j_i,j_i)}_{j_im_il_in_i}(h) 
\\ \nn & \hspace{1cm} = 
\sum_{i,k,k'} d_{i} d_{k} d_{k'}
 \left(\begin{array}{c} l_i \\ n_i \end{array}\right)^{(k)} \left(\begin{array}{c} l'_i \\ n'_i \end{array}\right)^{(k')}
B^\g_4(j_i,l_i;i,k) B^\g_4(j_i,l'_i;i,k').
\end{align}
When we glue the strands to the vertices, the $4jm$-symbols above with magnetic labels $n_i$ and $ n'_i$ contract to form SU(2) $nj$-symbols at the vertices,
and we are left with the following edge amplitude,
\begin{align}\label{A4}
d_{i_e} A^\g_e(j_f,i_e,l_{fv}, k_{ev}, k_{ev'}) &:= d_{i_e} d_{k_{ev}} d_{k_{ev'}} B^\g(j_f,l_{fv};i_e,k_{ev}) B^\g(j_f,l_{fv'};i_e,k_{ev'}).
\end{align}
As a result, the partition function for a 5-valent complex takes the factorised form
\be\label{Z1}
Z^{EPRL}_{\cal C} = \sum_{j_f, i_e, l_{fv}, k_{ev}}  \prod_f d_{j_f} \prod_e d_{i_e} A^\g_e(j_f,i_e,l_{fv}, k_{ev}, k_{ev'}) \prod_v \{15j\}_v(l_{fv}, k_{ev}).
\ee
These simple algebraic manipulations can be done compactly using a graphical notation, as shown in Figs.~\ref{Fig_edge} and \ref{Figfactorfoam}, and make it natural to refer to \Ref{B4} as a dipole graph amplitude, or dipole amplitude, for short.

The vertex structures are now purely SU(2), and the edge amplitude contains all the dependence on boosts and on the Immirzi parameter.  
Geometrically, one can think of this construction as SU(2) areas and shapes associated with individual polyhedra being boosted between adjacent polytopes.
The trade-off for this factorisation is that one has to make explicit the infinite summations over the spin labels $l_{fv}$.

A few comments are in order. First, 
the factorisation generalises to an arbitrary 2-complex, provided the amplitude is integrable: the edge amplitude now carries a set of $v_e-3$ intertwiner labels $(\{ i_e \}, \{k_{ev}\}, \{k_{ev'}\})$ per edge of valency $v_e$, and the SU(2) $\{15j\}$-symbol of Fig.~\ref{Figfactorfoam} is replaced by the relevant $\{nj\}$-symbol associated with the valence and combinatorial structure of the vertex. The partition function factorises as
\be\label{Zgen}
Z^{EPRL}_{\cal C} = \sum_{j_f, \{i_e\}, l_{fv}, \{k_{ev}\}, \{k_{ev'}\}}  \prod_f d_{j_f} \prod_e d_{\{i_e\}} A^\g_e(j_f,\{i_e\},l_{fv}, \{k_{ev}\}, \{k_{ev'}\}) \prod_v \{nj\}_v(l_{fv}, k_{ev}).
\ee
Again, the edge amplitude is the product of $d_{k}$ factors and two dipole amplitudes, each given by
\begin{align}\label{Bn}
B^\g(j_i,l_i;\{ i\},\{ k\}) &= \int d\m(r) \sum_{p_i}
 \left(\begin{array}{cccc} j_i \\ p_i \end{array}\right)^{(\{i\})}
 \left(\begin{array}{cccc} l_i \\ p_i \end{array}\right)^{(\{k\})}
  \bigotimes_i d^{(\g j_i,j_i)}_{j_il_ip_i}(r),
\end{align}
see Appendix \ref{AppSU2} for details on the generalised $njm$ symbols.

Second, 
we restricted attention so far to the simple representations \Ref{simple}, but the factorisations \Ref{Z1} and \Ref{Zgen} hold for arbitrary representations, with 
$(\g j,j)$ in the dipole amplitudes replaced in $(\r,k)$. For instance, a similar decomposition can be obtained also for the Lorentzian Barrett-Crane model \cite{BarrettCraneLor}, or in general for relativistic spin networks and projected spin networks \cite{EteraProj}. Indeed, this factorisation is just an example of a more general factorisation property of $\SL(2,\C)$ Clebsch-Gordan coefficients, to be reviewed in the next Section. 

Third, the partition function is real, and so must be the edge amplitudes, provided one works with the conventional, real SU(2) Clebsch-Gordan coefficients. The individual dipole amplitudes  \Ref{B4} and \Ref{Bn} are a priori complex, but can be made real if we choose the right 
phase conventions for the boost matrices $d^{(\r,k)}$, as we discuss in the next Section.

Summarising, applying the Cartan decomposition \Ref{g_par} to \Ref{Z}, one obtains a factorisation of the EPRL model in terms of SU(2) $nj$-symbols at the vertices, `boosted' by edges amplitudes carrying the non-compact integrations over the rapidities $r$'s. 
The edge amplitudes contain the dependence of the model on $\g$, and are a product of two dipole amplitudes \Ref{B4} or \Ref{Bn} in general.
Our next goal is to show how the integrals defining the edge amplitudes can be explicitly computed, using the explicit form of the boost matrix elements, and the precise relation between the dipole amplitudes and $\SL(2,\C)$ Clebsch-Gordan coefficients.

\subsection{Simplified EPRL model}

Before moving on, let us take advantage of \Ref{Zgen} to suggest the introduction of a simplified EPRL model, useful in certain approximations. 
The factorisation property of the EPRL model is somewhat hindered by the proliferation of spin labels, with as many new ones per face as its valence. This is of course unavoidable, as these summations are present from the start in \Ref{Av1}, just implicit in the matrix product. In the light of the known computational complexity of the EPRL model, it is tempting to consider a natural simplification
where all the new spins $l_{fv}$ are fixed to their minimal values $j_f$. 
This amounts to imposing an extra $Y$ map also between the two $\SL(2,\C)$ group elements $h_a$ and $h_b$ in the definition \Ref{Av1} of the vertex amplitude. Geometrically, it means allowing only the shapes of the polyhedra (namely the intertwiners) to be boosted, and not the areas (namely the spins). This simplification may appear drastic; but as we will see below, the largest contributions to the EPRL edge amplitude do come from configurations with $l_i=j_i$, that is the minimal admissible values for the $l_i$ spins, which suggests that the simplified model can still capture some relevant properties of the full model.

For this simpler model, there is a single spin per face, and the partition function reads
\be\label{Zs}
Z^{EPRLs}_{\cal C} = \sum_{j_f, i_e, k_{ev}}  \prod_f d_{j_f} \prod_e d_{i_e} A^\g_e(j_f,i_e,j_f,k_{ev}) \prod_v \{nj\}_v(j_f,i_e).
\ee
Furthermore, the associated dipole diagram amplitude is 
\be
B_s^\g(j_i;\{ i\},\{ k\}) := B^\g(j_i,l_i;\{ i\},\{ k\}) = \int d\m(r) \ \sum_{p_i}
 \left(\begin{array}{cccc} j_i \\ p_i \end{array}\right)^{(\{i\})}
 \left(\begin{array}{cccc} j_i \\ p_i \end{array}\right)^{(\{k\})}
\bigotimes_i d^{(\g j_i,j_i)}_{j_ij_ip_i}(r),
\ee
in which only the `spin-diagonal' boost matrix elements \Ref{dgj} enter. These are given by a single hypergeometric function (see \Ref{dsimple} below).
Thanks to the reduced spin summations, and the simpler boost matrix elements, the model so defined is significantly simpler than the original one, and much faster to evaluate. 

Below in Section \ref{SecSimplified} we will use the results on the large spin scaling of the dipole amplitudes to estimate the overall scaling of the simplified model \Ref{Zs}.
Another interesting property of the simplified model is that it can be entirely re-expressed in terms of spinors, using formulas like \Ref{B3spinors} below.

\section{Boost matrix elements and $\SL(2,\C)$ Clebsch-Gordan coefficients}

For generality, we consider in this Section arbitrary irreps. We will specialise to the $\g$-simple irreps \Ref{simple} relevant to the EPRL model in the next Section.

\subsection{Boost matrix elements for simple representations} 

The explicit $z$-boost matrix elements can be found in the literature (e.g. \cite{Naimark,Ruhl,Strom,vongDuc1967,Ruhl,RashidBoost}), but in forms that may differ by a phase.
This is a delicate point, as the phase determines the reality of Clebsch-Gordan coefficients and dipole amplitudes.
For this paper, we take
\begin{align}\nn
d^{(\r,k)}_{jlp}(r) &= (-1)^{\f{j-l}2} e^{i\Phi^{\r}_j} e^{-i\Phi^{\r}_l} \f{\sqrt{d_j}\sqrt{d_l}}{(j+l+1)!}  
\left[(j + k)! (j - k)! (j + p)! (j - p)! (l + k)! (l - k)! (l + p)! (l - p)!\right]^{1/2} 
\\ \nn & \times e^{-(k-i\r +p+1)r} \sum_{s,t} \f{  (-1)^{s+t} e^{- 2 t r} (k + p + s + t)! (j + l - k - p - s - t)!}{s! (j - k - s)! (j - p - s)! (k + p + s)! t! (l - k - t)! (l - p - 
    t)! (k + p + t)!} \\ \label{dgen} & \hspace{3cm} \times \ {}_2F_1[l+1-i\r,k+p+1+s+t,j+l+2,1-e^{-2r}],
\end{align}
where 
\be\label{Phidef}
e^{i\Phi^\r_j} =  \f{\G(j+i\r+1)}{|\G(j+i\r+1)|}.
\ee
The phase is chosen to have the following symmetry property,
\be\label{dbard}
\overline{d^{(\r,k)}_{jlp}(r)} = (-1)^{j-l}d^{(\r,k)}_{jl,-p}(r),
\ee 
which in turns implies the symmetry
\be\label{d-d}
d^{(-\r,-k)}_{jlm}(r) = d^{(\r,k)}_{jlm}(r)
\ee
and the reality of group-averaged tensor products and dipole amplitudes \Ref{Bn}, as proved in Appendix \ref{AppBoost}.
This phase convention differs from the one in Ruhl's monography \cite{Ruhl}, which lacks the first three factors in \Ref{dgen}.
The factors \Ref{Phidef} are already present in the conventions of \cite{Strom,vongDuc1967}, and used in the literature on Clebsch-Gordan coefficients \cite{Rashid70I,Rashid70II,Gavrilik1972}.\footnote{A similar phase is also considered by Ruhl, for what are referred to as functions of the second kind, 
\[
e_{\rm Ruhl}^{(\r,k)} := \left(\f{\G(j+i\r+1)}{\G(j-i\r+1)}\right)^{1/2} \left(\f{\G(l-i\r+1)}{\G(l+i\r+1)}\right)^{1/2} d_{\rm Ruhl}^{(\r,k)}.
\] 
However with the phase factor defined by the square root instead of the norm as in \Ref{Phidef}, these functions fail to be a representation of the group -- as appropriately pointed out in \cite{Ruhl} --, for the trivial reason of extra minus signs appearing in the multiplication law. This is also the wrong phase definition appearing in \cite{Kerimov}, as we will comment upon below in Section \ref{secKerimov}. \label{footeRuhl}}
They lead to dipole amplitudes which are either real of purely imaginary. The additional factor $(-1)^{\f{j-l}2}$ makes the dipole amplitudes always real.
All three choices \Ref{dgen}, \cite{Ruhl} and \cite{Strom,vongDuc1967} are related by a straightforward unitary transformation.
The precise phase choice is to some extent irrelevant for spin foam amplitudes, as the latter are real by construction: any complex phase in the dipole amplitudes would cancel out when these are glued together. Nonetheless, it is convenient to work with real dipole amplitudes as it make it easier to investigate analytically and numerically their properties, as we will do below in Sections \ref{SecB3} and \ref{SecB4}.

The expression \Ref{dgen} simplifies in the case of the $\g$-simple representations, in particular fixing $k=j$ kills one summation, and we have
\begin{align}\label{dsimple}
d^{(\g j,j)}_{jlp}(r) &=  
(-1)^{\f{j-l}2} e^{i\Phi^{\g j}_j} e^{-i\Phi^{\g j}_l} \f{\sqrt{d_j}\sqrt{d_l}}{(j+l+1)!}  
\left[(2j)!(l+j)!(l-j)!\f{(l+p)!(l-p)!}{(j+p)!(j-p)!}\right]^{1/2} \\ \nn &\ \times e^{-(j-i\g j +p+1)r}
\sum_{s} \f{  (-1)^{s} \, e^{- 2 s r} }{s!(l-j-s)!} \, _2F_1[l+1-i\g j,j+p+1+s,j+l+2,1-e^{-2r}].
\end{align}
A further simplification of the matrix elements occurs in the minimal, or `spin-diagonal' case $l=j$: only the term $s=0$ survives in the summation, and the matrix elements are given by a single hypergeometric function,
\begin{align}\label{dgj}
d^{(\g j,j)}_{jjp}(r) & = e^{-(j-i\g j+p+1)r} \, {}_2F_1[j+p+1,j(1-i\g)+1,2j+2,1-e^{-2r}].
\end{align}
Below in Section \ref{SecSimplified} we will consider a simplified version of the EPRL model in which only these matrix elements occur.

The complicate expressions of the above boost matrix elements can be compared with the much simpler ones appearing in the Lorentzian Barrett-Crane model \cite{BarrettCraneLor}. This is defined for space-like faces using the matrix elements $k=j=l=0$, for which we have the elementary form
\be
 d^{(\r,0)}_{000}(r) = \f{\sin(\r r)}{\r \sinh r},
\ee
and the relevant Clebsch-Gordan coefficients have a simple, compact expression, see \Ref{chiBC} below.
Of course, the Barrett-Crane model has the drawback that the large spin limit does not capture the (discrete) Levi-Civita condition for the connection \cite{Alesci:2007tx}, so it fails to reproduce approximate solutions of general relativity in this limit.\footnote{The Barrett-Crane model also describes time-like faces, for which the prescription is to take irreps of the category $d^{(0,k)}$. 
To treat this case in the EPRL model, one has to take the normal vector defining the linear simplicity constraints to be space-like.
This leaves the first, Lorentz-invariant restriction $\r=\g j$ untouched, but replaces the SU(2) label $j$ in the second, non-Lorentz-invariant restriction of $k$, by a representation of the $\SU(1,1)$ group stabilising the chosen space-like direction. The solution to the simplicity and closure constraints describes time-like polyhedra,
with 
both space-like and time-like faces possible, described by $\SU(1,1)$ unitary irreps respectively in the discrete or continuous series (see e.g. \cite{Rennert?}).
A version of the EPRL model including time-like faces is studied in \cite{ConradyHnybida}, and it would certainly be of much interest to study the factorisation exposed here in terms of the $\SU(1,1)$ subgroup instead of SU(2).
Notice that thanks to the freedom of choosing an arbitrary direction to be stabilised, the EPRL model can in principle treat also null polyhedra \cite{IoNull}, unlike the Barrett-Crane model.}

\subsection{$\SL(2,\C)$ Clebsch-Gordan coefficients and dipole amplitudes}

The spin foam vertex amplitude \Ref{Av1} has a similar structure also in Euclidean signature, 
with the irreps and integrals of $\SL(2,\C)$ replaced by those for the compact groups SU(2) or SO(4). It is then customary to perform explicitly the integrations in terms of Clebsch-Gordan coefficients, using formulas like \Ref{W4}, and obtaining state sum models described by Wigner's $nj$-symbols. The same can be done in the Lorentzian case, with the corresponding $\SL(2,\C)$ Clebsch-Gordan coefficients.  These may be less familiar to the reader than those for SU(2) or $\SU(N)$, but have been studied at length in the literature, and relevant results will be recalled in this Section.

As shown initially by Naimark \cite{Naimark}, tensor products of $\SL(2,\C)$  unitary irreps of the principal series decompose among themselves with each irrep in the decomposition appearing only once. This allows to define the Clebsch-Gordan coefficients as usual,
\be\label{CGdec1}
\ket{\r_1,k_1;j_1m_1}\otimes \ket{\r_2,k_2;j_2,m_2} = \int d\r \sum_{k,j,m} C^{\r k j m}_{\r_1k_1j_1m_1\r_2k_2j_2m_2} \ket{\r, k; j, m}.
\ee
The recoupling conditions \cite{Naimark} turn out to be the usual triangle inequalities for the SU(2) spins, 
\be\label{CGconditions}
|j_1-j_2|\leq j_3 \leq j_1+j_2
\ee
and so on cyclically, 
plus the trivial condition 
\be\label{sumk}
\sum_i k_i \in \Z
\ee
on the discrete irrep labels. It is sometimes convenient to split the Clebsch-Gordan coefficients into two classes, those for which the $k_i$ themselves satisfy triangle inequalities like \Ref{CGconditions}, and those that do not. 
There are on the other hand no restrictions on the continuous labels $\r_i$, a fact which will play a role below, so the integral in \Ref{CGdec1} runs on $\R$.

The factorisation \Ref{Zgen} of the spin foam amplitude obtained in the previous Section is a direct consequence of the following  general factorisation of 
$\SL(2,\C)$ Clebsch-Gordan coefficients in terms of SU(2) ones (see e.g. \cite{Rashid70II}),
\be\label{Cfactor}
C^{\r k j m}_{\r_1k_1j_1m_1\r_2k_2j_2m_2} = \chi(\r_1,\r_2,\r,k_1,k_2,k;j_1,j_2,j) \, C^{jm}_{j_1m_1j_2m_2},
\ee
a property which follows more or less directly from the use of the canonical basis.
For brevity of notation, we will keep track from now on of the dependence of $\chi$ on the SU(2) spins $j_i$ only, and write $\chi(j_1,j_2,j_3)$.

To apply the Clebsch-Gordan decomposition to the matrices \Ref{Dh}, recall that these provide a generalised (in the sense of distributions) orthogonal basis of $L_2[\SL(2,\C), d\mu_{\rm Haar}]$, with non-trivial Plancherel measure appearing in the orthogonality formula \cite{Ruhl},
\begin{align}\label{Dortho}
\int dh\,
D^{(\r_1,k_1)}_{j_1m_1l_1n_1}(h)D^{(\r_2,k_2)}_{j_2m_2l_2 n_2}(h) &= 
\f{\d(\r_1-\r_2) \d_{k_1k_2}}{4(\r_1^2+k_1^2)} \, \d_{j_1j_2}\d_{l_1l_2} \d_{m_1m_2}\d_{n_1n_2}.
\end{align}
Accordingly, the tensor product decomposes as
\begin{align}\nn
D^{(\r_1,k_1)}_{j_1m_1l_1n_1}(h) D^{(\r_2,k_2)}_{j_2m_2l_2 n_2}(h) &= 
\int_{-\infty}^\infty d\r\sum_{k} 4(\r^2+k^2) \sum_{j,l,m,n} \bar{C}^{\r k jm}_{\r_1k_1j_1m_1\r_2k_2j_2m_2} C^{\r k ln}_{\r_1k_1l_1m_1\r_2k_2l_2m_2} D^{(\r,k)}_{jmln}(h),
\end{align}
with the sum over $k$ restricted by \Ref{sumk} and the other sums by $(j,l)\geq |k|$ and the usual triangle inequalities.
From the Clebsch-Gordan decomposition and the orthogonality \Ref{Dortho} it follows that
\begin{align}\label{intDDDbar}
\int dh D^{(\r_1,k_1)}_{j_1m_1l_1n_1}(h)D^{(\r_2,k_2)}_{j_2m_2l_2 n_2}(h) \overline{D^{(\r_3,k_3)}_{j_3m_3l_3n_3}(h)}&= 
\bar{C}^{\r_3 k_3 j_3m_3}_{\r_1k_1j_1m_1\r_2k_2j_2m_2} C^{\r_3 k_3 l_3n_3}_{\r_1k_1l_1m_1\r_2k_2l_2m_2} \\\nn
&= \bar{ \chi}(j_1,j_2,j_3) \chi(l_1,l_2,l_3)  C^{j_3m_3}_{j_1m_1j_2m_2} C^{l_3n_3}_{l_1n_1l_2n_2}.
\end{align}
The left-hand side can be easily shown to be real, thanks to 
\be\label{DbarD}
\overline{D^{(\r,k)}_{jmln}} = (-1)^{j-l +m-n} D^{(\r,k)}_{j-ml-n},
\ee
which follows from known properties of the Wigner matrices and our previous phase choice \Ref{dbard} for the boost matrices.
See Appendix \ref{AppBoost} for an explicit proof. Nonetheless, this does not imply that the individual Clebsch-Gordan coefficients $\chi(j_i)$ appearing on the right-hand side are real; and indeed, if one works with Naimark's basis they are not, because the boost raising and lowering operators have complex coefficients. For instance, the literature \cite{Rashid70I,Rashid70II} uses phase conventions that make them either real or purely imaginary. It is however possible to make them always real, following \cite{Kerimov,Kerimov75} as we will do below. We nevertheless keep track of the complex conjugate in our formulas, so to make them applicable also for researchers using different phase conventions. The (non-trivial) relation between the phase conventions of \cite{Rashid70I,Rashid70II} and \cite{Kerimov,Kerimov75} is discussed in Appendix \ref{AppSL2C}.

Using \Ref{DbarD} in the left-hand side of \Ref{intDDDbar} and known symmetries of Wigner's $3jm$ symbols (see \Ref{sign-flip}),
we also have
\begin{align}\label{intDDD}
 \int dh \bigotimes_{i=1}^3 D^{(\r_i,k_i)}_{j_im_il_in_i}(h) & = 
(-1)^{j_1-j_2+j_3+l_1-l_2+l_3} \sqrt{d_{j_3}} \sqrt{d_{l_3}} 
\left(\begin{array}{c} j_i \\ m_i \end{array}\right)\left(\begin{array}{c} l_i \\ n_i \end{array}\right)
\bar{\chi}(j_1,j_2,j_3) \chi(l_1,l_2,l_3).
\end{align}
From this expression, it is immediate to find the precise relation between Clebsch-Gordan coefficients and the 3-valent dipole amplitude. The latter is defined as in \Ref{Bn}, 
\begin{align}\label{B3}
B_3(\r_i,k_i;j_i,l_i) &= \int d\m(r) \sum_{p_i}
 \left(\begin{array}{c} j_i \\ p_i \end{array}\right)
 \left(\begin{array}{c} l_i \\ p_i \end{array}\right)
\bigotimes_{i=1}^3 \ d^{(\r_i,k_i)}_{j_il_ip_i}(r),
\end{align}
and related to the left-hand side of \Ref{intDDD}  by
\begin{align}\label{DDDSL2C}
 \int dh \bigotimes_{i=1}^3 D^{(\r_i,k_i)}_{j_im_il_in_i}(h) & = 
\left(\begin{array}{c} j_i \\ m_i \end{array}\right)\left(\begin{array}{c} l_i \\ n_i \end{array}\right)
B_3(\r_i,k_i;j_i,l_i),
\end{align}
where we used the Cartan decomposition and the property \Ref{intD3App} of Winger's $3jm$ symbols.
Comparing the two, we read
\be\label{B3CG}
B_3(\r_i,k_i;j_i,l_i)  = (-1)^{j_1-j_2+j_3+l_1-l_2+l_3} \sqrt{d_{j_3}} \sqrt{d_{l_3}}\, \bar{\chi}(j_1,j_2,j_3) \chi(l_1,l_2,l_3).
\ee
Reality of this expression can be again proven using \Ref{DbarD}. 

This relation extends to dipole amplitudes \Ref{Bn} with $n$ strands, with the appropriate generalised $n$-valent Clebsch-Gordan coefficients on the right-hand side. These are defined in Appendix \ref{AppSL2C}, and as for the more familiar SU(2) case, can be re-expressed in terms of fundamental 3-valent coefficients, introducing a recoupling scheme with virtual links carrying irreps corresponding to the coupling of only two irreps at a time. That is, we first use the Clebsch-Gordan decomposition to reduce the tensor product of four matrices to three, and then apply \Ref{intDDD} to eliminate the group integral.
Choosing to fix ideas to couple first the channels 1 and 2, the intermediate step introduces virtual irreps $(\r_{12}, k_{12})$, and the final result reads
\begin{align}
\int dh \ \bigotimes_{i=1}^4 \, D^{(\g j_i,j_i)}_{j_im_il_in_i}(h) & = 
(-1)^{j_1-j_2+j_3-j_4-l_1+l_2-l_3+l_4} \sqrt{d_{j_4}} \sqrt{d_{l_4}} \int_{-\infty}^\infty d\rho_{12} \sum_{k_{12}} 4(\r_{12}^2+k_{12}^2) 
\\\nn & \hspace{-1cm} \times\,
\sum_{j_{12},l_{12}} 
\bar{\chi}(j_1,j_2,j_{12}) \bar{\chi}(j_{12},j_3,j_4)  \chi(l_1,l_2,l_{12}) \chi(l_{12},l_3,l_4)
\sqrt{d_{j_{12}}} \sqrt{d_{l_{12}}} \left(\begin{array}{c} j_i \\ m_i \end{array}\right)^{(j_{12})} \left(\begin{array}{c} l_i \\ n_i \end{array}\right)^{(l_{12})},
\end{align}
where we used the definition of Wigner's $4jm$ symbol, see Appendix \ref{AppSU2}.
In this expression, the sums over $k_{12}$ are restricted by \Ref{sumk}, those over $j_{12}$ and $l_{12}$ by the triangle inequalities and by $(j_{12},l_{12})\geq |k_{12}|$. Notice that since $j_i$ are fixed, this last condition means that the bounds over $j_{12}$ by its triangle inequalities translate into bounds on $k_{12}$, so that actually the sum over $k_{12}$ is finite. 
Hence, thanks to the finiteness of all the sums and the uniqueness of the Clebsch-Gordan decomposition, we can safely swap the sum over $k_{12}$ with those over $j_{12}$ and $l_{12}$, obtaining 
\be
\sum_{k_{12}=-\infty}^\infty \ \sum_{j_{12}={\rm max\{|k_{12}|,|j_1-j_2|\}}}^{j_1+j_2} \ \sum_{l_{12}={\rm max\{|k_{12}|,|l_1-l_2|\}}}^{l_1+l_2} 
 \ldots  = \sum_{j_{12}=|j_1-j_2|}^{j_1+j_2} \ \sum_{l_{12}=|l_1-l_2|}^{l_1+l_2} \ \sum_{k_{12}={\rm -min\{j_{12},l_{12}\}}}^{\rm min\{j_{12},l_{12}\}} \ldots
\ee 
Finally comparing the with the Cartan decomposition \Ref{B4calc}, we read the desired result,
\begin{align}\label{B4CG}
\sqrt{d_{j_{12}}} \sqrt{d_{l_{12}}} \, B_4(\r_i,k_i;j_i,l_i;j_{12}, l_{12}) &= (-1)^{j_1-j_2+j_3-j_4-l_1+l_2-l_3+l_4} \sqrt{d_{j_4}} \sqrt{d_{l_4}}
\\\nn & \hspace{-3.5cm} \times\,
\int_{-\infty}^\infty d\rho_{12} \sum_{k_{12}={\rm -min\{j_{12},l_{12}\}}}^{\rm min\{j_{12},l_{12}\}} 
4(\r_{12}^2+k_{12}^2) \, \bar{\chi}(j_1,j_2,j_{12}) \bar{\chi}(j_{12},j_3,j_4)  \chi(l_1,l_2,l_{12}) \chi(l_{12},l_3,l_4).
\end{align}
The 4-valent dipole amplitude includes thus a linear superposition of all admissible virtual irreps.
Notice that the function $F_{k_{12}}(\r_{12})$ defined by the second line of \Ref{B4CG}, basically its integrand-summand, satisfies 
\be\label{parity-intsum}
F_{k_{12}}(-\r_{12}) = F_{-k_{12}}(\r_{12}),
\ee
a non-trivial property which follows from \Ref{d-d}. Therefore, both the integrand and the summand are even functions, and the integration domain can be restricted to positive $\rho_{12}$ when numerically evaluating \Ref{B4CG}, saving precious computing time at the bargain price of a factor of 2.

In a similar manner, one can extend to higher valence the relation between dipole amplitudes and squares of generalised $\SL(2,\C)$ Clebsch-Gordan coefficients. Knowing these precise relations between the dipole amplitudes of interest in spin foam models and the $\chi$ coefficients will allow us to use existing results in the literature to evaluate the spin foam amplitudes.

\subsection{Kerimov-Verdiev's finite sums formula} \label{secKerimov}

Explicit evaluations of Clebsch-Gordan coefficients for unitary irreps of $\SL(2,\C)$ have been studied in the literature by many authors, and as we can see from the above construction, it boils down to evaluating the $r$ integral in \Ref{B3} to compute the norm of $\chi$, and to provide a scheme to determine the overall sign and phase, since they are not always real. 
A simple way to evaluate the $r$ integral, considered for instance in \cite{Rashid70I}, is to use \Ref{dgen} and the explicit definition of the hypergeometric function as an infinite series of monomials in $r$. The $r$ integral can then be performed analytically, and one obtains a triple infinite series of an Euler $\beta$ function. More interesting is an evaluation of the integral in terms of \emph{finite} sums of Gamma functions, which was derived by Kerimov and Verdiev \cite{Kerimov}.
Using their result requires some adaptations and some care, as we now describe.

Let us introduce the shortcut notation
\be
J=\sum_i j_i, \qquad K=\sum_i k_i, \qquad P=\sum_i \r_i.
\ee
The result of \cite{Kerimov}, properly adapted as explained below, reads
\begin{align}\label{chiKer}
 \chi(j_1,j_2,j_3) &= \f{(-1)^{\f{K+J}2} }{4\sqrt{2\pi} } {\it x}(\r_i,k_i) 
\Gamma(\tfrac{1-iP+K}{2}) \Gamma(\tfrac{1-iP-K}{2}) \sqrt{d_{j_1}d_{j_2}d_{j_3}} \, \kappa(\r_i,k_i;j_i),
\end{align}
where:
\begin{align}\label{kgen}
\kappa(\r_i,k_i;j_i) & = (-1)^{ - k_1 - k_2} e^{-i\big(\Phi^{\r_1}_{j_1}+\Phi^{\r_2}_{j_2}-\Phi^{\r_3}_{j_3}\big)}  \f{(-1)^{j_1-j_2+j_3}}{\sqrt{d_{j_3}}}
\left[\f{(j_1-k_1)!(j_2+k_2)!}{(j_1+k_1)!(j_2-k_2)!}\right]^{1/2}\\\nn & \times
\sum_{n=-j_1}^{{\rm min}\{j_1,k_3+j_2\}} \left[ \f{(j_1-n)!(j_2+k_3-n)!}{(j_1+n)!(j_2-k_3+n)!} \right]^{1/2}
\, \Wthree{j_1}{j_2}{j_3}{n}{k_3-n}{-k_3}
\\\nn & \times 
\sum_{s_1={\rm max}\{k_1,n\}}^{j_1} \sum_{ s_2={\rm max}\{-k_2,n-k_3\}}^{j_2} \f{(-1)^{s_1+s_2 -k_1+k_2}  \, (j_1+s_1)!(j_2+s_2)!}{(j_1-s_1)!(s_1-k_1)!(s_1-n)!(j_2-s_2)!(s_2+k_2)!(k_3-n+s_2)!} 
\\\nn & \times 
\frac{\G(\tfrac{1-i (\r_{1}- \r_{2}- \r_{3})-K+2 s_{1}}{2}) \G(\tfrac{1+i (\r_{1}- \r_{2}+ \r_{3})+K+2 s_{2}}{2}) 
\Gamma(\tfrac{1-i (\r_{1}+ \r_{2}- \r_{3})-k_{1}+k_{2}+k_{3}-2 n+2 s_{1}+2 s_{2}}{2})}{\Gamma (1-i \r_{1}+s_{1}) \Gamma (1-i \r_{2}+s_{2}) \Gamma (1+i \r_{3}+s_{1}+s_{2}) \Gamma(\tfrac{1-i P 
-k_{1}+k_{2}+k_{3}-2 n}{2})},
\end{align}
an expressions which contains only finite sums, and 
\begin{align}\label{xdef}
{\it x}(\r_i,k_i)  &= 
\f{\Gamma\big(\tfrac{1+ i P - K)}{2}\big)}{|\Gamma\big(\tfrac{1+i P- K)}{2}\big)|}
\\\nn & \qquad \times \frac{\Gamma\big(\tfrac{1-i (-\r_{1}+ \r_{2}+ \r_{3})-k_1+k_2+k_3}{2}\big)}{|\Gamma\big(\tfrac{1-i (-\r_{1}+ \r_{2}+ \r_{3})-k_1+k_2+k_3}{2}\big)|} 
\f{\Gamma\big(\tfrac{1-i (\r_{1}- \r_{2}+ \r_{3})+k_{1}-k_{2}+k_{3}}{2} \big)}{|\Gamma\big(\tfrac{1-i (\r_{1}- \r_{2}+ \r_{3})+k_{1}-k_{2}+k_{3}}{2} \big)|} 
\frac{ \Gamma\big(\tfrac{1-i (-\r_{1} - \r_{2} + \r_{3})-k_{1}-k_{2}+k_{3}}{2}\big)}{| \Gamma\big(\tfrac{1- i (-\r_{1} - \r_{2} + \r_{3}) -k_{1}-k_{2}+k_{3}}{2}\big)|}.
\end{align}
is an additional phase which makes the coefficients always real. 

We did not attempt to rederive this remarkable formula, based on previous results by Naimark on $\SL(2,\C)$ generating functions 
\cite{Naimark} and a series of non-trivial manipulations using properties of integrals of hypergeometric functions; we merely contented ourselves to perform a series of analytic and numerical checks.\footnote{For the interested reader, we performed extensive numerical checks of three types: first, \Ref{B3Kgen} versus the integral expression \Ref{B3}; this checks the absolute value of \Ref{kgen} and the relative phases with $l_i\neq j_i$. Second, ($i$-)reality of \Ref{kgen}; this checks the phase \Ref{xdef}. Third, behaviour under permutations.} Doing so we confirmed the general validity of the formula, but with a few minor changes which we now discuss in details. 

First of all, apart from trivial notational changes, our formula differs from the one in \cite{Kerimov} by an additional factor $1/(4\sqrt{2} \, {\pi^2} )$, which comes from different normalisations of the Haar measure.
Then, we needed three minor corrections to the Kerimov-Verdiev's formula.

The first concerns the choice of phases. In \cite{Kerimov} they use $N^\r_j :=  [{\G(j+i\r+1)}/{\G(j-i\r+1)} ]^{1/2}$
instead of our $\exp\{i\Phi^\r_j\}$ given by \Ref{Phidef}. 
In other words, their boost matrices are precisely those called $e^{(\r,k)}$ in Ruhl, see footnote \ref{footeRuhl}. But these, as discussed in that footnote and  in \cite{Ruhl}, are not really good representations matrices, albeit for rather trivial reasons of some wrong minus signs. One should rather use \Ref{Phidef} with the absolute value instead of the square root, as defined in \cite{Strom,vongDuc1967} and already used in the literature on Clebsch-Gordan coefficients \cite{Rashid70I,Rashid70II,Gavrilik1972}.
Accordingly, one has to correct also the phase term \Ref{xdef}, switching from the use of square roots in \cite{Kerimov} to the absolute values used here.
These changes in the phases affect the Kerimov-Verdiev's formula by an overall factor $e^{in(\r_i,k_i,j_i)\pi/2}$, and none of the steps in their derivation, as far as we can tell. 

The second correction concerns the reality of the coefficients. Neither the original formula with $N^\r_j$ nor the amended one with $\exp\{i\Phi^\r_j\}$ (and the corresponding $x(\r_i,k_i)$ phases in the two cases) give real values, but either real or purely imaginary values. To achieve reality, and preserve \Ref{B3CG} -- which fixes the $j$-dependence of the overall phase of the $\chi$'s --, we need to add to the formula of \cite{Kerimov} the phase $(-1)^{(K+J)/2}$, explicit in \Ref{chiKer}, plus an additional $(-1)^{-k_3}$, which cancels a similar term in the original formula, leading to \Ref{kgen}. This missing phase may simply have been an omission: \cite{Kerimov} refers to a previous paper \cite{Kerimov75} for the proof of reality, a paper in russian initially unavailable to us. There a phase factor  $(-1)^{(k_1+k_2-k_3)/2}$ does appear; it and the $(-1)^{J/2}$ factor may simply have been lost in the steps from \cite{Kerimov75} to \cite{Kerimov}, which include also a change of basis and different conventions.\footnote{With \cite{Kerimov75} initially unaccessible, the additional $(-1)^{-k_3}$ was missing in the first version of this paper, available on the arXives, resulting in Clebsch-Gordan coefficients which are real for $k_3\in\N$, and purely imaginary for $k_3\in\N+1/2$.} 

The third and final correction concerns the upper bound on the $n$ summation, which is given as $j_1$ in \cite{Kerimov}. This would actually be correct for the $\g$-simple irreps \Ref{simple}, but not in general: $(j_2+k_3-n)!$ may introduce a smaller bound, hence the amendment in \Ref{kgen} (a max$\{-j_1,-j_2+k_3\}$ could also be specified for the lower bound, but omitting this is harmless). This is certainly a minor mistake, but it can be frustrating to look for when obtaining wrong numerical results from such a complex formula, and we think that pointing it out is useful.

For the $\g$-simple irreps \Ref{simple}, one summation collapses to its maximal value $s_1=j_1$, and it is possible to use the explicit form of the $3jm$-symbol to simplify \Ref{kgen} to
\begin{align}
& \kappa(\g j_i, j_i; j_i) = e^{-i\big(\Phi^{\g j_1}_{j_1}+\Phi^{\g j_2}_{j_2}-\Phi^{\g j_3}_{j_3}\big)} \f{(-1)^{j_1-j_2+j_3}}{\sqrt{d_{j_3}}}
\f{\D(j_i) \, a_{12}! \, \G\big(\f{1-(1- i\g) a_{23}}2\big)}{\G(1+(1-i\g) j_1)}
\sum_{n=-j_1}^{j_1} \f{ (-1)^{j_1+n}  (j_2+j_3-n)!}{(j_1-n)!(j_2-j_3+n)!} 
\nn\\ & \hspace{2cm} \times
\sum_{s}  \f{(-1)^{j_2+s}}{(j_2-s)!(j_3-n+s)!} 
\f{\G\big(\f{1+J -i\g a_{12}}2+s-n\big)\G\big(\f{1+J + i\g a_{13}}2+s\big)}{\G\big(\f{1 + a_{23}-i\g J}2- n\big)\G(1-i\g j_2 +s)\G(1+i\g j_3 +j_1+s_2)},
\end{align}
where
\be
a_{12} =  j_1+j_2-j_3, \quad {\rm etc.}, \qquad \D(j_i) = \left(\f{\prod_i (2j_i)!}{(j_1+j_2+j_3+1)! \prod_{i<j} a_{ij}!}\right)^{1/2}.
\ee
Once again, we can compare the complexity of the EPRL model with the much simpler Barrett-Crane model, for which 
\begin{align}\label{chiBC}
\chi(\r_i,0;0) &= \f1{4\sqrt{2 \r_1\r_2\r_3}} \\\nn &
\times\left(\f{\sinh \pi\r_1 \sinh \pi\r_2\sinh \pi\r_3}
{  \cosh\f\pi2(\r_1+\r_2+\r_3)\cosh\f\pi2(-\r_1+\r_2+\r_3)\cosh\f\pi2(\r_1-\r_2+\r_3)\cosh\f\pi2(\r_1+\r_2-\r_3)}\right)^{1/2}.
\end{align}

Another interesting way of evaluating the coefficients is to use recursion relations, a procedure that can certainly also help speeding up spin foam calculations. This approach was developed in \cite{Rashid70II}. While the norm of the coefficients so obtained coincides with the ones above, the phase conventions are different, hence some care is needed in using the results of \cite{Rashid70II}. For the interested reader, we describe in Appendix \ref{AppSL2C} the detailed comparison.

\subsection{Dipole amplitudes with the finite sums formula} 
To complete this Section, we apply the amended Kerimov-Verdiev formula to express the dipole amplitudes with the $r$ integrals analytically solved.
This is a simple exercise, but it is useful to have explicit formulas handy. Specifically, some of the phases simplify, and we can use the basic identity
\be
\left|\Gamma\big(\tfrac{1-iP+K}2\big)\right|^2 \left|\Gamma\big(\tfrac{1-iP -K}2\big) \right|^2
= \f{2\pi^2}{\cosh(\pi P) + \cos(\pi K)}
\ee
to remove some Gamma functions.
For the 3-valent dipole amplitude we have, starting from \Ref{B3CG}, applying \Ref{chiKer} and paying attention to the various phase factors,
\begin{align}\label{B3Kgen}
B_3(\r_i,k_i;j_i;l_i) 
& = (-1)^{\f{J-L}2}\f\pi{16} \prod_{i=1}^3 \sqrt{d_{j_i}d_{l_i} } \, \f{\overline{\kappa(\r_i,k_i;j_i)} \kappa(\r_i,k_i;l_i)}{\cosh(\pi P)+\cos(\pi K)}.
\end{align}
The reality of this expression is by no means manifest, and indeed, the phase prefactor shows that the rest of the expression can still be purely imaginary. 

For the 4-valent one we proceed in the same way, starting from \Ref{B4CG} this time, and we get 
\begin{align}
B_4(\r_i,k_i;j_i,l_i;j_{12},l_{12}) &= 
(-1)^{\f{J-L}2} (-1)^{j_{12}-l_{12}}  \f{\pi^2}{2^5} d_{j_{12}} d_{l_{12}} 
\prod_{i=1}^4 \sqrt{d_{j_i}d_{l_i} } \int_{0}^\infty d\r_{12} \sum_{k_{12}={\rm -min\{j_{12},l_{12}\}}}^{\rm min\{j_{12},l_{12}\}} (\r_{12}^2+k_{12}^2)
\nn\\\label{B4Kgen} & \hspace{-3.2cm} \times
\f{ \overline{\kappa(j_1,j_2,j_{12}) \kappa(j_3,j_4,j_{12})} \kappa(l_1,l_2,l_{12}) \kappa(l_3,l_4,l_{12})}{[\cos(\pi (k_1+k_2+k_{12}))+\cosh(\pi(\r_1+\r_2+\r_{12}))][\cos(\pi (k_3+k_4+k_{12}))+\cosh(\pi(\r_3+\r_4+\r_{12}))]},
\end{align}
where we used the fact that the integrand is even, as explained above, and a permutation symmetry of the $\chi$'s to rearrange the terms, see \Ref{chiswap} and \Ref{AppPerm}. 

We see that the Kerimov-Verdiev formula allows us to express the 3-valent dipole amplitude in terms of finite sums. The 4-valent amplitude still requires an integration, the one over the virtual irrep $\r_{12}$. Trading the $r$ integral for the one over $\rho_{12}$ turns out to be convenient.
When we use these formulas in the EPRL model, we simply have to restrict the face irreps to the $\g$-simple conditions \Ref{simple}. However, nothing restricts the virtual irreps $\r_{12}$ and $k_{12}$, which are free to take arbitrary values admitted by the Clebsch-Gordan decomposition. 
Hence, we have introduced in this way new, virtual irreps which are off-shell of the simplicity constraints. The existence of non-simple irreps in the EPRL model, albeit virtual ones, was not expected to us, and it would be of valuable interest to understand their geometrical and physical meaning.  
Remarkably though, numerical investigations reported below show that both labels are strongly peaked on the corresponding simple values, $\r^o_{12}=\g j_{12}$ and $k^o_{12}=j_{12}$. The situation is thus somewhat reminiscent of off-shell propagation in Feynman amplitudes, where virtual particles can have off-shell momenta, but the amplitudes are peaked on the on-shell values. From a practical viewpoint, the peakedness has the important consequence that the expression \Ref{B4Kgen} remains much faster to evaluate than the $r$-integral one \Ref{B4}, in spite of containing an indefinite integration itself.

This concludes the main body of analytic results of the paper. We then proceeded to test numerically both formulas \Ref{B3Kgen} and \Ref{B4Kgen} against the integral expressions \Ref{B3} and \Ref{B4}.
These numerical studies highlighted many interesting properties of these amplitudes, and are reported in the next two Sections.

\section{Numerical studies: 3-valent case}\label{SecB3}

In this and the following Section, we report on numerical studies of the 3-valent and 4-valent dipole amplitudes \Ref{Bn}. We restrict attention here to the $\g$-simple irreps \Ref{simple}, those directly relevant for spin foams.
Although the 3-valent case is rarely considered in spin foams, as the quantum geometry associated to it has zero 3-volume, it is useful to consider it first for its simplicity:
There are no intertwiner degrees of freedom, and the dipole amplitude $B^\g_3(j_i,l_i) = B_3(\g j_i,j_i;j_i,l_i)$ is determined by the spins only.
At fixed spins, the $r$ integrals in \Ref{B3} can be performed with Wolfram's Mathematica, and give elementary trigonometric functions, for instance
\begin{align}\nn 
& B_{3}^\g(\tfrac{1}{2},\tfrac{1}{2},1;\tfrac{1}{2},\tfrac{1}{2},1) = \f{\g (1 + 4 \g^2)}{8 (1 + \g^2)^3} \coth^2(\f{\pi\g}2) \tanh(\pi \g),  \qquad
B_{3}^\g(1,1,1;1,1,1) = \f{27 \g (4 + 9 \g^2) \cosh^3(\f{\pi \g}2)}{256(1 + \g^2)^3 \sinh(\f32\pi \g)},
\end{align}
and
\be\nn 
B_{3}^\g(1,1,1;1,2,2) = \f{9\sqrt{5} \g (4 + 9 \g^2)}{2048(1 + \g^2)^3} \Big(3 \coth(\f{\pi \g}2) -\coth(\f{3\pi\g}2) \Big).
\ee
These expressions can be derived also from the finite sums formula \Ref{B3Kgen}.\footnote{To get a flavour for this, recall that 
$\Gamma$ functions are related to trigonometric functions,
for instance $|\G(i y)|^2 = \pi/y\sinh(\pi y)$,
\[
|\G(n+1+iy)|^2 = \f\pi{y\sinh(\pi y)} \prod_{k=0}^n(k^2+y^2), \qquad 
|\G(\tfrac{2n+3+iy}2)|^2 = \f\pi{2^{2n}\cosh(\pi y/2)} \prod_{k=0}^n [(2k+1)^2+y^2].
\]
}
As \Ref{B3Kgen} shows, the number of terms grows with increasing spins. 
In spite of its complexity, the expression in terms of finite sums is useful for analytical manipulations, and we found it to be much faster to evaluate numerically using Wolfram's Mathematica:
For instance, computing the first thousand non-zero values of $B_3$ with spins up to 6 takes us about 20 minutes using the integral representation \Ref{B3}, 
and only 20 seconds using the finite sums expression \Ref{B3Kgen}. Some explicit values are reported in Appendix \ref{AppTables}, comparing evaluation times with the two methods. This said about computing times, we should also add that we are by no means experts on Wolfram's Mathematica nor coding in general, and it is quite likely that the numerical integration can be largely optimised. The numerical evaluations are also a way to explicitly check the correct equivalence of \Ref{B3} and \Ref{B3Kgen}, including the overall sign. 
The two main results emerging from the numerical investigations are the peakedness of the amplitude on the minimal spin configurations $l_i=j_i$, and the asymptotic behaviour for homogeneously large spins.

\subsection{Peakedness on the minimal configurations}

\begin{figure}[ht!]
\begin{center} \includegraphics[width=16cm]{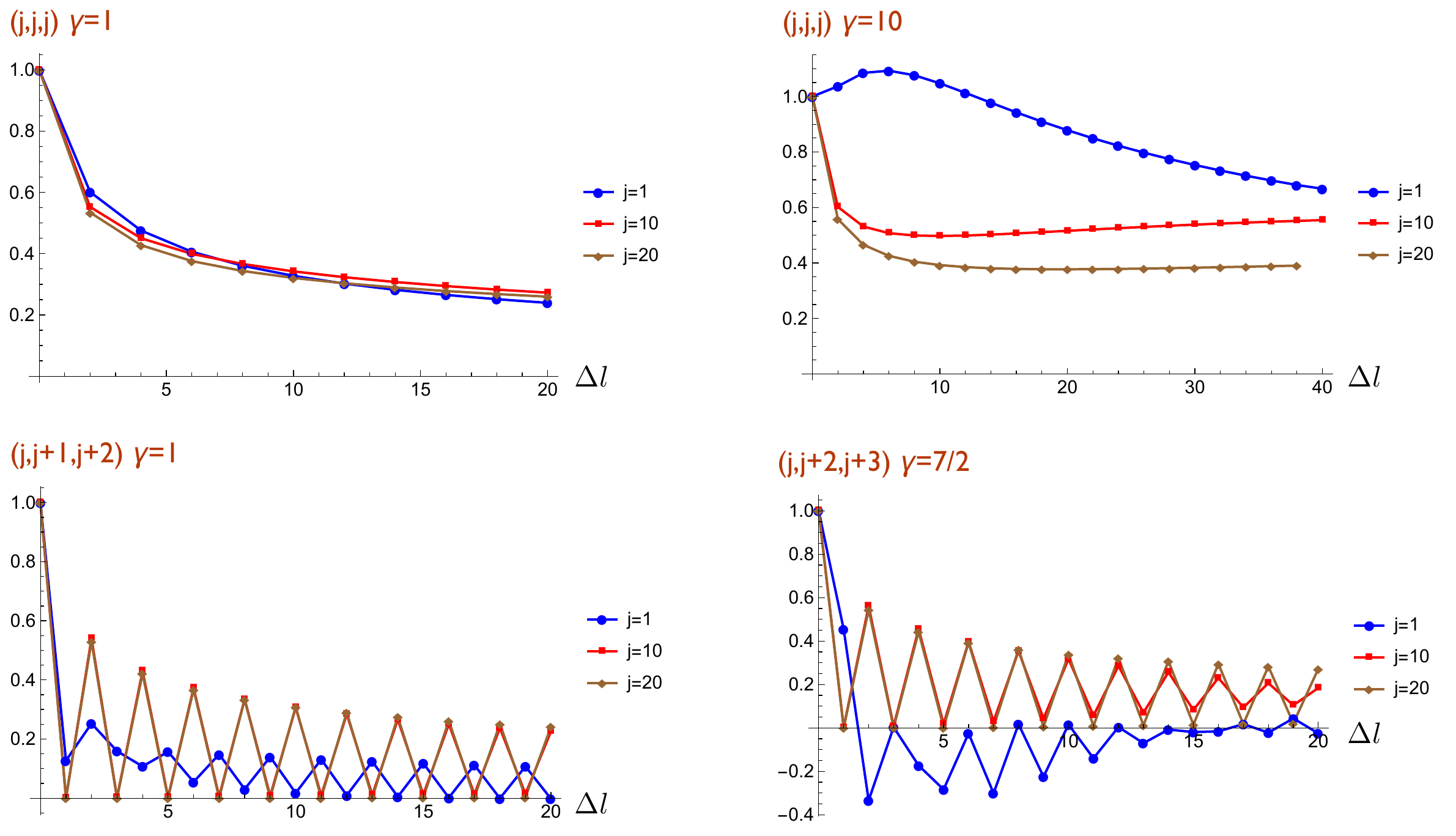}\end{center}
\caption{\label{Figdiagonal} {\small{\emph{Study of the peakedness of $B_3^\g(j_i,l_i)$ on the `spin-diagonal' configurations $l_i=j_i$. {\rm Top panels:}  
cases with equal spins, the plots show
$B_3^\g(j,j,j,j+\D l,j+\D l,j+\D l)/B_3^\g(j,j,j)$ with different values of $j$ (shown in different colours), $\g=1$ (left) and $\g=10$ (right). For $\g=1$, the spin-diagonal configurations dominate at all studied spins, and the decay is a power law, see Fig.~\ref{Fig-peak-minimalB3}. For smaller $\g$ the plots are very similar. On the other hand, local maxima can appear for $\g>1$. For small spins the maximum can even exceed 1, thus shifting the actual peak of the ratio. This we observed only for small spins. 
Notice that here the amplitude vanishes identically for $\D l$ odd, due to symmetries of the Clebsch-Gordan coefficients.
{\rm{Bottom panels}}: examples with non-equal spins, $B_3^\g(j,j+1,j+2,j+\D l,j+1+\D l,j+2+\D l)/B_3^\g(j,j+1,j+2)$ with $\g=1$ (left), 
and $B_3^\g(j,j+2,j+3,j+\D l,j+2+\D l,j+3+\D l)/B_3^\g(j,j+2,j+3)$ with $\g=7/2$ (right).
The values for odd increments are this time non-zero, although they can be 2 orders of magnitude smaller and indistinguishable on the plots. 
Oscillations and negative values are shown in these cases, and the power-law peakedness at least for large spins is manifest.}}}}
\end{figure}

Recall that in the dipole amplitudes we have $l_i = j_i + \D l_i, \D l_i\in \N, \forall i$. 
To investigate the relative weight of non-minimal configurations, we considered homogeneous shifts $\D l_i = \D l \ \forall i$, and numerically evaluated the ratio $B_3^\g(j_i;j_i+\D l)/B^\g_3(j_i;j_i)$ as a function of $\D l$, for various choices of $\g$ and $j_i$. In almost all cases considered, the behaviour shows a clear peak at the spin-diagonal values $\D l=0$, and a monotonic or oscillating decay. See Fig.~\ref{Figdiagonal} for a set of representative examples. 
\begin{figure}[ht!]
\begin{center} \includegraphics[width=7cm]{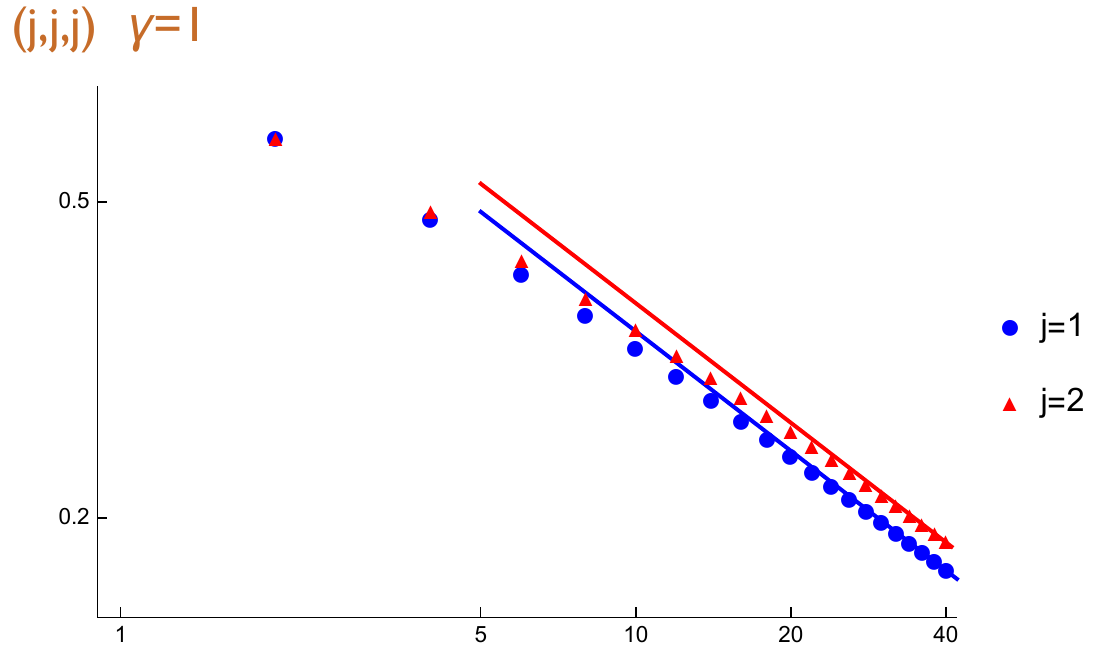}\end{center}
\caption{\label{Fig-peak-minimalB3} {\small{\emph{Fit of the fall-off in $\Delta l$ of $B_3^\g(j_i;j_i+\D l)$ for the simplest configuration, all spins equal. The same function of the top-left panel of Fig.~\ref{Figdiagonal} is here shown on a log-log plot, and together with the numerical best fits: $1.1 \D l^{-1/2}$ for $j=1$ and  $1.8 \D l^{-1/2}$ for $j=2$.}}}}
\end{figure}
The only minor exception we observed occurred at large $\g$ and small $j_i$, see top-right panel of Fig.~\ref{Figdiagonal}, where the peak is at a small but non-zero $\D l$, before the decaying behaviour sets in. Although we could not push the numerics beyond the values shown, it is reasonable to imagine that the observed behaviour of the $j=10$ and $j=20$ lines moves towards a local maximum, smaller that 1, before decreasing again, something like a long period oscillation. So that the maximum at non-zero $\D l$ moves to the right and well beneath the value 1 as $j_i$ are increased.
We found similar situations in all large $\g$ investigated case for which the shifted peak is initially present at small spins. 
Similarly for oscillating decays like at non equal spins in the Figure, increasing $\g$ can increase the amplitude of the oscillations for small $j_i$ (although we did not observe situations where it goes above 1), and increasing $j_i$ the effect is strongly reduced, restoring the same qualitative decay as in the pictures.
Hence, we conclude that the peakedness at the minimal-configurations is a generic feature of the amplitude at large spins, and valid also for small spins at at small $\g$.

Concerning the decay itself, it is in general roughly power-law.
For the simplest case with all spins equal, a best fit gives a power-law decay $\D l^{-1/2}$, see Fig.~\ref{Fig-peak-minimalB3}.
Another way to expose this peakedness is to study the large spin asymptotics for diagonal and non-diagonal configurations, as we show next.

\subsection{Large spin asymptotics}

\begin{figure}[ht!]
\begin{center} \includegraphics[width=14cm]{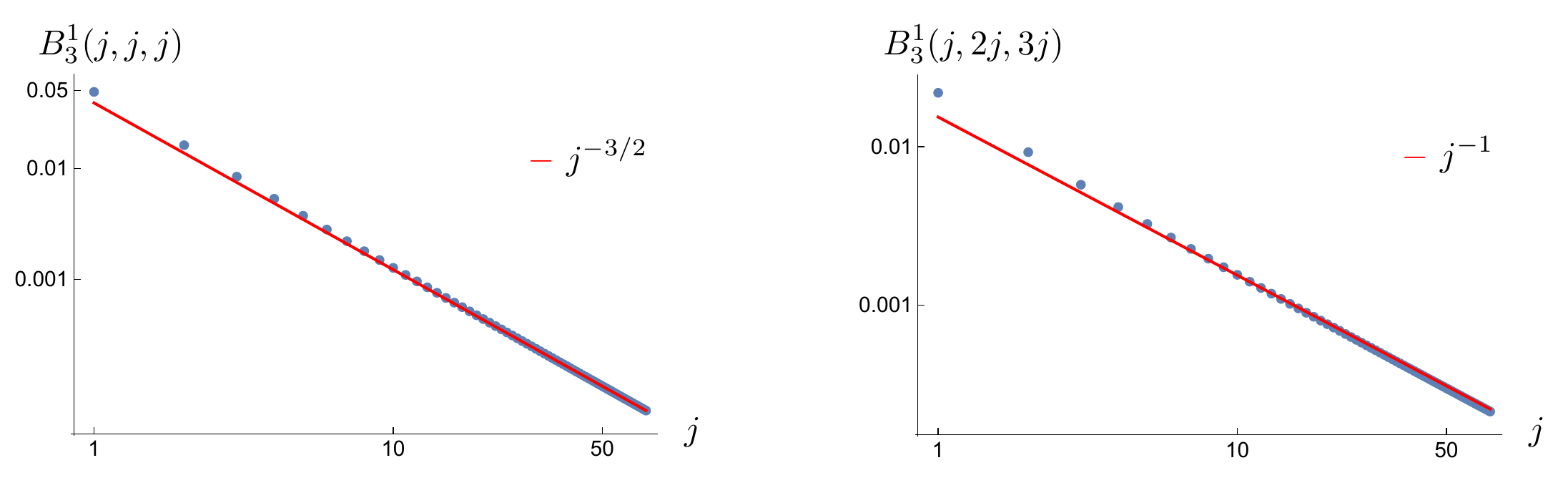}
\end{center}
\caption{\label{FigFit} {\small{Left panel: \emph{Asymptotic behaviour of $B_3^\g(j,j,j)$ as a function of $j$, for $\g=1$, on a log-log plot.  The numerical evaluations are the (blue) dots, the (red) line is a best fit using  data $j\in [50,70]$, which gives $0.0388 \, j^{-3/2}$.} 
Right panel: \emph{Asymptotic behaviour of $B_3^\g(j,2j,3j)$, again for $\g=1$, and fit $0.0154 \, j^{-1}$.}}}}
\end{figure}

We report first the asymptotic behaviour of the spin-diagonal configurations. Evaluation in this case is significantly faster,  especially using the finite sums formula, and it becomes manageable to go up to spins of order $10^2$ in less than an hour. These numerical investigations show that $B_3^\g(Nj_i;Nj_i)$ can have two different power law decays,
\begin{align}\label{B3asymp}
& B_3^\g(N j_i;Nj_i) \sim c_1(\g,j_i)N^{-3/2} \qquad {\rm if} \quad \sum_i j_i = 2n+1, \\
& B_3^\g(N j_i;Nj_i) \sim c_2(\g,j_i)N^{-1} \qquad {\rm if} \quad \sum_i j_i = 2n,
\end{align}
see Fig.~\ref{FigFit} for examples. 
\begin{figure}[h!t]
\begin{center}\includegraphics[width=7cm]{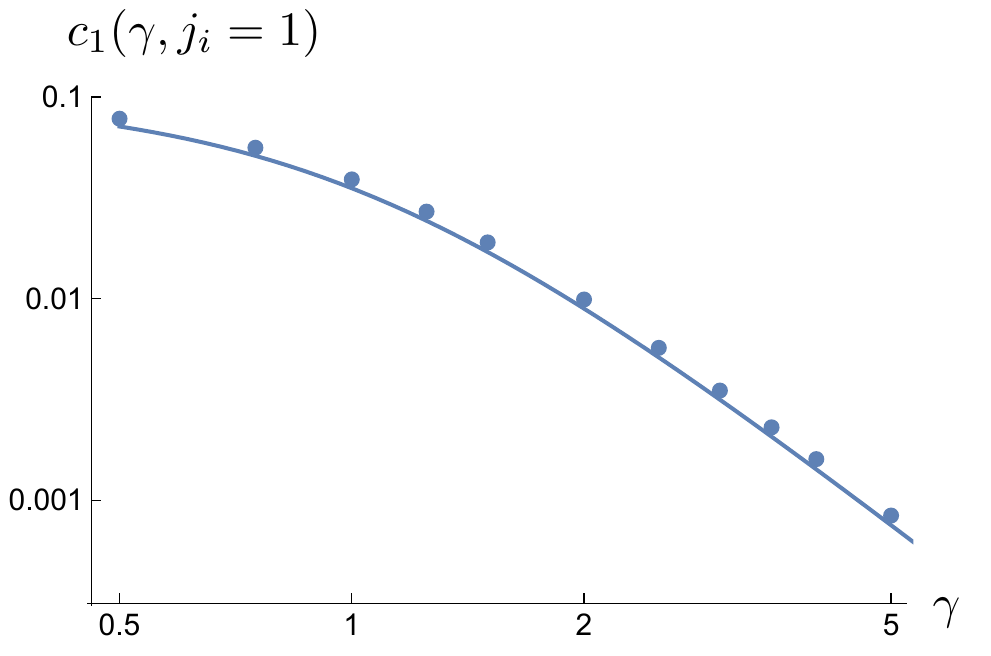}\end{center}
\caption{\label{Figgamma} {\small{\emph{Behaviour of the numerical coefficient of the fit for  $B_{3}^\g(j,j,j;j,j,j)$ at spins of order 40, for varying $\g$, versus the asymptotic formula found in \cite{Puchta:2013lza}. Log--log plot.}}}}
\end{figure}
Experience with saddle point analysis of this type of integrals (e.g. \cite{LS}) suggests that these two behaviours can be explained by a degenerate Hessian in the second case, probably related to the existence of aligned directions in the phase space. 
We also investigated the dependence of the coefficient $c_1$ on $\g$, see Fig. \ref{FigFit}. This matches the result $(1+\g^2)^{-3/2}$ of \cite{Puchta:2013lza} (see \Ref{Puchta} below), although the precise numerical agreement is slightly short. The discrepancy could be due to the spins in our numerics not being high enough.

\begin{figure}[ht!]
\begin{center}\includegraphics[width=16cm]{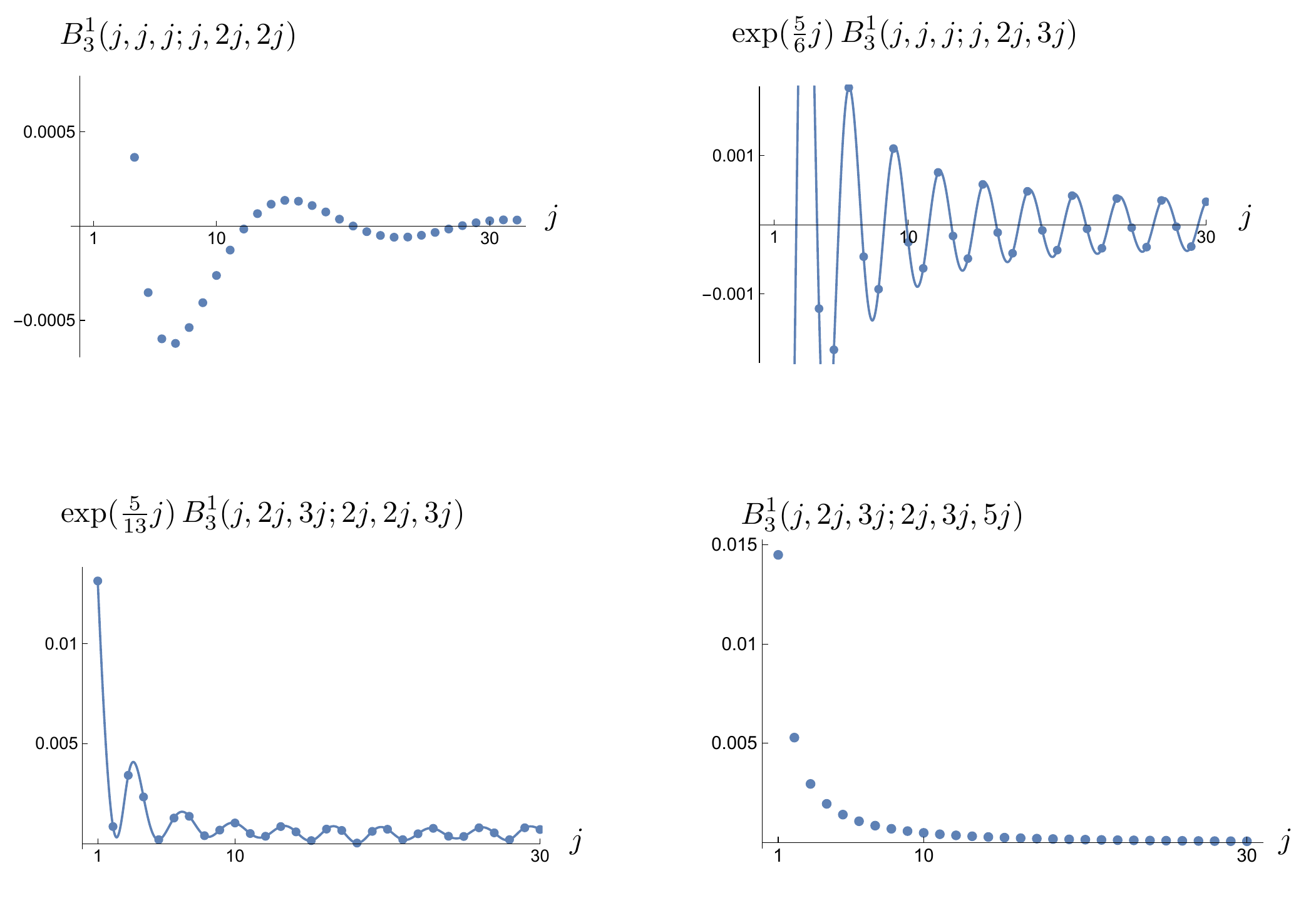}
\end{center}
\caption{\label{FigNonDiag} {\small{\emph{Examples of asymptotic behaviour in the non-diagonal case, showing a variety of different possibilities, mixing power laws and exponential decays with or without oscillations. In the former case, we have added an interpolating line to help the visualisation. {\rm Top-left:} power-law damping $\sim j^{-2}$ with oscillations; {\rm top-right:} oscillations and exponential decay (notice a rescaling by $\exp(5j/6)$ has been artificially added to enhance the visibility of the plot); {\rm bottom-left:} also oscillations and exponential decay, but always with positive values (and again an artificial rescaling); {\rm bottom-right:} exponential decay without oscillations.}}}}
\end{figure}

Next, we considered the asymptotic behaviour for non diagonal cases with different spins but all homogeneously rescaled, that is $B^\g_3(N j_i; Nl_i)$.
The numerical evaluations are a bit slower, but can still be done efficiently using the finite sums expression.
A variety of different behaviours is now possible, including power laws and exponential decays, with or without oscillations. 
In no case we found a behaviour as slow as $j^{-3/2}$, thus providing further evidence for the peakedness of the amplitude on the diagonal configurations.
A representative selection of various cases is shown in Fig.~\ref{FigNonDiag}.

Clearly, these numerical studies point out the existence of an interesting zoo of different behaviours, and plead for an analytic investigation. %
At least for the case of diagonal configurations, a useful tool would be the use of spinors. For instance, using the generating function of Wigner's $3jm$-symbols \cite{Varshalovich} (see also \cite{Bonzom:2012bn,Alesci:2016dqx}) and the spinorial formalism for $\SL(2,\C)$ as in \cite{BarrettLorAsymp,IoWolfgang}, \Ref{B3} can be written as
\be\label{B3spinors}
B^\g_3(j_i) =\D(j_i) \D(l_i) \int_0^\infty d\m(r) \prod_{i=1}^3  \int d\m(\z_i) \int d\m(\z'_i) 
\int \f{ d\m(\om_i)}{||\om_i||^2 \, d(r,|\om_i^1|^2/||\om_i||^2)} e^{S(r, \om_i, \z_i,\z'_i)},
\ee
with all the integrals over ${\C P^1}$ with its SU(2)-invariant measure $d\m(z)$, and
\begin{align}
S(r, \om_i, \z_i,\z'_i) 
& = \sum_{i=1}^3 j_i \left[ \ln \f{ \bra{\z_i} e^{\tfrac r2 \s_3} | \om_i\ra^2 \bra{\om_i}\z'_i\ra^2}{d(r,|\om_i^1|^2/||\om_i||^2)^{1+i\g}} + \ln \f{[\z_i\ket{\z_{i\hat +1}}[\z_{i\hat +2}\ket{\z_i}}{[\z_{i\hat +1}\ket{\z_{i\hat +2}}} +  \ln \f{\bra{\z'_{i\hat +1}}\z'_i] \bra{\z'_i}\z'_{i\hat +2}]}{ \bra{\z'_{i\hat +2}} \z'_{i\hat +1}]}  \right].
\end{align}
Here $\ket{\om}\in\C^2$ is a spinor, $||\om||^2=|\om^0|^2+|\om^1|^2$, $d(r,t)$ is defined in \Ref{ddef} in the Appendix, $\ket{\z}:=n(\z)\ket{\tfrac12,-}$, $|\z]:=n(\z)\ket{\tfrac12,+}$ are Perelemov's coherent states in the fundamental representation, and finally $\hat +$ is a sum modulo 3.
A more detailed analytic study of the asymptotic scaling is postponed to future work \cite{IoSL2Casymp}.

\section{Numerical studies: 4-valent case}\label{SecB4}

The 4-valent dipole amplitude \Ref{B4} is the case of most common interest in LQG and spin foams, and we study it in details in this Section, restricting again attention to the simple irreps \Ref{simple}. For short-hand, we will denote it $B^\g_4(j_i;l_i;j_{12},l_{12}):=B_4(\r_i,k_i;j_i;l_i;j_{12},l_{12})$, consistently with \Ref{B4}. It is the simplest amplitude with non-trivial intertwiners, and  
geometrically, it corresponds to a quantum tetrahedron being boosted among adjacent frames:
The two sets $j_i$ and $l_i$ describe the four areas of the tetrahedron in the two frames connected by a boost, and the two intertwiners, say $j_{12}$ and $l_{12}$, describe the quantum  intrinsic shape of the tetrahedron.\footnote{We recall this is half the information needed to characterise the classical intrinsic shape, see e.g. \cite{IoPoly} for more on the polyhedral picture of intertwiners.}
Again, we performed numerical simulations using both the $r$-integral expression \Ref{B4} and the one obtained using Kerimov-Verdiev's finite sums formula \Ref{B4Kgen}: our numerical studies served first as a test ground to check the equivalence of the two expressions, and then to study various properties and asymptotic behaviours of the amplitude. Anticipating on the results presented in details below, 
the amplitudes are dominated by the minimal $l_i=j_i$ configurations and by diagonal intertwiners, and show a power-law decay $N^{-3/2}$ for large spins:
\begin{align}\label{B4asymp}
& \sqrt{d_{j_{12}}} \sqrt{d_{l_{12}}} B_4^\g(N j_i;N j_i;j_{12},l_{12}) \sim c(\g,j_i,j_{12})N^{-3/2} \d_{j_{12}l_{12}}.
\end{align}
The dimensional factors on the left are added for convenience since our $4jm$-symbols used in $B_4$ are not normalised. 
These numerics confirm the results of \cite{Puchta:2013lza}, where this leading order estimate was obtained with a saddle point analysis. 
The decay away the non-minimal configurations is again roughly power law, and the square of the 3-valent case, that is
\begin{align}\label{B4asymp}
& B_4^\g(j_i; j_i + \D l;j_{12},l_{12}) \sim \D l^{-1}.
\end{align}
For the suppressed, non-minimal configurations $l_i\neq j_i$, the large spin asymptotics can have various behaviours, including exponential fall-offs and oscillations, and can be peaked at non-equal values of the intertwiners. The peak on the intertwiner labels, whether equal (minimal spins) or not (non-minimal spins), is generically sharper at small $\g$, and broader at large $\g$.

\subsection{Off-shell peakedness}

As anticipated below \Ref{B4Kgen}, the numerical evaluation of \Ref{B4Kgen} for simple irreps is still faster than that of \Ref{B4}, in spite the fact that both contain an infinite integration.
This is because the integrand in the case of \Ref{B4Kgen} is significantly localised. In fact, both the integrand and the summand in \Ref{B4Kgen} turn out to be strongly peaked on the values that would solve the $\g$-simple conditions, as we now report.

\begin{figure}[ht]
\begin{center}\includegraphics[width=15cm]{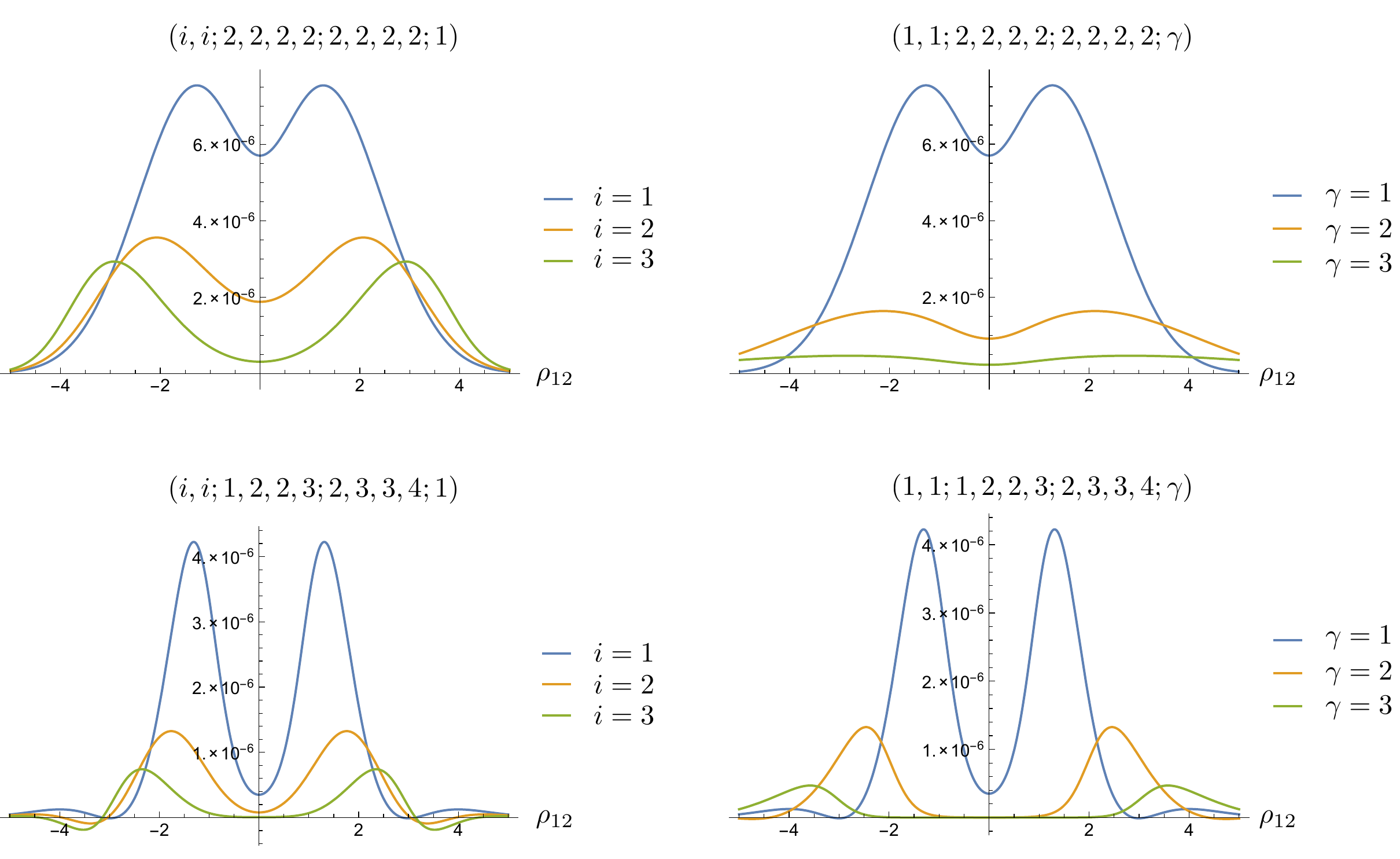} \end{center}
\caption{\label{Fig-peak-rho} {\small{\emph{Plots of the integrand of \Ref{B4Kgen} for simple irreps, showing the approximate peakedness of the virtual irrep $\r_{12}$ on $\g j_{12}$, for $l_{12}=j_{12}$, as well as the parity of the integrand in $\rho_{12}$, a consequence of \Ref{parity-intsum}. The spins, intertwiners and $\g$ values used are reported above each plot.} Top panels: \emph{A minimal configuration with all spins equal $2$ and intertwiners equal to $1$, moving the peak with $j_{12}$ (left, maxima at $1.27$, $2.07$, $2.93$) and $\g$ (right, maxima at $1.27$, $2.12$, $2.76$).} Bottom panels: \emph{A non-minimal configuration with all spins different, again moving the peak with $j_{12}$ (left, maxima at $1.31$, $1.76$, $2.35$) and $\g$ (right, maxima at $1.31$, $2.46$, $3.57$). For the suppressed non-diagonal configurations with $l_{12}\neq j_{12}$, not shown here, the position of the peak lies typically between the two values, much closer to the minimal one. } }}}
\end{figure}

To study the peakedness in $\rho_{12}$, we considered the integrand of \Ref{B4Kgen}, defined including the summation over $k_{12}$, for various configurations. See Fig.~\ref{Fig-peak-rho} for examples. We generically observed an oscillating behaviour of the integrand, with a clear principal peak. For diagonal intertwiners, $j_{12}=l_{12}$, the principal peak lies at approximately $\g j_{12}$; the precise location of the maximum and shape of the peak depend on the values of the spins and of $\g$: It is sharper for the minimal configurations $l_i = j_i$, and broader for very non minimal ones; It broadens also as $\g$ is increased. For non-diagonal intertwiners (that have suppressed amplitudes), the integrand is still peaked, this time on values which lie in between $j_{12}$ and $l_{12}$, typically closer to the smaller of the two, and the secondary peaks become more important.

\begin{figure}[ht]
\begin{center}\includegraphics[width=15cm]{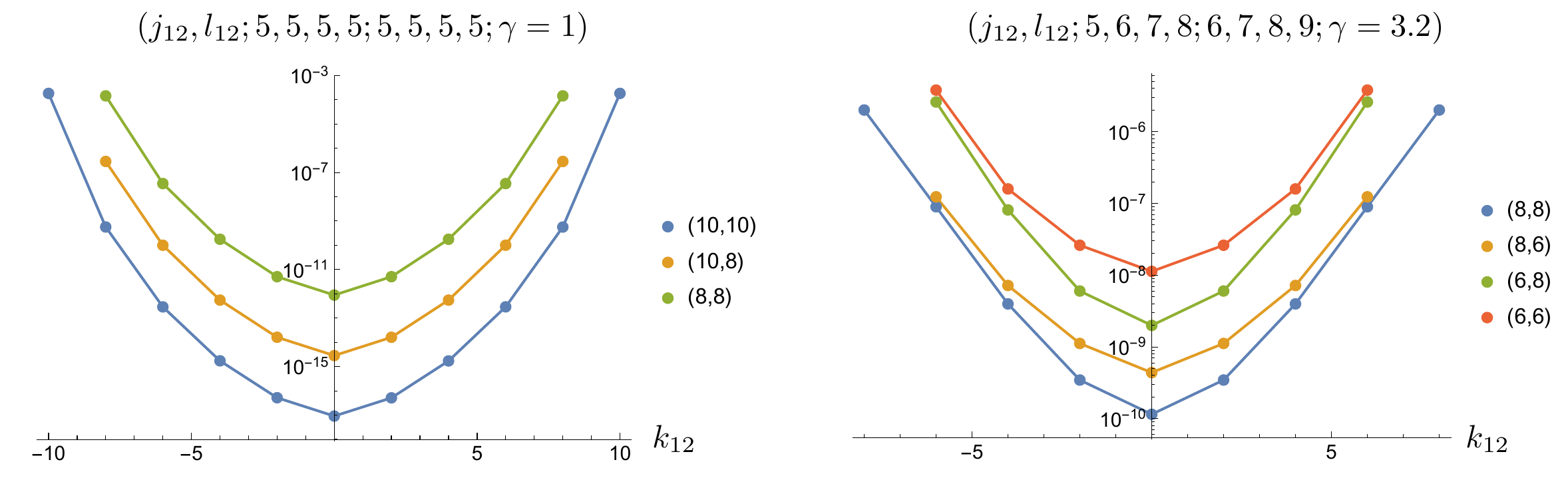} \end{center}
\caption{\label{Fig-peak-k12} {\small{\emph{Plots of the summand of \Ref{B4Kgen} for simple irreps, showing the peakedness of the virtual irrep $k_{12}$ on ${\rm min}\{j_{12},l_{12}\}$. With report a configuration with equal spins (left panel) and one with different spins (right panel), for both diagonal and non diagonal intertwiners. Both on a log plot, these examples show an exponential peakedness on the maximally allowed value.} }}} 
\end{figure}

To study the peakedness in $k_{12}$, we considered the summand of \Ref{B4Kgen}, defined including the integral over $\r_{12}$. For all considered cases, minimal or non-minimal spins, diagonal or non-diagonal intertwiners, small or large $\g$, the plots show an exponential peak at the maximally allowed value $k_{12} = {\rm min}\{j_{12},l_{12}\}$. See Fig.~\ref{Fig-peak-k12} for examples.

Finally, notice also from the pictures that both the integrand and the summand are even functions, as expected from \Ref{parity-intsum}.

\subsection{Peakedness on diagonal intertwiner labels}

\begin{figure}[ht]
\begin{center}\includegraphics[width=7cm]{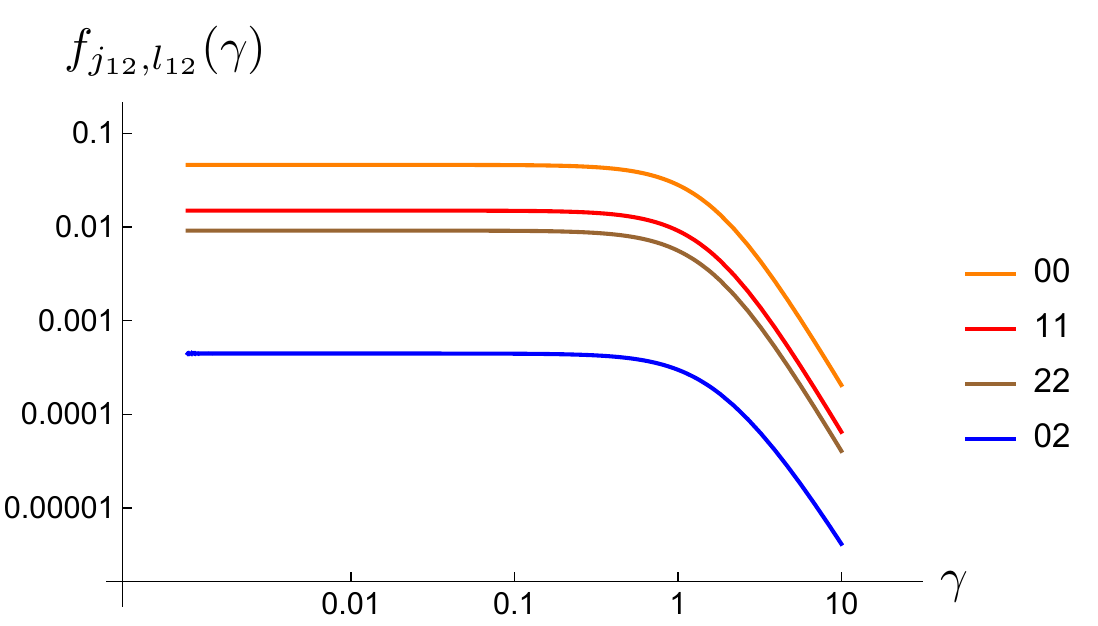} \hspace{1cm} \includegraphics[width=7cm]{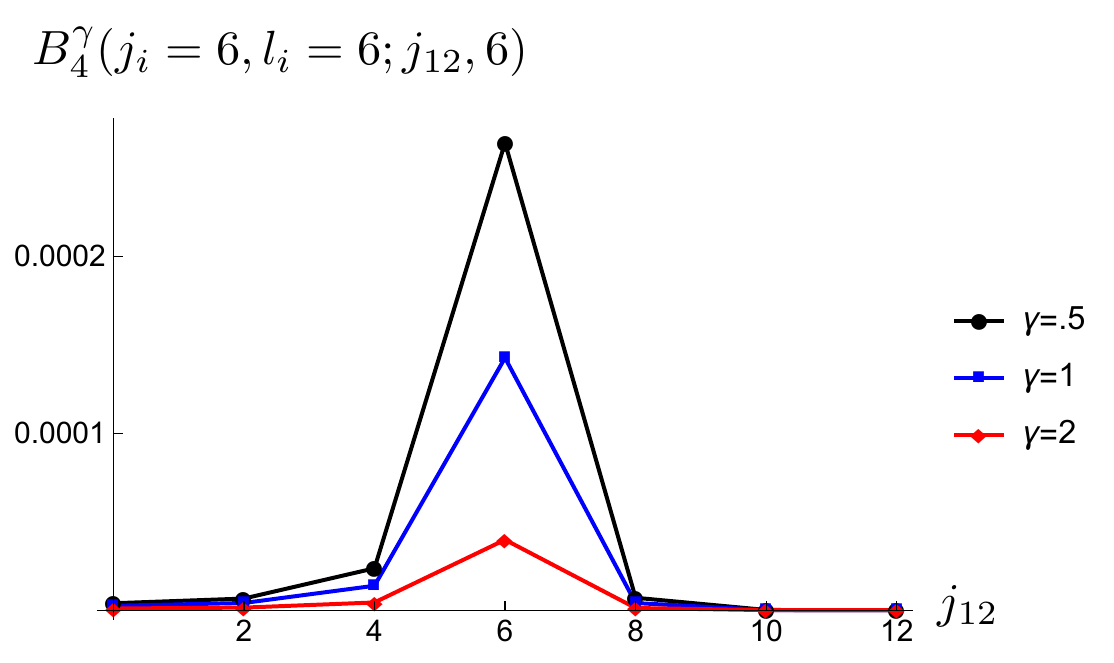}\end{center}
\caption{\label{FigB4-int-peak-1} {\small{\emph{Peakedness on equal intertwiners shown for $j_i\equiv l_i\equiv 6$.} Left panel: \emph{Behaviour of \Ref{f66} for different values of $(j_{12},l_{12})$, log--log plot. We see that the non-diagonal intertwiner configuration is suppressed. The apparent different magnitudes of the diagonal ones is mainly due to the lack of normalisation of the $4jm$-symbol.
} Right panel: \emph{Peakedness on equal intertwiners shown for the example $j_i\equiv l_i\equiv 6$, $l_{12}=6$, showing how it flattens as $\g$ increases.} }}}
\end{figure}

Again, it is possible to compute explicit analytic values of \Ref{B4} at fixed spins with Mathematica. For example, the first two  configurations with all spin equals, $j_i=l_i=j$, give
\begin{align}
B^\g_4(j_{12},l_{12};j_i=\tfrac12,l_i=\tfrac12) = \d_{j_{12}l_{12}}
\f4{15\pi} \frac{4 \pi \g (1-\g^4)  \left(\coth (\pi  \g) - 2 \tanh \left(\frac{\pi  \g }{2}\right) \right) + 15  \g^2 (2 - \g^2) - 3}{(1+\g^2)^4}
\end{align}
and
\begin{align}
B^\g_4(j_{12},l_{12};j_i=1,l_i=1) &= \d_{j_{12}l_{12}}\,  f_{j_{12},l_{12}}(\g) + \d_{j_{12},l_{12}\pm2} \, f_{j_{12},l_{12}}(\g),
\\\label{f66}  f_{j_{12},l_{12}}(\g) &= \frac{4\pi \g  \, p^1_{j_{12}l_{12}}(\g) \, \left(3 \coth (\pi  \g) - \tanh (\pi  \g) \right) + p^2_{j_{12}l_{12}}(\g)}{560 \pi \g^4 (1+\g^2)^4 (j_{12}+1)(2j_{12}+1)(l_{12}+1)(2l_{12}+1)},
\end{align}
where $p^i_{j_{12}l_{12}}(\g)$ are even polynomials of order 8 in $\g$ which we do not report here. A plot of $f_{j_{12}l_{12}}(\g)$, see left panel of Fig.~\ref{FigB4-int-peak-1}, shows that these coefficients are much larger for diagonal intertwiner labels, $j_{12}=l_{12}$, than for the only admissible non-diagonal value. Notice also from the picture that the diagonal values appear to have a clear hierarchy among them. This reflects the lack of normalisation of the intertwiners, and it would mainly wash out if we multiply $B_4$ by $\sqrt{d_{j_{12}}}\sqrt{d_{l_{12}}}$.
The peakedness on diagonal intertwiner labels, sharp at small $\g$ and broader at large $\g$, turns out by numerical explorations to be a generic characteristic of the minimal configurations, see right panel of Fig.~\ref{FigB4-int-peak-1} for an example.
This diagonal behaviour is suggestive that the integral over $r$ has a saddle point at $r=0$, 
as indeed estimated in \cite{Puchta:2013lza} and used to derive 
\Ref{B4asymp}  analytically.
Notice that the Kronecker delta would be the exact result for an ordinary SU(2) dipole amplitude.
Since our numerical investigations show that the next-to-leading order to \Ref{B4asymp} is suppressed for small $\g$, the EPRL model appears to resembles the more and more an SU(2) theory in this limit.

\begin{figure}[ht]
\begin{center}\includegraphics[width=15cm]{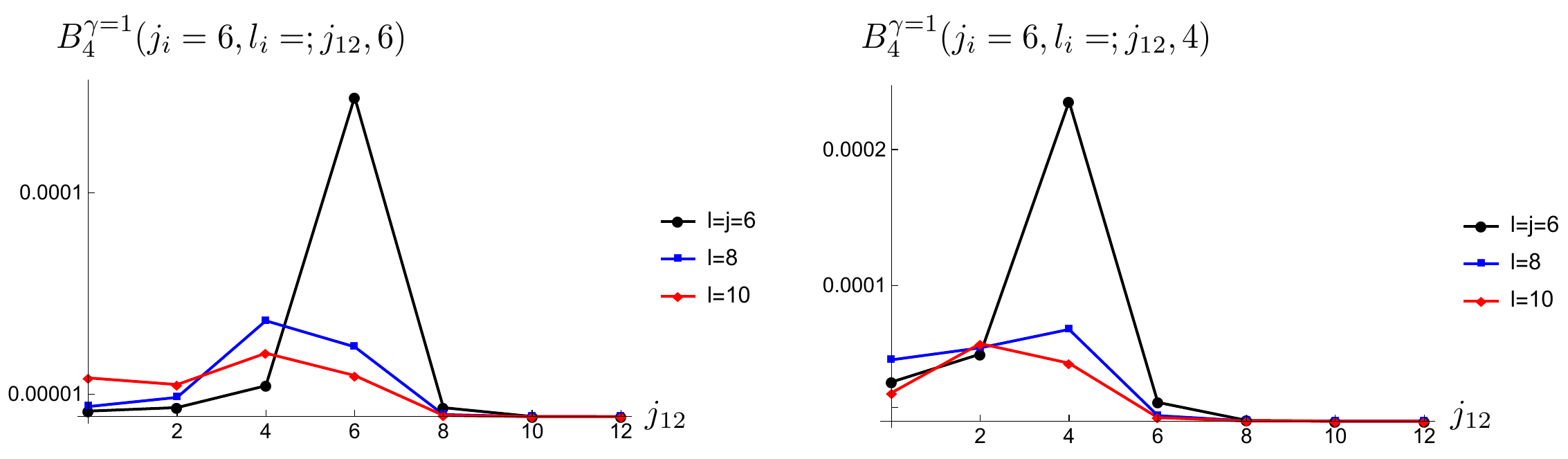}\end{center}
\caption{\label{FigB4-peak-int-6-4} {\small{\emph{Variation of the peak as non minimal configurations are considered: Increasing $l_i$ from their minimal value the peak lowers in magnitude and may also shift to non diagonal values of the intertwiners.}}}}
\end{figure}

On the other hand, the peak can move away from diagonal intertwiners for non-minimal configurations, see Fig.~\ref{FigB4-peak-int-6-4}.
But since the non-minimal configurations are sub-dominant (see next Subsection), these contributions are also next-to-leading order, and the overall dominant behaviour remains \Ref{B4asymp}.

\subsection{Peakedness on the minimal configurations}

\begin{figure}[ht!]
\begin{center} \includegraphics[width=16cm]{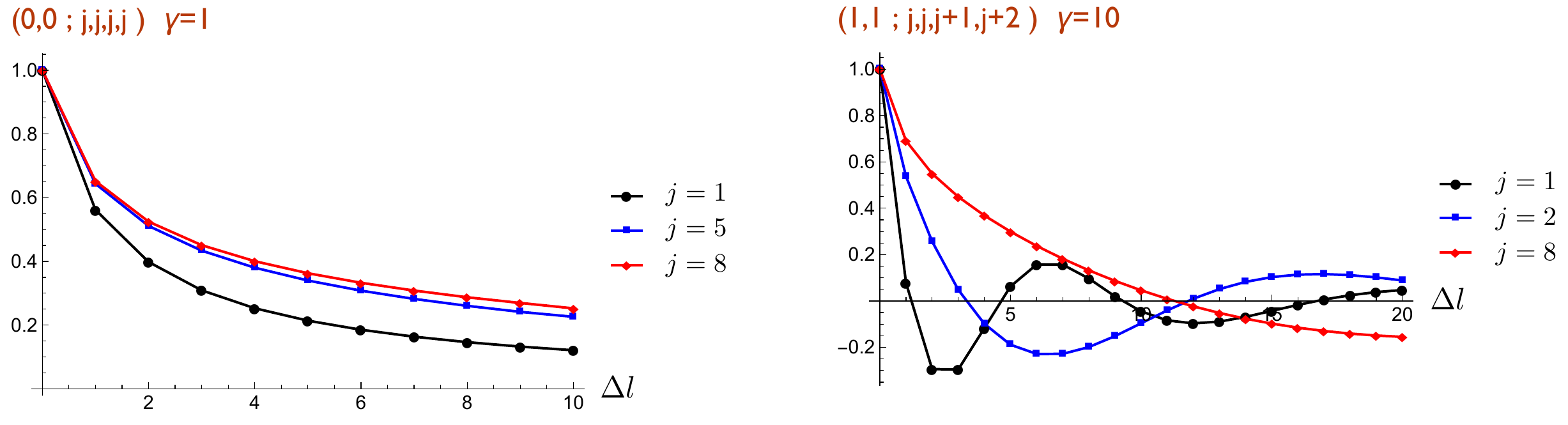}\end{center}
\caption{\label{FigB4diagonal} {\small{\emph{Peakedness of $B_4^\g(i,k;j_i;l_i)$ on the minimal configurations $l_i=j_i$, with fixed intertwiners. {\rm Left panel:}  
cases with equal spins, the plots show
$B_4^\g(0,0;j,j,j,j;j+\D l,j+\D l,j+\D l,j+\D l)/B_4^\g(0,0;j,j,j,j;j,j,j,j)$ with different values of $j$ (shown in different colours), $\g=1$. 
{\rm{Right panel}}: example with non-equal spins, $B_4^\g(1,1;j,j,j+1,j+2;j+\D l,j+\D l,j+1+\D l,j+2+\D l)/B_4^\g(1,1;j,j,j+1,j+2;j,j,j+1,j+2)$ with $\g=10$. }}}}
\end{figure}

\begin{figure}[ht!]
\begin{center} \includegraphics[width=7cm]{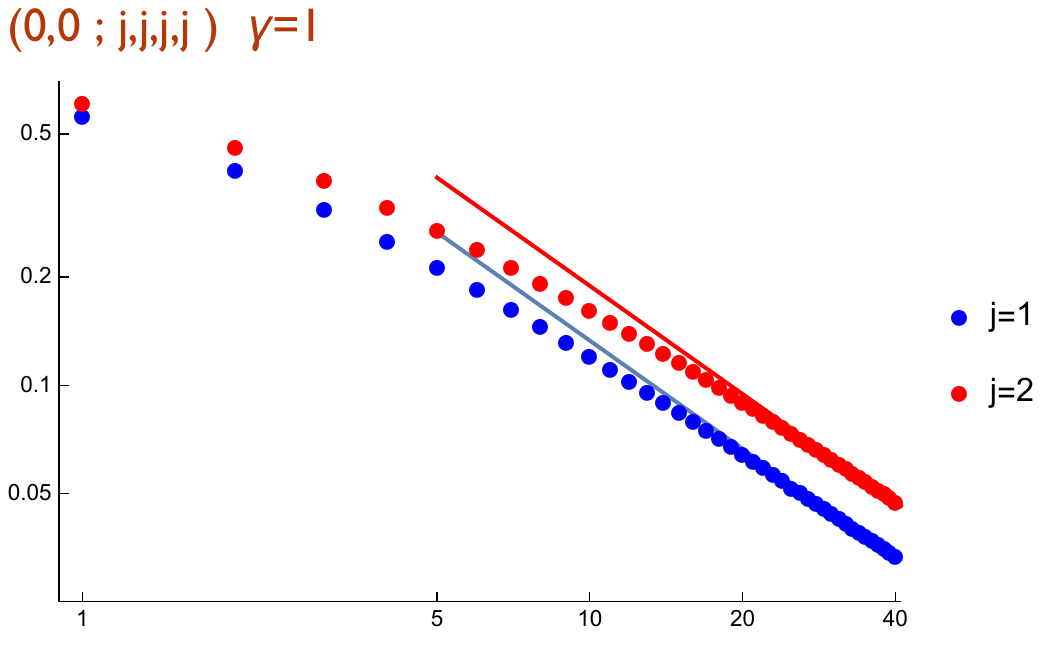}\end{center}
\caption{\label{Fig-peak-minimalB4} {\small{\emph{Fit of the fall-off in $\Delta l$ of $B_4^\g(j_i;j_i+\D l;0,0)$ for the simplest configuration, all spins equal and zero intertwiners. The same function of the left panel of Fig.~\ref{FigB4diagonal} is here shown on a log-log plot, and together with the numerical best fits:  $1.3 \D l^{-1}$ for $j=1$ and  $1.9 \D l^{-1}$ for $j=2$.}}}}
\end{figure}

The next result we want to show is that as for the 3-valent amplitude, the 4-valent amplitude is also dominated by the minimal configurations $l_i=j_i$. See Fig.~\ref{FigB4diagonal}, where for convenience we kept the interwiner labels fixed. The observed behaviour is qualitatively similar to the 3-valent case. In particular, we have a nice power-law drop-off in the case of equal spins, and oscillations for non equal configurations. For the equal spin case, 
taking the smallest possible spins to keep reasonable computing times, the plots in Fig.\ref{Fig-peak-minimalB4} show a decay like $\D l^{-1}$, consistent with the square of the 3-valent case.
This time we did not observe shifted maxima for large $\g$ as in the 3-valent case, see Fig.~\ref{Figdiagonal}. On the other hand, qualitatively similar figures were observed for non-diagonal intertwiners. Since these are subdominant contributions anyways, we refrain from showing these plots here.

\subsection{Large spin asymptotics}

\begin{figure}[h!t]
\begin{center}\includegraphics[width=15cm]{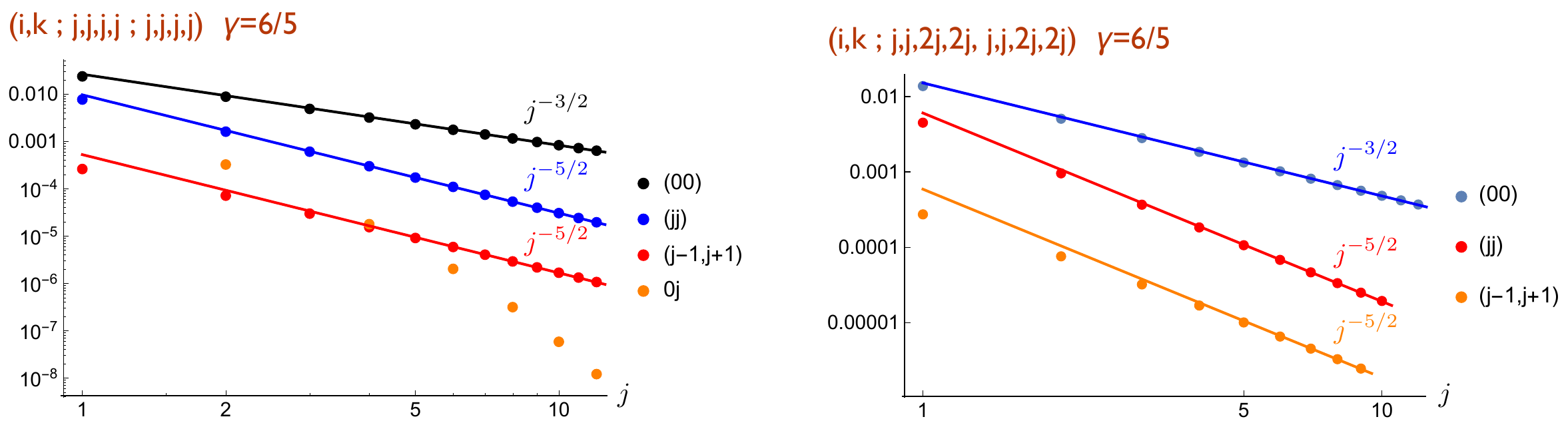}\end{center}
\caption{\label{FigB4-asymp} {\small{\emph{Asymptotic behaviour of $B_4^\g$ for two minimal spin configurations $l_i\equiv j_i$, for different intertwiner labels $j_{12}$ and $l_{12}$, and corresponding fits. The exponential suppressed case $(0j)$ is not reported on the second plot. Other configurations show similar behaviours.
}}}}
\end{figure}

For equal spins and at fixed intertwiners, the large spin behaviour of the amplitude shows a clear power-law fall-off, see Fig.~\ref{FigB4-asymp}.
The different powers shown in the plots reflect the lack of normalisation of the $4jm$ symbol, see \Ref{CG4ortho}; if we rescale $B_4$ as in \Ref{B4asymp}, all diagonal intertwiner contributions scale like $N^{-3/2}$. The figure also shows clearly the suppression of amplitudes with non-diagonal intertwiner labels, which are at least one order of magnitude smaller.

\begin{figure}[h!t]
\begin{center}\includegraphics[width=6cm]{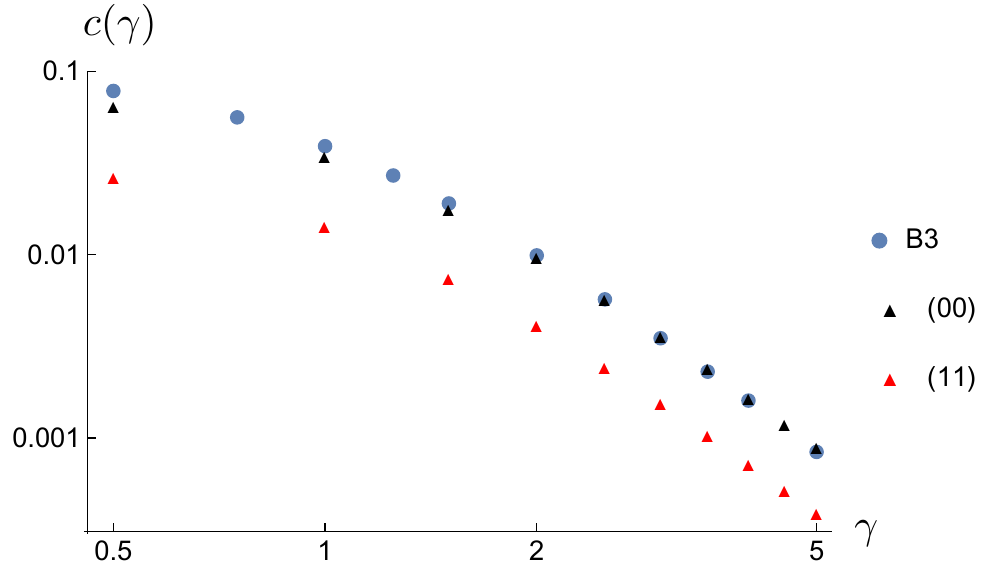}
\hspace{1cm}
\includegraphics[width=7cm]{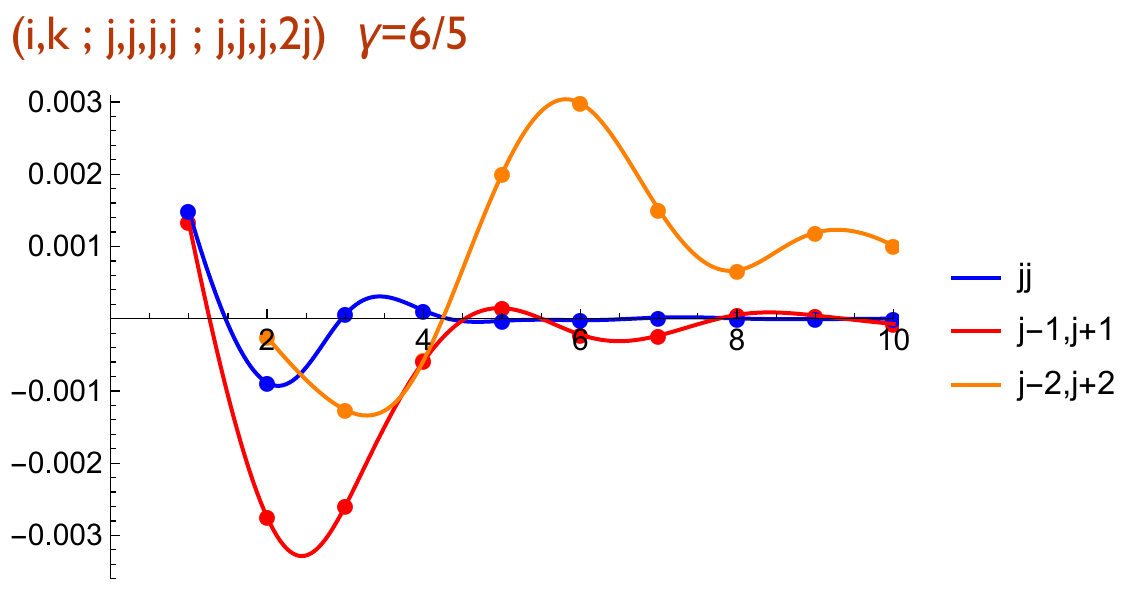}\end{center}
\caption{\label{FiggammaB4} {\small{Left panel: \emph{Behaviour of the numerical coefficient of the fit for  $B_{4}^\g(j_i=j;l_i=j;j_{12},l_{12})$ on a log--log plot, for varying $\g$ and two different diagonal interwiner configurations, in triangles. For comparison we plotted $c_1(\g)$ for $B_{3}^\g(j_i=j;l_i=j)$, round dots.}
Right panel: \emph{Example of oscillating, exponential decay for a non-minimal spin configuration. The plot shows $\exp(3/2j)\, B_4^\g(j_i=j;j,j,j,2j;j_{12},l_{12})$ as a function of $j$, for $\g=6/5$ and three different values of intertwiners. The artificial rescaling by $\exp(3/2j)$ is made for ease of visualisation only. We observe that in this subdominant non-minimal case, diagonal or non-diagonal intertwiner labels can have the same scaling.}}}}
\end{figure}
We also studied the dependence of the coefficient $c(\g,j_i,j_{12})$ of \Ref{B4asymp} on $\g$, for equal spins and different values of the intertwiners, see left panel of Fig.~\ref{FiggammaB4}. For $j_{12}=l_{12}=0$ this is basically the same as for the 3-valent asymptotics \Ref{B3asymp}, reported for comparison in the picture. Increasing the intertwiners appears to only introduce a constant shift, thus suggesting that $c(\g,j_i,j_{12})= \tl c(\g,j_i) f(j_{12})$. 
In \cite{Puchta:2013lza} the following estimate was given:
\be\label{Puchta}
\sqrt{d_{j_{12}}} \sqrt{d_{l_{12}}} B_4^\g(N j_i;N j_i;j_{12},l_{12}) \sim \f1{(4\pi)^2}\left[\f{6\pi}{(1+\g^2)N\sum_{i}j_i}\right]^{3/2} \, \d_{j_{12}l_{12}}.
\ee
Our numerical analysis clearly confirms the $N^{-3/2}$ scaling and peak on diagonal intertwiners. As for the $(\g,j_i,j_{12})$ dependence, while this formula worked with good accuracy in the (non-degenerate) $B_3$ case, it appears to be less accurate in the case of $B_4$, at least at relatively small spins: We performed checks with spins of order 30, and the formula was typically off by a factor of order 1. See for instance in the left panel of Fig.~\ref{FiggammaB4} the matching of the equal spins scalings of $B_3$ and $B_4$ with $j_{12}=l_{12}=0$, whereas a relative ratio $(4/3)^{3/2}\sim 1.54$ would be expected from \Ref{Puchta}. The $(1+\g^2)^{-3/2}$ appears to be in good qualitative agreement, but further numerics also suggest an additional explicit dependence on $j_{12}$, with the accuracy of the formula improving for large $j_{12}$. More work will be needed to establish whether this is a limitation of \Ref{Puchta}, or simply the fact that we did not push the asymptotic behaviour to high enough spins.\footnote{The estimate \Ref{Puchta} was on the other hand derived for arbitrary valence, and we run some numerical tests with the 5-valent dipole amplitude, confirming that for the minimal configurations $l_i=j_i$, the leading order large spin behaviour is diagonal in the two pairs of intertwiners, and scales like $N^{-3/2}$.}

For the non-minimal spin configurations, again the situation is richer, with exponential decays and oscillations also present. Furthermore, in this sub-dominant configurations the amplitudes are no more necessarily peaked at diagonal values of the intertwiners, see right panel of Fig.~\ref{FiggammaB4}.

\section{Scaling behaviour of the simplified EPRL model}\label{SecSimplified}

To control the properties of a spin foam model under refinement and rescaling, it is necessary to know its scaling properties when all spins are large. Not much is known in this context, with explicit results basically boiling down to the Euclidean estimate at $\g=0$ of \cite{IoCarloClaudio}, the Lorentzian self-energy analysis of \cite{Riello13}. As an application of the results here exposed, in particular the factorisation property \Ref{Zgen} and the observed asymptotic scalings \Ref{B3asymp} and \Ref{B4asymp}, we provide in this Section a simple estimate of the large spin scaling of the simplified model on a fixed foam. The result is useful to study divergences and renormalisation properties, and preliminary to the more refined estimate of the degree of divergence, which further requires studying how many of the bulk spin summations are unbounded and diverge. Here we are merely studying how the whole amplitude rescales as all the spins become large, and not the way the regularised amplitude scales with a cut-off.
We restrict attention to the simplified model: extending our analysis the the full model will require deepening and extending the control on the sub-dominance of the non-minimal configurations. 
We discuss two specific examples: first, the case of a foam with all edges 3-valent, a case possibly of not direct relevance to quantum gravity models for the vanishing of quantum 3-volumes, but simpler and of interest of tensor models. Then, all 4-valent, the standard case. In both cases we will consider a closed 2-complex, the generalisation to an open one with boundary is straightforward.
 
If we assume that all edges are 3-valent, all vertices of the foam must have even-valued valence, say $2n_v$, to allow a correct routing of all the strands. A vertex of valence $2n_v$ has amplitude given by a $3n_vj$-symbol,\footnote{The precise symbol will depend on the specific routing of the strands; however the scaling behaviour is always the same.} whose asymptotic behaviour is \cite{Varshalovich} 
\be
\{3n_v j\} \sim (j^{-3/2})^{n_v-1}.
\ee
Then, if we restrict attention to the simplified model, we can use the asymptotic behaviour exposed in \Ref{B3asymp}. Clearly it is the slower behaviour that matters when investigating the overall scaling, therefore we can assume the second term to be dominant. Each edge contains two dipole amplitudes, and thus tributes a scaling factor $N^{-2}$. 
Collecting the results, we have
\be
Z_{{\cal C}_3}^{EPRLs} \sim N^{F-2E-3/2\sum_{v}(n_v-1)} = N^{\chi - 5/2E+V/2},
\ee
where we used the fact that $\sum_v 2n_v = 2E$, and introduced the Euler characteristic of the 2-complex, \linebreak $\chi=F-E+V$. In the special case of all 4-valent vertices, $E=2V$ and the scaling reduces to
\be
Z_{{\cal C}_3}^{EPRLs} \sim N^{\chi - 9/2V}.
\ee

In the case of a standard foam with 4-valent edges, there is no restriction on the valence of the vertices, and any $n_vj$-symbol can appear. 
To estimate the edge amplitude. Using \Ref{B4asymp}, it is easy to see that the two sums over the $k$ intertwiners in \Ref{A4} have a single dominant contribution, scaling like $N^{-3}$ in the spins $j_f$. As for the intertwiner label $i_e$,
we trade this for coherent states following \cite{LS}, and keep track only of the rescaling of the spins.
Hence,
\be
Z_{{\cal C}_4}^{EPRLs} \sim N^{F-3E-3/2\sum_{v}(n_v-1)} = N^{\chi - 5E+V/2},
\ee
Assuming all 5-valent vertices as in the simplicial case \Ref{Z}, $2E=5V$ and the scaling reduces to
\be
Z_{{\cal C}_4}^{EPRLs} \sim N^{\chi-12V}.
\ee

These scalings have immediate applications to estimate the relative weights and divergences of different foams, and we expect them together with their extension to the full EPRL model to play an important role in future developments concerning, refining, resumming and renormalising spin foams.

\section{Conclusions}

Spin foam models for Euclidean signature are typically described performing all the group integrals and expressing the amplitudes in terms of contracted Clebsch-Gordan coefficients. In this paper we have shown how to extend this reformulation to Lorentzian models based on $\SL(2,\C)$, and more specifically to the EPRL model. The procedure has two steps. First, using the Cartan decomposition, part of the integrals can be immediately performed in terms of SU(2) coefficients, generating familiar SU(2) $nj$-symbols at the vertices, and isolating the non-compactness in boost integrals associated with the edges of the spin foam. This gives a geometric picture of the Lorentzian EPRL model in terms of Euclidean polytopes linked by boosts -- weighted by the Immirzi parameter, and makes manifest and exact the separation of intrinsic and extrinsic variables emerging at the saddle point analysis of \cite{BarrettLorAsymp}. Second, using a formula derived in \cite{Kerimov} to explicitly perform the boost integrals, in terms of finite sums of Gamma functions.
The formulation makes the numerical evaluation of the model much faster, but also allows to develop new analytic tools, exploiting the explicit factorisation in SU(2) amplitudes and the off-shell behaviour of the virtual Lorentz irreps, or asymptotic properties of the Gamma functions resolving the edge amplitude, or using formulas like \Ref{B3spinors} to apply the spinorial techniques that have already led to a number of useful results in spin foam models \cite{EteraHoloEucl,IoHolo,IoWolfgang,FreidelPachner14,Hnybida:2015ioa}.

In the course of our analysis we also considered a simplified model in which only the intertwiners are boosted, and not the spins. This is a strong truncation, motivated mainly by the desire to streamline some calculations, but the numerical investigations performed show that the main contribution to the amplitudes come from the minimal `spin-diagonal' configurations, thus suggesting that the simplified model may be a good approximation to the full model in certain regimes.
Another result of our numerics is that the half-edge dipole amplitude in the `spin-diagonal' configurations is peaked on diagonal intertwiners and falls off like $N^{-3/2}$, for arbitrary valence (except 3 when a degenerate case also appears), confirming an analytic result obtained in \cite{Puchta:2013lza} using a saddle point calculation. We then used these scalings to estimate the overall scaling of two different classes of generic foams.
The various peakedness properties are found to depend sensibly on the Immirzi parameter $\g$, and the generic trend is that the peaks are sharper for small $\g$, and broader for large $\g$.

The analysis pursued here has been also the occasion to review and at places extend results of the literature on $\SL(2,\C)$ coefficients for unitary irreps of the principal series. We have showed how the phases for the boost matrix elements can be chosen to guarantee reality of the edge amplitudes and more in general of integrals of tensor products, and characterise and compare the various phase choices in the literature. We have provided a definition of generalised $\SL(2,\C)$ Clebsch-Gordan coefficients, and the $\SL(2,\C)$ equivalent of Wigner's $3jm$ symbol. 

We hope that our results provide new stimulus to improve the numerical and analytic understanding of the EPRL model and of Lorentzian models in general, with physical calculations like radiative corrections \cite{Riello13} or tunnelling amplitudes \cite{IoCarloTunnelling16} in mind. 


\subsection*{Acknowledgments}
I am thankful to Sergei Alexandrov, Andrea Giusti, Jonathan Engle, Hongguang Liu, Etera Livine, Alejandro Perez, Antonia Zipfel and Carlo Rovelli for many helpful discussions, Marco Fanizza and Pietro Don\`a for the improvements leading to v2 of the paper, in particular understanding the reality of the Clebsch-Gordan conditions derived in \cite{Kerimov75}, and finally Art\"em Starodubtsev for providing me with a copy of \cite{Kerimov75}.

\appendix
\setcounter{equation}{0}
\renewcommand{\theequation}{\Alph{section}.\arabic{equation}}

\section{Boost matrix elements for general unitary irreps \label{AppBoost}} 

To provide an explicit parametrization of the $\SL(2,\C)$ unitary irreps, it is convenient to use the Cartan decomposition 
$h = u e^{\f r2\s_3} v^{-1},$ and write
\be
D^{(\r,k)}_{jmln}(h) = \sum_p D^{(j)}_{mp}(u) d^{(\r,k)}_{jlp}(r) D^{(l)}_{pn}(v^{-1}),
\ee
where $D^{(j)}_{mn}(u)$ are the Wigner matrices for SU(2), and 
\be
D^{(\r,k)}_{jmln}(e^{\f r2\s_3}) \equiv \d_{mn} D^{(\r,k)}_{jmlm}(e^{\f r2\s_3}):=\d_{mn} d^{(\r,k)}_{jlm}(r)
\ee
are the matrix elements of a boost in the $z$ direction. The SU(2) Wigner matrices can be found in all SU(2) literature. The $z$-boost matrices have also been extensively studied (e.g. \cite{Naimark,Strom,Ruhl,RashidBoost}), 
and are canonical up to a phase. 
This phase difference turns out to be important when evaluating Clebsch-Gordan coefficients, in particular 
the phase can be chosen as to guarantee that all dipole amplitudes $B_n$, and in general all group-average integrals of tensor products, are real.
Since this is a delicate point, we discuss it in details, reviewing the choices present in the literature.

The $d^{(\r,k)}$ matrix elements can be written in integral form (see e.g. \cite{Ruhl}), 
\be\label{dRuhl}
d^{(\r,k)}_{jlp}(r) = e^{i\psi^\r_{jl}} \sqrt{d_j}\sqrt{d_l}\int_0^1\f{dt}{d(r,t)^{1-i\r}} d^{(j)}_{kp}(2t-1) d^{(l)}_{kp}\left(2\f{t e^{-r}}{d(r,t)}-1\right),
\ee
where $e^{i\psi^\r_{jl}}$ parametrises the freedom in the phase,
\be\label{ddef}
d(r,t) = te^{-r} + (1-t)e^r,
\ee
and $d^{(j)}_{mn}(2t-1)$ are the little Wigner matrices with $2t-1 = \cos\b$, explicitly 
\be\nn
d^{(j)}_{mn}(\cos\b) = \sum_k (-1)^{k+j-n} \f{\sqrt{(j+m)!(j-m)!(j+n)!(j-n)!}}{k!(k+m+n)!(j-m-k)!(j-n-k)!} (\cos\f\b2)^{2k+m+n}(\sin\f\b2)^{2j-2k-m-n}.
\ee
Using this expression, the integral over $t$ can be given in terms of hypergeometric functions:
\begin{align}\label{dhyper}
d^{(\r,k)}_{jlp}(r) &= e^{i\psi^\r_{jl}}(-1)^{j-l}\f{\sqrt{d_j}\sqrt{d_l}}{(j+l+1)!}  
\left[(j + k)! (j - k)! (j + p)! (j - p)! (l + k)! (l - k)! (l + p)! (l - p)!\right]^{1/2} 
\\ \nn & \times e^{-(k-i\r +p+1)r} \sum_{s,t} \f{  (-1)^{s+t} e^{- 2 t r} (k + p + s + t)! (j + l - k - p - s - t)!}{s! (j - k - s)! (j - p - s)! (k + p + s)! t! (l - k - t)! (l - p - 
    t)! (k + p + t)!} \\ \nn & \hspace{3cm} \times \ {}_2F_1[l+1-i\r,k+p+1+s+t,j+l+2,1-e^{-2r}],
\end{align}
which can also be rewritten as an infinite series in powers of $e^{-r}$ \cite{RashidBoost}. 
As used in the main text, there is only one summation in the simple case $j=k$ (corresponding to $s=0$), and only one term in the ``diagonal" elements $l=j$ (corresponding to $t=0$).

The matrix elements have a number of symmetries under conjugation, sign flips and permutations, see \cite{Ruhl}.\footnote{The phase $\psi^\r_{jl}=0$ in \cite{Ruhl}. We also adapted the notation of Ruhl's monograph to the one most used in the modern literature, that is $\r=\r_{\rm Ruhl}/2, k=m_{\rm Ruhl}/2$. For experts in spin foams, we point out that the $k$ mapping is opposite to the one used in \cite{BarrettLorAsymp}, leading to a $\g\mapsto -\g$ flip in the amplitudes. Finally, notice also that Ruhl's parametrization of Wigner's $d^{(j)}$ matrices differs by a factor $(-1)^{m-n}$ from the one most commonly used nowadays \cite{Varshalovich}. This does not affect however the explicit evaluation \Ref{dhyper}.}
The one that is relevant to report here for the reality of the dipole amplitudes is the symmetry linking complex conjugation to the parity map $m\mapsto -m$: 
\be\label{brdd-}
\overline{d^{(\r,k)}_{jlm}(r)} = e^{-2i\psi^\r_{jl}} \, \f{\G(j+i\r+1)}{\G(j-i\r+1)}\f{\G(l-i\r+1)}{\G(l+i\r+1)}\, d^{(\r,k)}_{jl-m}(r).
\ee
Using the property of the Wigner matrices relating conjugation to parity, namely
\be\label{DbarSU2App}
\overline{{D}^{(j)}_{mn}(g)} = (-1)^{m-n} D^{(j)}_{-m-n}(g),
\ee  
\Ref{brdd-} implies 
\be\label{DbarDApp}
\overline{D^{(\r,k)}_{jmln}} = e^{2i\a^\r_{jl}} (-1)^{m-n} D^{(\r,k)}_{j-ml-n}
\ee
with
\be
e^{2i\a^\r_{jl}} = e^{-2i\psi^\r_{jl}} \, \f{\G(j+i\r+1)}{\G(j-i\r+1)}\f{\G(l-i\r+1)}{\G(l+i\r+1)}.
\ee

Let us now study the reality of the dipole amplitudes.
Starting from \Ref{intDDDbar} in the main text, and taking its complex conjugate, we get
\begin{align}\label{CCreal}
\int \overline{D^{(\r_1,k_1)}_{j_1m_1l_1n_1}} \overline{D^{(\r_2,k_2)}_{j_2m_2l_2 n_2}} D^{(\r_3,k_3)}_{j_3m_3l_3n_3} &= 
e^{2i\left(\a^{\r_1}_{j_1l_1}+\a^{\r_2}_{j_2l_2}-\a^{\r_3}_{j_3l_3}\right)} (-1)^{m_1+m_2-m_3-n_1-n_2+n_3} \\\nn
& \qquad \times \int D^{(\r_1,k_1)}_{j_1-m_1l_1-n_1} D^{(\r_2,k_2)}_{j_2-m_2l_2 -n_2} \overline{D^{(\r_3,k_3)}_{j_3-m_3l_3-n_3}} = \\\nn
& \hspace{-3cm} = e^{2i\left(\a^{\r_1}_{j_1l_1}+\a^{\r_2}_{j_2l_2}-\a^{\r_3}_{j_3l_3}\right)}  (-1)^{m_1+m_2-m_3-n_1-n_2+n_3} 
\, \bar{ \chi}(j_1,j_2,j_3) \chi(l_1,l_2,l_3)  C^{j-m}_{j_1-m_1j_2-m_2} C^{l_3-n_3}_{l_1-n_1l_2-n_2}
\\\nn
& \hspace{-3cm} = e^{2i\left(\a^{\r_1}_{j_1l_1}+\a^{\r_2}_{j_2l_2}-\a^{\r_3}_{j_3l_3}\right)}  (-1)^{J-L} 
\, \bar{ \chi}(j_1,j_2,j_3) \chi(l_1,l_2,l_3)  C^{jm}_{j_1m_1j_2m_2} C^{l_3n_3}_{l_1n_1l_2n_2},
\end{align}
where in the last line we used \Ref{sign-flip} and the recoupling conditions on the magnetic indices.
Looking at the right-hand side of \Ref{intDDDbar} we see that
the integrals are generically complex, and that a sufficient and minimal choice for their reality is
\be\label{alfawanted}
e^{2i\a^\r_{jl}} = (-1)^{j-l}.
\ee
This property can be obtained choosing the phase 
\be\label{alfa}
e^{i\psi^\r_{jl}} = (-1)^{-\f{j-l}2} e^{i\Phi^\r_j}e^{-i\Phi^\r_l} = (-1)^{-\f{j-l}2} \f{\G(j+i\r+1)}{|\G(j+i\r+1)|}\f{\G(l-i\r+1)}{|\G(l+i\r+1)|}, 
\ee
which leads to the expression \Ref{dsimple} used in the main text.\footnote{The minus sign in front of $\f{j-l}2$ here may look confusing at first, compared to \Ref{dsimple}, but notice the factor $(-1)^{j-l}$ that pops up in going from the integral representation \Ref{dRuhl} to the hypergeometric representation.}
Accordingly,
\be\label{bardd-}
\overline{d^{(\r,k)}_{jlm}(r)} = (-1)^{j-l} d^{(\r,k)}_{jl-m}(r),
\ee
and
\be\label{DbarDApp1}
\overline{D^{(\r,k)}_{jmln}} = (-1)^{j-l +m-n} D^{(\r,k)}_{j-ml-n}.
\ee
Using then \Ref{DbarDApp1}, one shows in the same way that all invariant tensors obtained from group averaging are real, in particular the dipole amplitudes \Ref{Bn} defined in the main text.
With this choice, it also follows immediately (specialising \Ref{CCreal} to $j_i=l_i$) that 
$\chi(j_i)$ is either real or purely imaginary.

On the other hand, the phase choice in Ruhl's monography is $\psi^\r_{jl}=0$, in  agreement with the original phase conventions by Naimark \cite{Naimark}.
While this parametrization has a simpler integral expression \Ref{dRuhl}, it has the disadvantage of giving complex dipole amplitudes.
It is immediate to see that the two phase choices $\psi^{\r}_{jl}=0$ and \Ref{alfa} are related by a unitary transformation, thus preserving the faithfulness of the representation and making our phase choice \Ref{alfa} perfectly admissible. 
Furthermore, our choice merely adds a factor $(-1)^{-\f{j-l}2}$ to the one of \cite{Strom,vongDuc1967}, 
\be\label{phaseStrom}
e^{i\psi^\r_{jl}} = e^{i\Phi^\r_j} e^{-i\Phi^\r_l} =  
\f{\G(j+i\r+1)}{|\G(j+i\r+1)|}  \f{\G(l-i\r+1)}{|\G(l-i\r+1)|},
\ee
already largely used in the literature on Clebsch-Gordan coefficients \cite{Rashid70I,Rashid70II,Gavrilik1972,Kerimov}.
This choice is sufficient to make the $\chi_i$ either real or purely imaginary (more on this below in Section B.2), on the other hand, $e^{2i\a^\r_{jl}} = 1$ in \Ref{DbarDApp} and the dipole amplitudes are also real or purely imaginary.

Finally, notice that the choice \Ref{DbarDApp1} corresponds to taking the parity tensor to be real and coincident with the SU(2) one, that is,
\be
\eps^{(\r,k)}_{jmj'm'} = (-1)^{j-m} \d_{jj'} \d_{m,-m'},
\ee
unlike in \cite{Ruhl} where it carries the phase here reabsorbed in $\psi^\r_{jl}$.

For completeness, we report also the symmetry properties relating negative irrep labels to the positive ones:
\be\label{Dsyms}
\qquad d^{(-\r,k)}_{jlm}(r) = (-1)^{j-l} \overline{d^{(\r,k)}_{jlm}(r)},
\qquad d^{(\r,-k)}_{jlm}(r) = d^{(\r,k)}_{jl-m}(r).
\ee
We respect to Ruhl's choice $\psi^\r_{jl}=0$, the first has picked up an extra phase $(-1)^{j-l} $, and the second is unchanged.
Using these two properties and \Ref{bardd-}, the property \Ref{d-d} in the main text follows.

\section{Clebsch-Gordan coefficients, definitions and conventions}

\subsection{(Generalised) SU(2) Clebsch-Gordan coefficients}\label{AppSU2}
To fix our notations and conventions, we briefly review here some properties of Clebsch-Gordan coefficients for SU(2).
We restrict attention to those more directly relevant to the calculations of the paper, and we refer the reader to \cite{Varshalovich} for a complete list of symmetries and properties of these symbols.

We use the standard \cite{Varshalovich} phase conventions for the SU(2) Clebsch-Gordan coefficients $C^{jm}_{j_1m_1j_2m_2}$, and their relation to Wigner's $3jm$-symbols given by
\be
\Wthree{j_1}{j_2}{j}{m_1}{m_2}{m}  = \f{(-1)^{j_1-j_2-m}}{\sqrt{d_j}} C^{j-m}_{j_1m_1j_2m_2},
\ee
which has the advantage of more symmetric behaviour under permutations and sign flips. For instance, a property we will use below is
\be\label{sign-flip}
\Wthree{j_1}{j_2}{j}{-m_1}{-m_2}{-m} = (-1)^{j_1+j_2+j_3} \Wthree{j_1}{j_2}{j}{m_1}{m_2}{m}.
\ee
The generalised  coefficients for the coupling of three angular momenta are given by
\be
C^{j_{12}jm}_{j_1m_1j_2m_2j_3m_3} 
 = C^{j_{12}m_{12}}_{j_1m_1j_2m_2} C^{jm}_{j_{12}m_{12}j_3m_3}, 
\ee
where $m_{12} = m_1+m_2$, and correspondingly one can define a generalised Wigner $4jm$-symbol as
\begin{align}\label{CG4}
\Wfour{j_1}{j_2}{j_3}{j}{m_1}{m_2}{m_3}{m}{j_{12}}  &= 
\f{(-1)^{j_1-j_2+j_3+m}}{\sqrt{d_{j_{12}}}\sqrt{d_j}} C^{j_{12}j,-m}_{j_1m_1j_2m_2j_3m_3} \\\nn &
= \sum_{m_{12}} (-1)^{j_{12}-m_{12}}  \Wthree{j_1}{j_2}{j_{12}}{m_1}{m_2}{m_{12}} \Wthree{j_{12}}{j_3}{j}{-m_{12}}{m_3}{m}.
\end{align}
The scheme can be straightforwardly extended to an arbitrary number $n$ of external legs, iterating the step above and introducing $n-3$ extra recoupling labels $j_{12}, j_{123},$ etc. We can define in this way the general symbol 
\begin{align}\label{3jm-gen}
 \left(\begin{array}{c} j_i \\ m_i \end{array}\right)^{(\{i\})} =& \sum_{m_{12},m_{123},\ldots}
  (-1)^{j_{12}-m_{12}+j_{123}-m_{123}+\ldots}  \\\nn &\qquad \times
  \Wthree{j_1}{j_2}{j_{12}}{m_1}{m_2}{m_{12}} \Wthree{j_{12}}{j_3}{j_{123}}{-m_{12}}{m_3}{m_{123}} \Wthree{j_{123}}{j_4}{j}{-m_{123}}{m_4}{m} \ldots,
\end{align}
where  $\{i\}$ stands for the set of $n-3$ virtual spins. In the main text we used this symbol also for the cases $n=3,4$, as a shorthand notation.

With these definitions and the conventional definitions of Wigner's matrices for SU(2), 
$D^{(j)}_{mn}(g)$,
we have
\begin{align}\label{intD3App}
& \int dg D^{j_1}_{m_1n_1} D^{j_2}_{m_2 n_2} D^{j_3}_{m_3 n_3} = \Wthree{j_1}{j_2}{j_3}{m_1}{m_2}{m_3} \Wthree{j_1}{j_2}{j_3}{n_1}{n_2}{n_3}, 
\\ & \label{intD4}
\int dg D^{j_1}_{m_1n_1} D^{j_2}_{m_2 n_2} D^{j_3}_{m_3 n_3} D^{j_4}_{m_4 n_4} = 
\sum_{j_{12}} d_{j_{12}} \Wfour{j_1}{j_2}{j_3}{j_4}{m_1}{m_2}{m_3}{m_4}{j_{12}} \Wfour{j_1}{j_2}{j_3}{j_4}{n_1}{n_2}{n_3}{n_4}{j_{12}},
\end{align}
and so on for more general symbols. 

We will also use the fact that the symbols satisfy the following orthogonality properties,
\begin{align}
& \sum_{m_1,m_2} \Wthree{j_1}{j_2}{j_3}{m_1}{m_2}{m_3} \Wthree{j_1}{j_2}{l_3}{m_1}{m_2}{n_3} = \f{\d_{j_{3}l_{3}}\d_{m_3n_3}}{d_{j_3}}, \\ \label{CG4ortho}
& \sum_{m_i} \Wfour{j_1}{j_2}{j_3}{j}{m_1}{m_2}{m_3}{m}{j_{12}} \Wfour{j_1}{j_2}{j_3}{j}{m_1}{m_2}{m_3}{m}{l_{12}} = \f{\d_{j_{12}l_{12}}}{d_{j_{12}}}.
\end{align}
Notice from the second equation above that unlike the basic $3jm$-symbol, the $4jm$-symbol we defined is not normalised. 
In the literature one also finds a normalised $4jm$-symbol, obtained multiplying the right-hand side of \Ref{CG4} by $\sqrt{d_{j_{12}}}$. We chose the non-normalised convention because it is the one that corresponds to a 4-valent node in the SU(2) graphical calculus (e.g. \cite{Varshalovich}).

\subsection{(Generalised) $\SL(2,\C)$ Clebsch-Gordan coefficients}\label{AppSL2C}

%
As in the SU(2) case, the $\SL(2,\C)$ Clebsch-Gordan coefficients have limited symmetries, and dimensional factors may appear under permutations of the labels.
For instance, from \Ref{intDDD} in the main text, it follows that
\be\label{chiswap}
\chi(j_1,j_3,j_2) = (-1)^{J+f(\r_i, k_i)} \sqrt{\f{d_{j_3}}{d_{j_2}}} \chi(j_1,j_2,j_3),
\ee
with $f(\r_i,k_i)$ depending on the phase conventions for the $\chi$'s. 
It can be then useful to define a more symmetric symbol, an $\SL(2,\C)$ version of Wigner's $3jm$-symbol, as
\be\label{WignerSL2C}
\Wthree{\r_1,k_1}{\r_2,k_2}{\r_3,k_3}{j_1}{j_2}{j_3} = (-1)^{j_1-j_2+j_3} \sqrt{d_{j_3}} \, \chi(j_1,j_2,j_3).
\ee
This symbol is invariant under permutations up to a phase, for instance,
\be
\Wthree{\r_1,k_1}{\r_3,k_3}{\r_2,k_2}{j_1}{j_3}{j_2} = (-1)^{-j_1+j_2+j_3+f(\r_i,k_i)} \Wthree{\r_1,k_1}{\r_2,k_2}{\r_3,k_3}{j_1}{j_2}{j_3}.
\ee
With this definition, starting from \Ref{intDDDbar} and using \Ref{DbarDApp1}, we have
\begin{align}\label{intDDDApp}
\int dh D^{(\r_1,k_1)}_{j_1m_1l_1n_1}D^{(\r_2,k_2)}_{j_2m_2l_2 n_2} D^{(\r_3,k_3)}_{j_3m_3l_3n_3} &= 
\Wthree{\r_1,k_1}{\r_2,k_2}{\r_3,k_3}{j_1}{j_2}{j_3} \Wthree{\r_1,k_1}{\r_2,k_2}{\r_3,k_3}{l_1}{l_2}{l_3} \\\nn &\hspace{.5cm} \times
\Wthree{j_1}{j_2}{j_3}{m_1}{m_2}{m_3} \Wthree{l_1}{l_2}{l_3}{n_1}{n_2}{n_3},
\end{align}
in analogy with \Ref{intD3App}.

For the generalised coefficient coming from the recoupling of three irreps, we proceed as in the SU(2) case, and define
\begin{align}
C^{\r_{12} k_{12} \r k j m}_{\r_1k_1j_1m_1\r_2k_2j_2m_2\r_3k_3j_3m_3} &:= \nn
\sum_{j_{12}} C^{\r_{12} k_{12} j_{12} m_{12}}_{\r_1k_1j_1m_1\r_2k_2j_2m_2} C^{\r k j m}_{\r_{12} k_{12} j_{12} m_{12}\r_3k_3j_3m_3}\\
&= \sum_{j_{12}} \chi(j_1,j_2,j_{12}) \chi(j_{12},j_3,j) C^{j_{12}m_{12}}_{j_1m_1j_2m_2} C^{jm}_{j_{12}m_{12}j_3m_3},
\end{align}
where $j_1+j_2\geq j_{12} \geq {\rm max}\{|k_{12}|,|j_1-j_2|\}$. 
Changing the recoupling basis goes as for SU(2), with the relevant $\{6\r,6k\}$-symbol given using \Ref{intDDDApp} and \Ref{WignerSL2C} by
\begin{align}\nn
\{6\r,6k\} = \sum_{j_i} \{6j_i\} &
\Wthree{\r_1,k_1}{\r_2,k_2}{\r_3,k_3}{j_1}{j_2}{j_3}\Wthree{\r_1,k_1}{\r_5,k_5}{\r_6,k_6}{j_1}{j_5}{j_6} \\\nn & \times
\Wthree{\r_4,k_4}{\r_2,k_2}{\r_6,k_6}{j_4}{j_2}{j_6}\Wthree{\r_4,k_4}{\r_5,k_5}{\r_3,k_3}{j_4}{j_5}{j_3}.
\end{align}
The factorisation of this symbol is just a toy example of the factorisation \Ref{Zgen} for the case of a  tetrahedral vertex graph.

The relation between the generalised coefficient and the group-averaged tensor product is obtained as in the SU(2) case,
\begin{align}\nn
& \int dh D^{(\r_1,k_1)}_{j_1m_1l_1n_1}D^{(\r_2,k_2)}_{j_2m_2l_2 n_2}D^{(\r_3,k_3)}_{j_3m_3l_3 n_3} \overline{D^{(\r_4,k_4)}_{j_4m_4l_4n_4}} = 
\int_{-\infty}^\infty d\rho_{12} \sum_{k_{12}} 4(\r_{12}^2+k_{12}^2) 
 \\\nn &\qquad \times\,
\sum_{j_{12},l_{12}} \bar{C}^{j_{12}m_{12}}_{j_1m_1j_2m_2} \bar{C}^{j_4m_4}_{j_{12}m_{12}j_3m_3}
C^{l_{12}n_{12}}_{l_1n_1l_2n_2} C^{l_4n_4}_{l_{12}n_{12}l_3n_3}
\overline{\chi}(j_1,j_2,j_{12}) \overline{\chi}(j_{12},j_3,j_4)  \chi(l_1,l_2,l_{12}) \chi(l_{12},l_3,l_4),
 \end{align}
from which \Ref{B4CG} in the main text follows using \Ref{DbarDApp1}. 

We conclude with a discussion on the phases of the coefficients $\chi$.
With the choice \Ref{chiKer} based on \cite{Kerimov}, the coefficients are always real. The downside is that we do not know explicitly the behaviour under permutations \Ref{chiswap}. Numerical investigations show that the function $f(\r_i,k_i)$ depends non-trivially on $\r_i$, and we did not try to evaluate it analytically, because it is irrelevant to our scopes: it never enter permutations of the dipole amplitudes, where each $\chi$ appears with a conjugate $\chi$ with same $(\r_i,k_i)$ labels.
The only permutation that is easy to identify is the swap of the first two entries, which gives
\begin{equation}\label{AppPerm}
\chi(j_2,j_1,j_3) = (-1)^{J+K} \chi(j_1,j_2,j_3).
\end{equation}

An alternative procedure to fix the phase was proposed in \cite{Rashid70II}, based on the use of recursion relations to generate the $\chi$'s, and on fixing by hand the phase of the seed coefficients so that they are always real and, when possible, positive. The authors of \cite{Rashid70II} use Naimark's basis amended by \Ref{Phidef}. With this choice, the boost raising and lowering coefficients are generically complex. Hence, even choosing real seeds, they obtain $\chi$'s which are either real or purely imaginary. The phases under permutations are then explicitly known, they do not depend on $\r_i$, but they do depend on whether the $k_i$'s are triangular or not, because this changes the choice of seed (For triangular $k_i$'s, one can take the minimal coefficients $k_i=j_i$ as seed, whereas for non-triangular $k_i$'s one minimises the $j$ of the exceeding $k$ then completes to one of the smallest possible spin configurations available). Notice that with this procedure it is not possible to give a closed formula for the phase.

Finally, real raising and lowering boost matrix elements can be obtained rescaling the basis vectors by $(-1)^{(k+j)/2}$, as we did in our paper.
This leads to a straightforward modification of the recursion relations derived in \cite{Rashid70II}, and once this is taken into account, the latter are applicable to the expression \Ref{chiKer}. Of course, the phase generated in this way will be consistent with the phase of \Ref{chiKer} and not with that of \cite{Rashid70II}.

\newpage
%
\section{Some explicit values and comparative times}\label{AppTables}
%
\begin{table}[ht!]
\begin{center}\includegraphics[width=15cm]{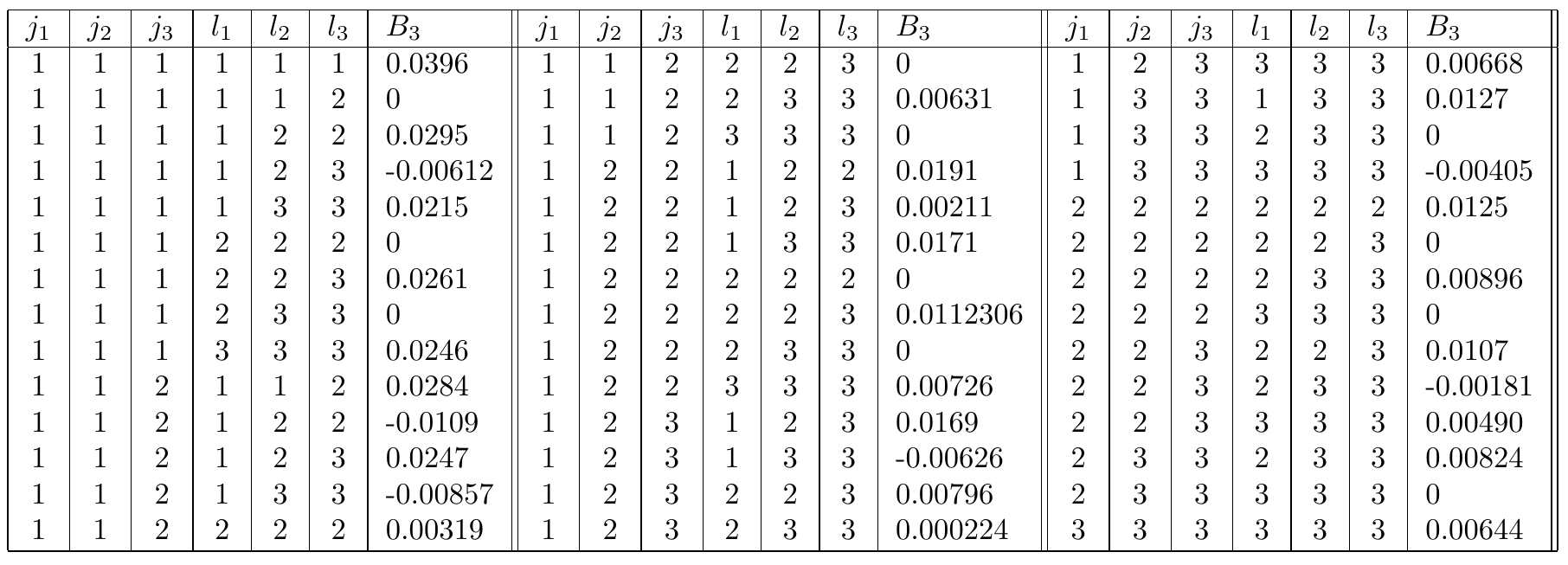}\end{center}
\caption{ {\small{\emph{Some explicit integer values of $B_3(\g j_i, j_i; j_i,l_i)$, for $\g=1.2$, obtained with Mathematica. Generating this table takes about 10 seconds with the integral formula and less than half a second with the finite sums formula.}}}}
\end{table}
\begin{table}[ht!]
\begin{center}\begin{tabular}{|c|c|c|c|c|c|l||c|c|c|c|c|c|l|} \hline
 $j_1$ & $j_2$ & $j_3$ & $l_1$ & $l_2$ & $l_3$ & $B_3$	&	 $j_1$ & $j_2$ & $j_3$ & $l_1$ & $l_2$ & $l_3$ & $B_3$ \\
\hline
 1/2  & 1/2 & 1 & 1/2 & 1/2 & 1 & 0.0765	&	 1/2 & 3/2 & 2 & 3/2 & 3/2 & 3 & 0.0223 \\
 1/2 & 1/2 & 1 & 1/2 & 3/2 & 2 & 0.0676	&	 3/2 & 3/2 & 2 & 3/2 & 3/2 & 2 & 0.0177 \\
 1/2 & 1/2 & 1 & 3/2 & 3/2 & 2 & 0		&	 3/2 & 3/2 & 2 & 3/2 & 3/2 & 3 & 0 \\
 1/2 & 1/2 & 1 & 3/2 & 3/2 & 3 & 0.0671	&	 3/2 & 3/2 & 3 & 3/2 & 3/2 & 3 & 0.0164 \\
 1/2 & 3/2 & 2 & 1/2 & 3/2 & 2 & 0.0312 	&	 1/2 & 3/2 & 2 & 3/2 & 3/2 & 2 & 0.0116  	\\
\hline
\end{tabular}\end{center}
\caption{\label{TableCG3h} {\small{\emph{Some explicit half-integer values of $B_3(\g j_i, j_i; j_i,l_i)$, for $\g=1.2$, obtained with Mathematica. Generating this table takes about 1 second with the integral formula and less than $.1$ seconds with the finite sums formula.}}} }
\end{table}
\begin{table}[ht!]
\begin{center}\includegraphics[width=16cm]{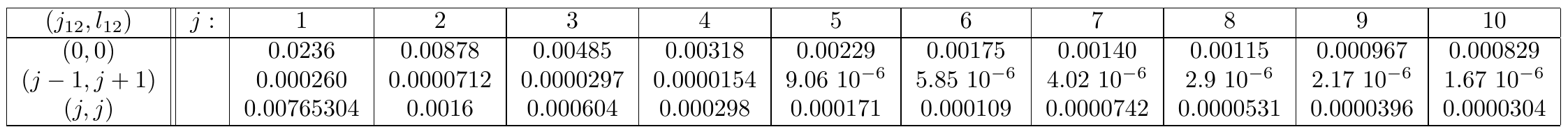}\end{center}
\caption{ {\small{\emph{Some explicit integer values of $B_4(\g j_i, j_i; j_i,j_i)$, for $\g=1.2$, obtained with Mathematica. Generating this table takes about 90 minutes with the integral formula and less than 6 minutes with the finite sums formula.} }}}
\end{table}

\providecommand{\href}[2]{#2}\begingroup\raggedright\endgroup


\begin{thebibliography}{10}


\bibitem{PerezLR}
A.~Perez, {\it {The Spin Foam Approach to Quantum Gravity}},  Living Rev.Rel.
  {\bf 16} (2013) 3 [\href{http://arXiv.org/abs/1205.2019}{{\tt 1205.2019}}].

\bibitem{RovelliVidotto}
C.~Rovelli and F.~Vidotto, {\em {Covariant Loop Quantum Gravity}}.
\newblock Cambridge Monographs on Mathematical Physics. Cambridge University
  Press, 2014.

\bibitem{EPRL}
J.~Engle, E.~Livine, R.~Pereira and C.~Rovelli, {\it {LQG vertex with finite
  Immirzi parameter}},  Nucl.Phys. {\bf B799} (2008) 136--149
  [\href{http://arXiv.org/abs/0711.0146}{{\tt 0711.0146}}].

\bibitem{EPR}
J.~Engle, R.~Pereira and C.~Rovelli, {\it {The Loop-quantum-gravity
  vertex-amplitude}},  Phys.Rev.Lett. {\bf 99} (2007) 161301
  [\href{http://arXiv.org/abs/0705.2388}{{\tt 0705.2388}}].

\bibitem{LS}
E.~R. Livine and S.~Speziale, {\it {A New spinfoam vertex for quantum
  gravity}},  Phys.Rev. {\bf D76} (2007) 084028
  [\href{http://arXiv.org/abs/0705.0674}{{\tt 0705.0674}}].

\bibitem{LS2}
E.~R. Livine and S.~Speziale, {\it {Consistently Solving the Simplicity
  Constraints for Spinfoam Quantum Gravity}},  Europhys.Lett. {\bf 81} (2008)
  50004 [\href{http://arXiv.org/abs/0708.1915}{{\tt 0708.1915}}].

\bibitem{FK}
L.~Freidel and K.~Krasnov, {\it {A New Spin Foam Model for 4d Gravity}},
  Class.Quant.Grav. {\bf 25} (2008) 125018
  [\href{http://arXiv.org/abs/0708.1595}{{\tt 0708.1595}}].

\bibitem{BarrettEPRAsymp}
J.~W. Barrett, R.~Dowdall, W.~J. Fairbairn, H.~Gomes and F.~Hellmann, {\it
  {Asymptotic analysis of the EPRL four-simplex amplitude}},  J.Math.Phys. {\bf
  50} (2009) 112504 [\href{http://arXiv.org/abs/0902.1170}{{\tt 0902.1170}}].

\bibitem{BarrettLorAsymp}
J.~W. Barrett, R.~Dowdall, W.~J. Fairbairn, F.~Hellmann and R.~Pereira, {\it
  {Lorentzian spin foam amplitudes: Graphical calculus and asymptotics}},
  Class.Quant.Grav. {\bf 27} (2010) 165009
  [\href{http://arXiv.org/abs/0907.2440}{{\tt 0907.2440}}].

\bibitem{KKL}
W.~Kaminski, M.~Kisielowski and J.~Lewandowski, {\it {Spin-Foams for All Loop
  Quantum Gravity}},  Class.Quant.Grav. {\bf 27} (2010) 095006
  [\href{http://arXiv.org/abs/0909.0939}{{\tt 0909.0939}}].

\bibitem{CarloGenSF}
Y.~Ding, M.~Han and C.~Rovelli, {\it {Generalized Spinfoams}},  Phys.Rev. {\bf
  D83} (2011) 124020 [\href{http://arXiv.org/abs/1011.2149}{{\tt 1011.2149}}].

\bibitem{RovelliMagic}
C.~Rovelli, {\it {Discretizing parametrized systems: the magic of
  Ditt-invariance}},  \href{http://arXiv.org/abs/1107.2310}{{\tt 1107.2310}}.

\bibitem{Dittrich:2014mxa}
B.~Dittrich, S.~Mizera and S.~Steinhaus, {\it {Decorated tensor network
  renormalization for lattice gauge theories and spin foam models}},  New J.
  Phys. {\bf 18} (2016), no.~5 053009
  [\href{http://arXiv.org/abs/1409.2407}{{\tt 1409.2407}}].

\bibitem{Dittrich:2014ala}
B.~Dittrich, {\it {The continuum limit of loop quantum gravity - a framework
  for solving the theory}},  \href{http://arXiv.org/abs/1409.1450}{{\tt
  1409.1450}}.

\bibitem{BahrSteinhaus15}
B.~Bahr and S.~Steinhaus, {\it {Investigation of the Spinfoam Path integral
  with Quantum Cuboid Intertwiners}},  Phys. Rev. {\bf D93} (2016), no.~10
  104029 [\href{http://arXiv.org/abs/1508.07961}{{\tt 1508.07961}}].

\bibitem{Bonzom:2012hw}
V.~Bonzom, R.~Gurau and V.~Rivasseau, {\it {Random tensor models in the large N
  limit: Uncoloring the colored tensor models}},  Phys. Rev. {\bf D85} (2012)
  084037 [\href{http://arXiv.org/abs/1202.3637}{{\tt 1202.3637}}].

\bibitem{Benedetti:2011nn}
D.~Benedetti and R.~Gurau, {\it {Phase Transition in Dually Weighted Colored
  Tensor Models}},  Nucl. Phys. {\bf B855} (2012) 420--437
  [\href{http://arXiv.org/abs/1108.5389}{{\tt 1108.5389}}].

\bibitem{Carrozza:2013wda}
S.~Carrozza, D.~Oriti and V.~Rivasseau, {\it {Renormalization of a SU(2)
  Tensorial Group Field Theory in Three Dimensions}},  Commun. Math. Phys. {\bf
  330} (2014) 581--637 [\href{http://arXiv.org/abs/1303.6772}{{\tt
  1303.6772}}].

\bibitem{Geloun:2016qyb}
J.~Ben~Geloun, R.~Martini and D.~Oriti, {\it {Functional Renormalisation Group
  analysis of Tensorial Group Field Theories on $\mathbb{R}^d$}},  Phys. Rev.
  {\bf D94} (2016), no.~2 024017 [\href{http://arXiv.org/abs/1601.08211}{{\tt
  1601.08211}}].

\bibitem{IoCarloTunnelling16}
M.~Christodoulou, C.~Rovelli, S.~Speziale and I.~Vilensky, {\it {Realistic
  Observable in Background-Free Quantum Gravity: the Planck-Star
  Tunnelling-Time}},  \href{http://arXiv.org/abs/1605.05268}{{\tt 1605.05268}}.

\bibitem{Naimark}
M.~A. Naimark, {\em Linear representations of the Lorentz group}.
\newblock Elsevier, 2014 (1st Ed. Pergamon Press, 1964).

\bibitem{Ruhl}
W.~Ruhl, {\em {The Lorentz Group and Harmonic Analysis}}.
\newblock W. A. Benjamin, 1970.

\bibitem{Rashid70I}
R.~L. Anderson, R.~Raczka, M.~A. Rashid and P.~Winternitz, {\it {Clebsch-gordan
  coefficients for the coupling of sl(2,c) principal-series representations}},
  J. Math. Phys. {\bf 11} (1970) 1050--1058.

\bibitem{Rashid70II}
R.~L. Anderson, R.~Raczka, M.~A. Rashid and P.~Winternitz, {\it {Recursion and
  symmetry relations for the clebsch-gordan coefficients of the homogeneous
  lorentz group}},  J. Math. Phys. {\bf 11} (1970) 1059--1068.

\bibitem{Kerimov}
G.~A. Kerimov and I.~A. Verdiev, {\it {Clebsch-Gordan Coefficients of the
  SL(2,c) Group}},  Rept. Math. Phys. {\bf 13} (1978) 315--326.

\bibitem{Kerimov75}
  I.~A.~Verdiev, G.~A.~Kerimov and Y.~A.~Smorodinsky,
  ``Clebsch-Gordan Coefficients of the Lorentz Group,''
  Yad.\ Fiz.\  {\bf 21} (1975) 1351.

\bibitem{Puchta:2013lza}
J.~Puchta, {\it {Asymptotic of Lorentzian Polyhedra Propagator}},
  \href{http://arXiv.org/abs/1307.4747}{{\tt 1307.4747}}.

\bibitem{IoCarloCov}
C.~Rovelli and S.~Speziale, {\it {Lorentz covariance of loop quantum gravity}},
   Phys.Rev. {\bf D83} (2011) 104029
  [\href{http://arXiv.org/abs/1012.1739}{{\tt 1012.1739}}].

\bibitem{IoTwistorNet}
E.~R. Livine, S.~Speziale and J.~Tambornino, {\it {Twistor Networks and
  Covariant Twisted Geometries}},  Phys. Rev. {\bf D85} (2012) 064002
  [\href{http://arXiv.org/abs/1108.0369}{{\tt 1108.0369}}].

\bibitem{IoMiklos}
M.~Langvik and S.~Speziale, {\it {Twisted geometries, twistors and conformal
  transformations}},  Phys. Rev. {\bf D94} (2016), no.~2 024050
  [\href{http://arXiv.org/abs/1602.01861}{{\tt 1602.01861}}].

\bibitem{Bonzom:2009wm}
V.~Bonzom, {\it {From lattice BF gauge theory to area-angle Regge calculus}},
  Class.Quant.Grav. {\bf 26} (2009) 155020
  [\href{http://arXiv.org/abs/0903.0267}{{\tt 0903.0267}}].

\bibitem{BaratinOriti}
A.~Baratin and D.~Oriti, {\it {Group field theory and simplicial gravity path
  integrals: A model for Holst-Plebanski gravity}},  Phys.Rev. {\bf D85} (2012)
  044003 [\href{http://arXiv.org/abs/1111.5842}{{\tt 1111.5842}}].

\bibitem{EteraHoloEucl}
M.~Dupuis and E.~R. Livine, {\it {Holomorphic Simplicity Constraints for 4d
  Spinfoam Models}},  Class.Quant.Grav. {\bf 28} (2011) 215022
  [\href{http://arXiv.org/abs/1104.3683}{{\tt 1104.3683}}].

\bibitem{HanZhangLor}
M.~Han and M.~Zhang, {\it {Asymptotics of Spinfoam Amplitude on Simplicial
  Manifold: Lorentzian Theory}},  Class. Quant. Grav. {\bf 30} (2013) 165012
  [\href{http://arXiv.org/abs/1109.0499}{{\tt 1109.0499}}].

\bibitem{HellmannFlatness}
F.~Hellmann and W.~Kaminski, {\it {Holonomy spin foam models: Asymptotic
  geometry of the partition function}},  JHEP {\bf 1310} (2013) 165
  [\href{http://arXiv.org/abs/1307.1679}{{\tt 1307.1679}}].

\bibitem{Thiemann:2013lka}
T.~Thiemann and A.~Zipfel, {\it {Linking covariant and canonical LQG II: Spin
  foam projector}},  Class. Quant. Grav. {\bf 31} (2014) 125008
  [\href{http://arXiv.org/abs/1307.5885}{{\tt 1307.5885}}].

\bibitem{BahrOperatorSF}
B.~Bahr, F.~Hellmann, W.~Kaminski, M.~Kisielowski and J.~Lewandowski, {\it
  {Operator Spin Foam Models}},  Class.Quant.Grav. {\bf 28} (2011) 105003
  [\href{http://arXiv.org/abs/1010.4787}{{\tt 1010.4787}}].

\bibitem{Engle:2015zqa}
J.~Engle, I.~Vilenskiy and A.~Zipfel, {\it {The Lorentzian proper vertex
  amplitude: Asymptotics}},  \href{http://arXiv.org/abs/1505.06683}{{\tt
  1505.06683}}.

\bibitem{Wieland:2014nka}
W.~M. Wieland, {\it {A new action for simplicial gravity in four dimensions}},
  Class. Quant. Grav. {\bf 32} (2015), no.~1 015016
  [\href{http://arXiv.org/abs/1407.0025}{{\tt 1407.0025}}].

\bibitem{twigeo}
L.~Freidel and S.~Speziale, {\it {Twisted geometries: A geometric
  parametrisation of SU(2) phase space}},  Phys.Rev. {\bf D82} (2010) 084040
  [\href{http://arXiv.org/abs/1001.2748}{{\tt 1001.2748}}].

\bibitem{IoWolfgang}
S.~Speziale and W.~M. Wieland, {\it {The twistorial structure of loop-gravity
  transition amplitudes}},  Phys. Rev. {\bf D86} (2012) 124023
  [\href{http://arXiv.org/abs/1207.6348}{{\tt 1207.6348}}].

\bibitem{Haggard:2015yda}
H.~M. Haggard, M.~Han, W.~Kami{\'n}ski and A.~Riello, {\it {Four-dimensional
  Quantum Gravity with a Cosmological Constant from Three-dimensional
  Holomorphic Blocks}},  Phys. Lett. {\bf B752} (2016) 258--262
  [\href{http://arXiv.org/abs/1509.00458}{{\tt 1509.00458}}].

\bibitem{Baez:2001fh}
J.~C. Baez and J.~W. Barrett, {\it {Integrability for relativistic spin
  networks}},  Class. Quant. Grav. {\bf 18} (2001) 4683--4700
  [\href{http://arXiv.org/abs/gr-qc/0101107}{{\tt gr-qc/0101107}}].

\bibitem{Kaminski:2010qb}
W.~Kaminski, {\it {All 3-edge-connected relativistic BC and EPRL spin-networks
  are integrable}},  \href{http://arXiv.org/abs/1010.5384}{{\tt 1010.5384}}.

\bibitem{BarrettCraneLor}
J.~W. Barrett and L.~Crane, {\it {A Lorentzian signature model for quantum
  general relativity}},  Class.Quant.Grav. {\bf 17} (2000) 3101--3118
  [\href{http://arXiv.org/abs/gr-qc/9904025}{{\tt gr-qc/9904025}}].

\bibitem{DupuisLivine}
M.~Dupuis and E.~R. Livine, {\it {Lifting SU(2) Spin Networks to Projected Spin
  Networks}},  Phys.Rev. {\bf D82} (2010) 064044
  [\href{http://arXiv.org/abs/1008.4093}{{\tt 1008.4093}}].

\bibitem{Reisenberger:1997sk}
M.~P. Reisenberger, {\it {A Lattice world sheet sum for 4-d Euclidean general
  relativity}},  \href{http://arXiv.org/abs/gr-qc/9711052}{{\tt
  gr-qc/9711052}}.

\bibitem{EteraProj}
E.~R. Livine, {\it {Projected spin networks for Lorentz connection: Linking
  spin foams and loop gravity}},  Class.Quant.Grav. {\bf 19} (2002) 5525--5542
  [\href{http://arXiv.org/abs/gr-qc/0207084}{{\tt gr-qc/0207084}}].

\bibitem{Strom}
S.~Str{\"o}m, {\it On the matrix elements of a unitary representation of the
  homogeneous lorentz group},  Arkiv f. Fysik {\bf 29} (1965) 467--483.

\bibitem{vongDuc1967}
D.~vong Duc and N.~van Hieu, {\it On the theory of unitary representations of
  the sl(2c) group},  Acta Physica Academiae Scientiarum Hungaricae {\bf 22}
  (1967), no.~1 201--219.

\bibitem{RashidBoost}
M.~A. Rashid, {\it {Boost Matrix Elements Of The Homogeneous Lorentz Group}},
  J. Math. Phys. {\bf 20} (1979) 1514--1519.

\bibitem{Gavrilik1972}
A.~M. Gavrilik, {\it Clebsch-gordan coefficients for sl(2, c)},  Theoretical
  and Mathematical Physics {\bf 11} (1972), no.~1 342--351.

\bibitem{Alesci:2007tx}
E.~Alesci and C.~Rovelli, {\it {The Complete LQG propagator. I. Difficulties
  with the Barrett-Crane vertex}},  Phys.Rev. {\bf D76} (2007) 104012
  [\href{http://arXiv.org/abs/0708.0883}{{\tt 0708.0883}}].

\bibitem{Rennert?}
J.~Rennert, {\it {talk at Helsinki Workshop on Quantum Gravity}},  2016.

\bibitem{ConradyHnybida}
F.~Conrady and J.~Hnybida, {\it {A spin foam model for general Lorentzian
  4-geometries}},  Class.Quant.Grav. {\bf 27} (2010) 185011
  [\href{http://arXiv.org/abs/1002.1959}{{\tt 1002.1959}}].

\bibitem{IoNull}
S.~Speziale and M.~Zhang, {\it {Null twisted geometries}},  Phys.Rev. {\bf D89}
  (2014) 084070 [\href{http://arXiv.org/abs/1311.3279}{{\tt 1311.3279}}].

\bibitem{Varshalovich}
D.~A. Varshalovich, A.~N. Moskalev and V.~K. Khersonsky, {\em {Quantum Theory
  of Angular Momentum: Irreducible Tensors, Spherical Harmonics, Vector
  Coupling Coefficients, 3nj Symbols}}.
\newblock World Scientific, Singapore, 1988.

\bibitem{Bonzom:2012bn}
V.~Bonzom and E.~R. Livine, {\it {Generating Functions for Coherent
  Intertwiners}},  Class. Quant. Grav. {\bf 30} (2013) 055018
  [\href{http://arXiv.org/abs/1205.5677}{{\tt 1205.5677}}].

\bibitem{Alesci:2016dqx}
E.~Alesci, J.~Lewandowski and I.~M{\"a}kinen, {\it {Coherent 3j-symbol
  representation for the loop quantum gravity intertwiner space}},
  \href{http://arXiv.org/abs/1606.06561}{{\tt 1606.06561}}.
   
\bibitem{IoSL2Casymp}
S.~Speziale~et al., {\it {Asymptotics of SL(2,C) Clebsch-Gordan coefficients}},
   in preparation.

\bibitem{IoPoly}
E.~Bianchi, P.~Dona and S.~Speziale, {\it {Polyhedra in loop quantum gravity}},
   Phys.Rev. {\bf D83} (2011) 044035
  [\href{http://arXiv.org/abs/1009.3402}{{\tt 1009.3402}}].

\bibitem{IoCarloClaudio}
C.~Perini, C.~Rovelli and S.~Speziale, {\it {Self-energy and vertex radiative
  corrections in LQG}},  Phys. Lett. {\bf B682} (2009) 78--84
  [\href{http://arXiv.org/abs/0810.1714}{{\tt 0810.1714}}].

\bibitem{Riello13}
A.~Riello, {\it {Self-energy of the Lorentzian Engle-Pereira-Rovelli-Livine and
  Freidel-Krasnov model of quantum gravity}},  Phys.Rev. {\bf D88} (2013),
  no.~2 024011 [\href{http://arXiv.org/abs/1302.1781}{{\tt 1302.1781}}].

\bibitem{IoHolo}
M.~Dupuis, L.~Freidel, E.~R. Livine and S.~Speziale, {\it {Holomorphic
  Lorentzian Simplicity Constraints}},  J. Math. Phys. {\bf 53} (2012) 032502
  [\href{http://arXiv.org/abs/1107.5274}{{\tt 1107.5274}}].

\bibitem{FreidelPachner14}
A.~Banburski, L.-Q. Chen, L.~Freidel and J.~Hnybida, {\it {Pachner moves in a
  4d Riemannian holomorphic Spin Foam model}},  Phys. Rev. {\bf D92} (2015),
  no.~12 124014 [\href{http://arXiv.org/abs/1412.8247}{{\tt 1412.8247}}].

\bibitem{Hnybida:2015ioa}
J.~Hnybida, {\it {Spin Foams Without Spins}},
  \href{http://arXiv.org/abs/1508.01416}{{\tt 1508.01416}}.
  
\end{thebibliography}
\end{document}